\documentclass{aa}  

\usepackage{graphicx}
\usepackage{caption}
\usepackage[varg]{txfonts}
\usepackage[colorlinks=true,linkcolor=red,citecolor=blue,urlcolor=magenta]{hyperref}
\makeatletter
\renewcommand*\aa@pageof{, page \thepage{} of \pageref*{LastPage}}
\makeatother

\usepackage{color}

\usepackage{nicefrac}

\usepackage{textcomp}

\bibpunct{(}{)}{;}{a}{}{,} 
\usepackage[dvipsnames]{xcolor}

\begin{document}

\title{Probing inner and outer disk misalignments in transition disks\thanks{Based on observations collected at the European Organisation for Astronomical Research in the Southern Hemisphere under ESO programs 098.D-0488(A), 099.B-0162(F), 099.C-0667(B), 0100.C-0278(E), 0101.C-0311(B), 0101.C-0281(A,B), 0102.C-0210(A), 0102.C-0408(A,D), 0103.C-0097(A), 0103.C-0347(C), 0104.C-0567(A,C), and 106.21JR.001. ALMA Program IDs are provided in the acknowledgments.
All codes used for the data analysis, as well as example data files, are  available at \url{https://github.com/18alex96/disk_misalignments}}}
\subtitle{Constraints from VLTI/GRAVITY and ALMA observations}

\author{
A.~J.~Bohn\inst{1,2}
\and M.~Benisty\inst{3,2}
\and K.~Perraut\inst{2}
\and N.~van~der~Marel\inst{4,5,1}
\and L.~Wölfer\inst{6,10}
\and E.~F.~van~Dishoeck\inst{1,6}
\and S.~Facchini\inst{7,8}
\and C.~F.~Manara\inst{8}
\and R.~Teague\inst{9}
\and L.~Francis\inst{4}
\and J-P.~Berger\inst{2}
\and R. Garcia-Lopez\inst{11,12}
\and C.~Ginski\inst{1}
\and T.~Henning\inst{12}
\and M.~Kenworthy\inst{1}
\and S.~Kraus\inst{13} 
\and F.~M\'enard\inst{2} 
\and A.~M\'erand\inst{8}
\and L.~M.~P\'erez\inst{14}
}

\institute{Leiden Observatory, Leiden University, PO Box 9513, 2300 RA Leiden, The Netherlands\\
\email{ajbohn.astro@gmail.com}
\and Univ. Grenoble Alpes, CNRS, IPAG, F-38000 Grenoble, France  
\and Unidad Mixta Internacional Franco-Chilena de Astronomía (CNRS, UMI 3386), Departamento de Astronomía, Universidad de Chile, Camino El Observatorio 1515, Las Condes, Santiago, Chile \\
\email{Myriam.Benisty@univ-grenoble-alpes.fr}
\and Physics \& Astronomy Department, University of Victoria, 3800 Finnerty Road, Victoria, BC, V8P 5C2, Canada
\and Banting Research fellow
\and Max-Planck Institut für Extraterrestrische Physik (MPE), Gießenbachstraße 1, 85748 Garching, Germany
\and Universit\`a degli Studi di Milano, via Giovanni Celoria 16, 20133 Milano, Italy
\and European Southern Observatory, Karl-Schwarzschild-Str. 2, 85748 Garching, Germany
\and Center for Astrophysics | Harvard \& Smithsonian, 60 Garden Street, Cambridge, MA 02138, USA
\and Universitäts-Sternwarte München, Scheinerstr. 1, 81679 München, Germany
\and School of Physics, University College Dublin, Belfield, Dublin 4, Ireland
\and Max Planck Institute for Astronomy, Königstuhl 17, D-69117, Heidelberg, Germany
\and University of Exeter, School of Physics \& Astronomy, Exeter, EX4 4QL, UK
\and Departamento de Astronomía, Universidad de Chile, Camino El Observatorio 1515, Las Condes, Santiago, Chile
}

\date{Received \today / Accepted <date>}

\abstract 
{Transition disks are protoplanetary disks with dust-depleted cavities, possibly indicating substantial clearing of their dust content by a massive companion. For several known transition disks, dark regions interpreted as shadows have been observed in scattered light imaging and are hypothesized to originate from misalignments between distinct regions of the disk.}%
{
We aim to investigate the presence of misalignments in transition disks. We study the inner disk ($<$1\,au) geometries of a sample of 20 well-known transition disks with Very Large Telescope Interferometer (VLTI) GRAVITY observations and use complementary $^{12}$CO and $^{13}$CO molecular line archival data from the Atacama Large Millimeter/submillimeter Array (ALMA)  to derive the orientation of the outer disk regions ($>$10\,au).  
} 
{
We fit simple parametric models to the visibilities and closure phases of the GRAVITY data to derive the inclination and position angle of the inner disks.
The outer disk geometries were derived from Keplerian fits to the ALMA velocity maps and compared to the inner disk constraints. 
We also predicted the locations of expected shadows for significantly misaligned systems. 
} 
{
Our analysis reveals six disks to exhibit significant misalignments between their inner and outer disk structures.
The predicted shadow positions agree well with the scattered light images of HD~100453 and HD~142527, and we find supporting evidence for a shadow in the south of the disk around CQ~Tau. In the other three targets for which we infer significantly misaligned disks, V1247~Ori, V1366~Ori, and RY~Lup, we do not see any evident sign of shadows in the scattered light images.  The scattered light shadows observed in DoAr~44, HD~135344~B, and HD139614 are consistent with our observations, yet the underlying morphology is likely too complex to be described properly by our models and the accuracy achieved by our observations. 
} 
{
The combination of near infrared and submillimeter interferometric observations allows us to assess the geometries of the innermost disk regions and those of the outer disk. Whereas we can derive precise constraints on the potential shadow positions for well-resolved inner disks around Herbig Ae/Be stars, the large statistical uncertainties for the marginally resolved inner disks around the T Tauri stars of our sample make it difficult to extract conclusive constraints for the presence of shadows in these systems.
}


\maketitle

\section{Introduction}
\label{sec:introduction}
Studying protoplanetary disks around pre-main sequence stars allows us to constrain the early stages of planet formation.
Observations with the Atacama Large Millimiter/submillimeter Array \citep[ALMA;][]{alma_partnership2015} have provided unprecedented insights in the large-scale dust and gas distributions in protoplanetary disks \citep[e.g.,][]{vandermarel2013,perez2014,andrews2018,vanterwisga2019,Teague2019}.
Adaptive-optics-assisted high-contrast imagers, such as the Spectro-Polarimetric High-contrast Exoplanet REsearch \citep[SPHERE;][]{beuzit2019} instrument, 
have provided complementary scattered-light images, revealing the 3D geometries of these disks \citep[e.g.,][]{monnier2017,uyama2020b}.
In both wavelength regimes, observations demonstrated that substructures such as rings, gaps, and spiral arms are ubiquitous in these disks \citep{garufi2018,andrews2020} and they might be linked to recently formed giant planets \citep[][]{bae2017,huang2018}.

While the outer disk regions can be probed with these techniques, the innermost disk regions at scales below 1\,au remain unresolved. However, the processes that are shaping these inner regions are of great interest, as they might play an important role for the formation and evolution of terrestrial planets. At such distances from the star, dust grains sublimate \citep[at T$\sim$1300-1500\,K;][]{kama2009} and the dust sublimation front (rim) is thought to be directly irradiated by the central star, and as a consequence, to puff up and  predominantly emit in the near-infrared regime \citep[e.g.,][]{natta2001,isella2005,dullemond2010}. Models of the rim indicate that it is a radially extended region, rather than a sharp edge, with its exact morphology depending on the properties and composition of the dust grains \citep{kama2009}. Observationally, the innermost disk regions can be studied with infrared interferometry, which enables milli-arcsecond (subau) resolution. Early observations of Herbig AeBe stars indicated a correlation between the inner disk radii and the stellar luminosity, supporting the presence of a rim at the dust sublimation radius \citep{monnier2005}.  Subsequent studies of specific objects \cite[e.g.,][]{tannirkulam2008,Benisty2010,Setterholm2018,Davies2020,bouarour2020}, or snapshot observations of large disk samples \citep{menu2015,lazareff2017, perraut2019}, enabled us to get more insights on the rim morphology. Detailed analysis of individual disks can now be achieved over a broader wavelength regime by combining all Very Large Telescope Interferometer (VLTI) instruments  \citep[e.g., GRAVITY, MATISSE;][]{Sanchez2021, Varga2021}. 


In this paper, we focus on connecting the geometry of the inner disk with that of the outer disk in a subclass of protoplanetary disks, the transition disks. Transition disks were originally identified through their spectral energy distributions (SEDs) that show a characteristic absence of excess infrared emission \citep[e.g.,][]{strom1989,skrutskie1990,calvet2002}. Such a dip in the SEDs indicates that the inner regions are (partially) cleared of dust material \citep{espaillat2014}, with a cavity possibly due to the dynamical clearing by massive companions or planets \citep[e.g.,][]{zhu2011, Bae2019}. These cavities can either be probed by (sub)millimeter interferometric imaging in both dust and gas tracers  \citep[e.g.,][]{vandermarel2016,vandermarel2018,dong2017}, mid-infrared interferometry \citep[e.g.,][]{kraus2013,kluska2018,Menu2014}, or in scattered light images \citep[e.g.,][]{bohn2019,deBoer2020}.
Several transition disk scattered light images present dark regions  \citep[e.g.,][]{stolker2016, casassus2018}, which are interpreted as shadows resulting from a misalignment between inner and outer disk regions \citep[e.g.,][]{marino2015,facchini2018,Nealon2019}. Depending on the misalignment angle, the shadows can appear as narrow lines \citep{benisty2017}, or as very broad areas \citep{benisty2018,muro_arena2020}. Such a misalignment between disk regions might not be uncommon, and could be induced by various mechanisms detailed in the discussion section. 

In this paper, we search for evidence of misalignments in a sample of 20 transition disks with observations from VLTI/GRAVITY, probing the geometry of the dust in the inner disk, and ALMA, probing the geometry of the gas in the outer disk. Sect.~\ref{sec:observations_and_data_reduction} presents our sample of transition disk hosting stars, and we describe the data that were collected on these targets and the basic data reduction.
The data analysis is detailed in Sect.~\ref{sec:analysis_gravity} and ~\ref{sec:analysis_alma} for GRAVITY and ALMA data, respectively. From these results, we derived misalignment angles for our targets in Sect.~\ref{sec:analysis_misalignments}. In Sect.~\ref{sec:discussion} we discuss the misalignments between inner and outer disks and compare our results to scattered light images of these transition disks. We present our conclusions in Sect.~\ref{sec:conclusions}.

\section{Observations and data reduction}
\label{sec:observations_and_data_reduction}

\subsection{Stellar properties}
\label{sec:sample_selection}
We observed 20 transition disks previously studied in near-infrared scattered light and at submillimeter wavelengths with ALMA. 
We compiled the stellar properties of our input sample from previous literature.
A list of all targets, their spectral types, effective temperatures, and parallactic distances is presented in Table~\ref{tbl:sample_selection}.
Stellar luminosities that had been calculated with pre-Gaia distance estimates were updated considering the latest parallax measurements of \textit{Gaia} EDR3 and corresponding distances \citep[][]{gaia2020,bailerjones2021}.
We derived stellar masses based on these updated luminosities and the effective temperatures of the objects by comparison to the isochronal models of \citet{feiden16}. We used the nonmagnetic tracks as in \citet{pascucci2016}.
Following \citet{manara2012}, the uncertainties on both quantities were modeled with a Monte Carlo approach, for which we calculated the stellar mass 1000 times while drawing $T_\mathrm{eff}$ and $L_\star$ randomly from their uncertainty distribution. 

Our sample comprises various spectral types from M0 to B9.5 with associated masses in the range $0.6\,M_\sun$ to $3.1\,M_\sun$. Ten of our targets are T~Tauri stars (DoAr~44, GM~Aur, IP~Tau, LkCa~15, LkH$\alpha$~330, PDS~70, RX~J1615, RY~Lup, SZ~Cha, and UX~Tau~A), three are intermediate-mass T~Tauri stars (CQ~Tau, HD~135344~B, and HD~142527) and seven are Herbig AeBe stars (HD~139614, HD~100453, HD~100546, HD~169142, HD~97048, V1247~Ori, and V1366~Ori). 

\subsection{Photometry}
We collected \textit{B}, \textit{V}, $G_\mathrm{BP}$, \textit{G}, $G_\mathrm{RP}$, \textit{R}, \textit{I}, \textit{J}, \textit{H}, and $K$ band photometric data from the Tycho-2 \citep[][]{hog2000}, Gaia EDR3 \citep{gaia2020}, USNO-B \citep[][]{monet2003}, and 2MASS \citep[][]{cutri2003,skrutskie2006} catalogs.
The Gaia EDR3 magnitudes for objects with 6-parameter astrometric solutions were corrected as described by \citet{riello2020}.
An overview of all photometric measurements is compiled in Table~\ref{tbl:target_photometry}.

\begin{table*}[]
\caption{
Stellar properties of our sample.
}
\label{tbl:sample_selection}
\def\arraystretch{1.2}
\setlength{\tabcolsep}{10pt}
\centering
\begin{tabular}{@{}llllllll@{}}
\hline\hline
Star & SpT & $T_\mathrm{eff}$ & $L_\star$ & $M_\star$\tablefootmark{a} & $D$\tablefootmark{b} & $k_\mathrm{s}$\tablefootmark{c} & Reference(s) \\
 &  & (K) & (L$_{\odot}$) & (M$_{\odot}$) & (pc) & & \\
\hline
\object{CQ~Tau} & F5 & $6750\pm 300$ & $6.17\pm 2.12$ & $1.49\pm 0.11$ & $149.4\pm1.3$ & 1.43 & (1,2) \\
\object{DoAr~44} & K2 & $5100\pm 150$ & $0.87\pm 0.34$ & $1.13\pm 0.16$ & $146.3\pm0.5$ & 1.23 & (3) \\
\object{GM~Aur} & K5 & $4440\pm 150$ & $1.25\pm 0.32$ & $0.97\pm 0.16$ & $158.1\pm1.2$ & 1.11 & (3) \\
\object{HD~100453} & A9 & $7250\pm 250$ & $6.15\pm 1.07$ & $1.59\pm 0.06$ & $103.8\pm0.2$ & 1.51 & (4,5) \\
\object{HD~100546} & A0 & $9750\pm 500$ & $24.16\pm 5.99$ & $2.13\pm 0.11$ & $108.1\pm0.4$ & 1.63 & (5) \\
\object{HD~135344~B} & F8 & $6375\pm 125$ & $6.54\pm 1.49$ & $1.56\pm 0.11$ & $135.0\pm0.4$ & 1.39 & (6,5) \\
\object{HD~139614} & A9 & $7750\pm 250$ & $5.97\pm 1.22$ & $1.57\pm 0.06$ & $133.6\pm0.5$ & 1.51 & (5) \\
\object{HD~142527} & F6 & $6500\pm 250$ & $10.18\pm 0.36$ & $1.75\pm 0.10$ & $159.3\pm0.7$ & 1.41 & (5) \\
\object{HD~169142} & F1 & $10700\pm 850$ & $20.58\pm 8.04$ & $2.11\pm 0.14$ & $114.9\pm0.4$ & 1.48 & (7,2) \\
\object{HD~97048} & A0 & $10500\pm 50$ & $36.56\pm 20.03$ & $2.36\pm 0.19$ & $184.4\pm0.8$ & 1.63 & (5) \\
\object{IP~Tau} & M0 & $3850\pm 100$ & $0.07\pm 0.00$ & $0.59\pm 0.02$ & $129.4\pm0.3$ & 0.96 & (8,9) \\
\object{LkCa~15} & K2 & $5100\pm 150$ & $1.51\pm 0.31$ & $1.40\pm 0.11$ & $157.2\pm0.7$ & 1.23 & (3) \\
\object{LkH$\alpha$~330} & G4 & $5680\pm 50$ & $22.81\pm 0.62$ & $3.05\pm 0.07$ & $318.2\pm3.5$ & 1.33 & (3) \\
\object{PDS~70} & K7 & $3972\pm 36$ & $0.35\pm 0.09$ & $0.76\pm 0.02$ & $112.4\pm0.2$ & 1.02 & (10,11,12) \\
\object{RX~J1615} & K7 & $4100\pm 150$ & $0.62\pm 0.18$ & $0.73\pm 0.14$ & $155.6\pm0.6$ & 1.02 & (3) \\
\object{RY~Lup} & K2 & $5100\pm 150$ & $1.72\pm 0.74$ & $1.47\pm 0.21$ & $153.5\pm1.4$ & 1.23 & (13,14) \\
\object{SZ~Cha} & K2 & $5100\pm 150$ & $1.65\pm 0.35$ & $1.45\pm 0.11$ & $190.2\pm0.9$ & 1.23 & (3) \\
\object{UX~Tau~A} & K2 & $5270\pm 100$ & $1.62\pm 0.08$ & $1.38\pm 0.05$ & $142.2\pm0.7$ & 1.23 & (9) \\
\object{V1247~Ori} & F0 & $7875\pm 375$ & $16.33\pm 6.23$ & $1.88\pm 0.14$ & $401.3\pm3.2$ & 1.49 & (5) \\
\object{V1366~Ori} & B9.5 & $9500\pm 250$ & $10.37\pm 4.42$ & $1.90\pm 0.06$ & $308.6\pm2.2$ & 1.66 & (5) \\
\hline
\end{tabular}
\tablefoot{%
\tablefoottext{a}{The masses are derived from the effective temperatures and updated luminosities as detailed in Sect.~\ref{sec:sample_selection}.}
\tablefoottext{b}{Distances are derived from the parallax measurements provided by Gaia EDR3 \citep{gaia2020} as $D=1/\varpi$.}
\tablefoottext{c}{Stellar spectral index as defined in Eq.~\eqref{eqn:spectral_index}.
Evaluated at $\lambda_0=2.25$\,\textmu m.}
}
\tablebib{
(1)~\citet{mora2001};
(2)~\citet{vioque2018};
(3)~\citet{manara2014};
(4)~\citet{vieira2003};
(5)~\citet{fairlamb2015};
(6)~\citet{coulson1995};
(7)~\citet{murphy2015};
(8)~\citet{herbig1986};
(9)~\citet{herczeg2014};
(10)~\citet{pecaut2016};
(11)~\citet{keppler2018};
(12)~\citet{muller2018};
(13)~\citet{Gahm1989};
(14)~\citet{alcala2017}
}
\end{table*}


\subsection{VLTI/GRAVITY observations}
\label{subsec:obs_and_dr_gravity}
GRAVITY operates in the near-infrared K-band between 2.0\,\textmu m and 2.4\,\textmu m and combines the light of four telescopes, either the 8-m Unit Telescopes (UTs) or the 1.8-m Auxiliary Telescopes (ATs).
The interferometric fringes on the six baselines are recorded simultaneously on the scientific instrument (SC) and on the fringe tracker (FT) that stabilizes the fringes at a frequency of 900~Hz or 300~Hz \citep{Lacour2019}. This makes possible long integration exposures (of 10 or 30~s) on the SC detector, and therefore to observe faint objects, as T Tauri stars, with the ATs. Data were obtained either in medium resolution ($\sim$~500)  or in high resolution ($\sim$~4000) spectral resolution modes. For all targets, we recorded several 5-min long files on the object itself, and interleaved these observations with observations of interferometric calibrators. These calibrators were selected to be unresolved single stars with a magnitude and a color similar as those of the stars.  A detailed list of the observation setup and weather conditions is presented in Table~\ref{tbl:gravity_observation_setup}.

We reduced all our data with the GRAVITY data reduction pipeline \citep{GRAVITY_DRS}.
For each file, we obtained six squared visibilities and four closure phases in each spectral channel of the FT and of the SC. We used the calibrator observations to determine the atmospheric transfer function for each night and to calibrate the interferometric observables. 
We checked that these interferometric quantities are consistent between SC and FT, that is to say that the fringes are not blurred on the SC due to a bad fringe tracking or a fast turbulence. This was the case for all but one target, RY~Lup. During the observations of RY Lup, the small coherence time of $\sim2$\,ms or less caused blurring effects of the SC fringes during the long-exposure images ($DIT$~=~30~s). For that reason, we used the FT data with $DIT$ of 0.85~ms for the analysis of RY~Lup. For all remaining targets, we used the SC data and binned them spectrally to have, for each dataset, six calibrated squared visibilities and four calibrated closure phases in five spectral channels covering the whole K-band. To mitigate weighting effects due to different exposure times and to facilitate a proper combination of data from several epochs, we performed a temporal binning into observing blocks of 30\,min. When binning the data both spectrally and spatially, we calculated the weighted average (using the inverse squared uncertainties as weights) and corresponding uncertainties.
We checked by visual inspection that the $(u,v)$ plane rotation within these 30\,min intervals was small. 

\begin{figure*}
\centering
\includegraphics[width=0.9\textwidth]{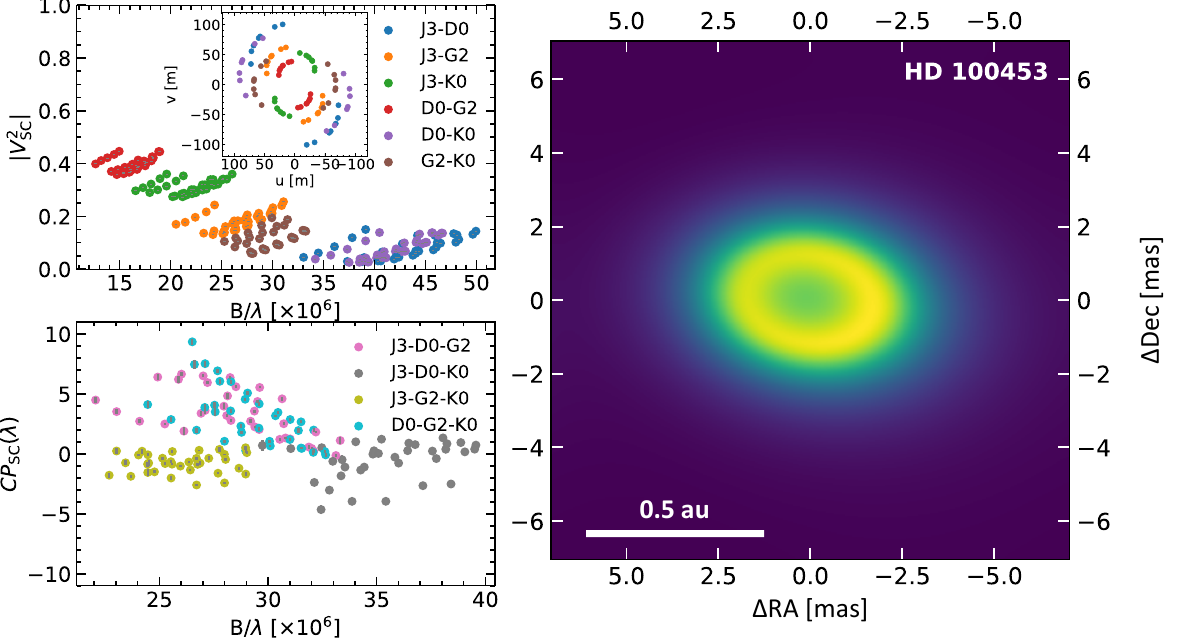}
\caption{
GRAVITY observations and best-fit model for HD~100453.
\textit{Top~left}:
Squared visibilities as a function of spatial frequency.
Colors refer to the different baselines. 
The inset presents the $(u, v)$ plane coverage of the observations.
\textit{Bottom~left}:
Closure phases as a function of the average spatial frequency.
The various colors refer to the different triplets.
\textit{Right panel}:
Best-fit model. 
}
\label{fig:vis2_cp_example}
\end{figure*}

An example of the $(u,v)$ plane coverage and the binned squared visibilities and closure phases for our data on HD~100453 is presented in the left panel of Fig.~\ref{fig:vis2_cp_example}. The plot comprises data from seven epochs (see Table~\ref{tbl:gravity_observation_setup}). The inner disk around this Herbig star is well resolved as indicated by the squared visibilities that go down to zero for the longest baselines. The brightness distribution of the disk appears asymmetric as the closure phases depart from zero. The spectral variation of the visibilities will be used to constrain the dust spectral index.

\subsection{ALMA data}
\label{subsec:obs_and_dr_alma}

We collected ALMA molecular line data of $^{12}$CO or $^{13}$CO 3--2 or 2--1 data cubes for all our targets from the ALMA archive, in order to derive the outer disk orientations from the velocity maps. The data cubes were obtained from published works or by running the calibration scripts provided with ALMA archival datasets, followed by imaging. A detailed list of the datasets that we use and the properties of the data cubes are presented in Table~\ref{tbl:alma_observation_setup}. The methodology employed for their analysis, as well as an example plot of ALMA data, are developed further in Sect.~\ref{sec:analysis_alma}. 

\section{Inner disks probed by VLTI/GRAVITY}
\label{sec:analysis_gravity}
The analysis of the inner disk geometries relies on the framework introduced by \citet{lazareff2017}. First, we fit the stellar SED to derive the disk $K$-band flux (Sect.~\ref{subsubsec:analysis_sed_modeling}).fit
This parameter is required as an input for our parametric disk models that we fit to the GRAVITY data in order to derive the geometry of the inner disk of particular interest in this paper, namely its inclination and position angle (Sect.~\ref{subsubsec:analysis_parametric_modeling}). 

\subsection{SED modeling}
\label{subsubsec:analysis_sed_modeling}
 For each target we fit the photometric data points with two blackbodies $B(\nu, T)$ representing the stellar flux and the inner disk flux. The total flux density at frequency $\nu_k$ is given by
\begin{equation}
    F(\nu_k)=\left(F_{\mathrm{s}V}\frac{B(\nu_k,T_\mathrm{eff})}{B(\nu_V,T_\mathrm{eff})} + F_{\mathrm{d}K}\frac{B(\nu_k,T_\mathrm{dp})}{B(\nu_K,T_\mathrm{dp})}\right)\times10^{-0.4A_Vr_k}\;,\label{eqn:flux_density}
\end{equation}
where $k\in\{1,\dots,10\}$ is an index for the photometric bandpasses; $\nu_k$, $\nu_V$, and $\nu_K$ are the mean frequencies of the $k^\mathrm{th}$, $V$, and $K$ band filter profiles, respectively; $T_\mathrm{eff}$ is the stellar effective temperature as presented in Table~\ref{tbl:sample_selection}. 
$r_k=\nicefrac{A_k}{A_V}$ is the extinction coefficient based on the extinction law of \citet{cardelli1989}. We adopt $R_V=3.1$. 

The fit parameters therefore are: $F_{\mathrm{s}V}$, the stellar flux in the $V$ band; $F_{\mathrm{d}K}$, the thermal dust emission in the $K$ band; the dust temperature $T_\mathrm{dp}$, and $A_V$, the total extinction in the $V$ band. 

\begin{figure}
\centering
\includegraphics[width=0.43\textwidth]{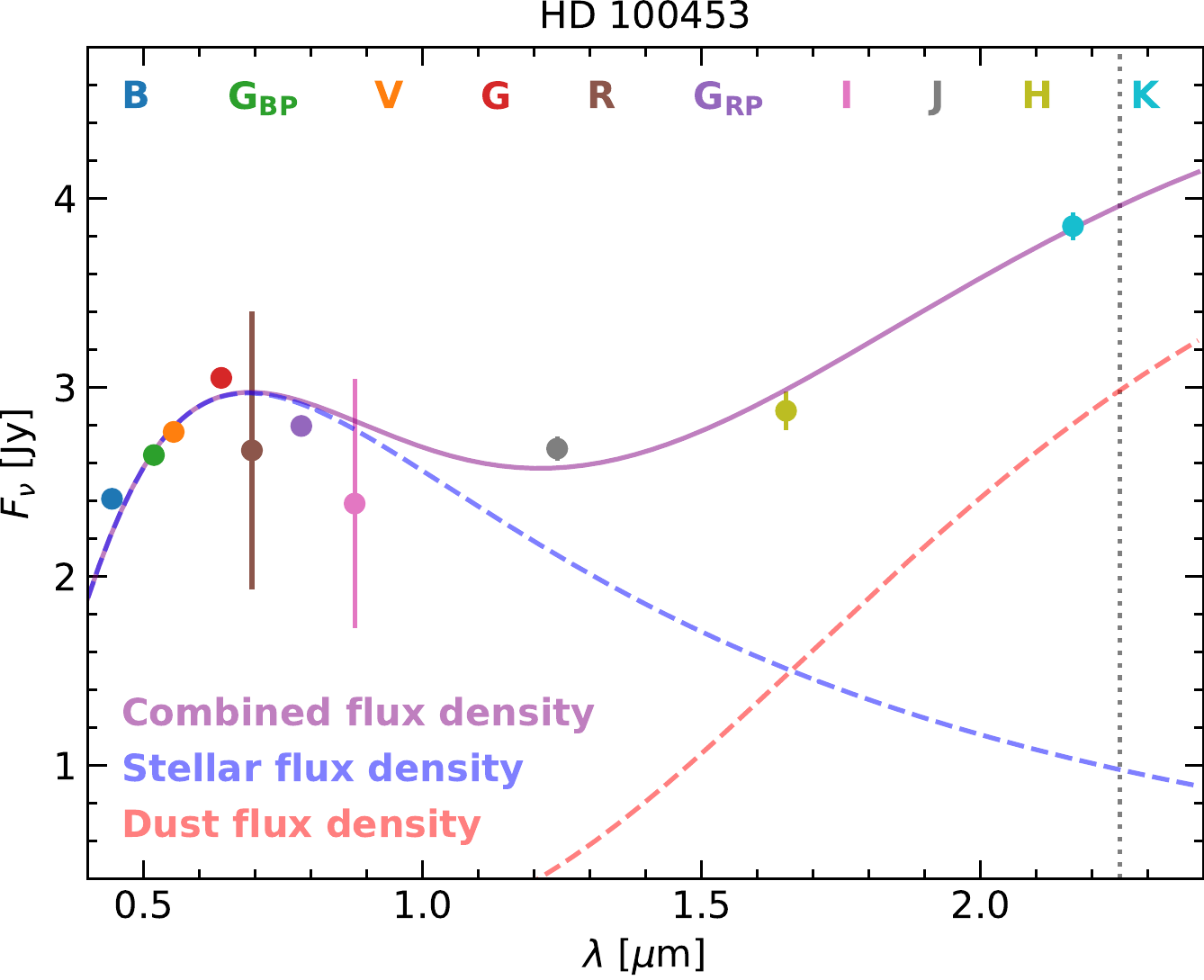}
\caption{%
SED fit for HD~100453. The circles represent the photometry and the solid purple curve, the best SED model from the MCMC posterior distribution.
The model consists of two individual blackbody components that are visualized by the dashed lines:
the blue curve represents the stellar flux  density whereas the red curve shows the flux density of the circumstellar dust.
The dotted line indicates the GRAVITY reference wavelength of 2.25\,\textmu m, at which we evaluate the dust flux contribution. } 
\label{fig:sed_fit_example}
\end{figure}

\begin{table*}[t]
\caption{
Results of our SED fits. 
}
\label{tbl:sed_fitting}
\def\arraystretch{1.4}
\setlength{\tabcolsep}{12pt}
\centering
\begin{tabular}{@{}llllll@{}}
\hline\hline
Target & $\log_{10}\left[F_{\mathrm{s}V}\right]$ & $\log_{10}\left[F_{\mathrm{d}K}\right]$ & $A_V$ & $\log_{10}\left[T_{\mathrm{dp}}\right]$ & $f_\mathrm{d}$ (2.25$\mu$m)\\    
& $\left(\log_{10}\left[\mathrm{Jy}\right]\right)$ & $\left(\log_{10}\left[\mathrm{Jy}\right]\right)$ & (mag) & $\left(\log_{10}\left[\mathrm{K}\right]\right)$ &  \\
\hline
CQ~Tau & $-0.010^{+0.097}_{-0.110}$ & $0.337^{+0.009}_{-0.010}$ & $1.65^{+0.26}_{-0.30}$ & $3.242^{+0.020}_{-0.026}$ & $0.829^{+0.033}_{-0.035}$ \\
DoAr~44 & $-0.638^{+0.016}_{-0.017}$ & $-0.290^{+0.014}_{-0.014}$ & $1.97^{+0.04}_{-0.05}$ & $3.235^{+0.015}_{-0.016}$ & $0.697^{+0.011}_{-0.012}$ \\
GM~Aur & $-1.331^{+0.029}_{-0.017}$ & $-0.581^{+0.008}_{-0.009}$ & $0.04^{+0.06}_{-0.03}$ & $3.385^{+0.006}_{-0.006}$ & $0.772^{+0.008}_{-0.014}$ \\
HD~100453 & $0.448^{+0.003}_{-0.003}$ & $0.448^{+0.011}_{-0.011}$ & $0.00^{+0.01}_{-0.00}$ & $3.155^{+0.014}_{-0.014}$ & $0.753^{+0.005}_{-0.006}$ \\
HD~100546 & $0.894^{+0.002}_{-0.002}$ & $0.446^{+0.015}_{-0.015}$ & $0.00^{+0.00}_{-0.00}$ & $3.117^{+0.018}_{-0.019}$ & $0.640^{+0.009}_{-0.009}$ \\
HD~135344~B & $0.108^{+0.011}_{-0.008}$ & $0.369^{+0.010}_{-0.010}$ & $0.03^{+0.03}_{-0.02}$ & $3.208^{+0.012}_{-0.012}$ & $0.777^{+0.005}_{-0.006}$ \\
HD~139614 & $0.255^{+0.002}_{-0.002}$ & $-0.173^{+0.021}_{-0.021}$ & $0.00^{+0.00}_{-0.00}$ & $3.037^{+0.031}_{-0.024}$ & $0.545^{+0.013}_{-0.013}$ \\
HD~142527 & $0.501^{+0.014}_{-0.015}$ & $0.750^{+0.010}_{-0.010}$ & $0.68^{+0.04}_{-0.04}$ & $3.230^{+0.014}_{-0.014}$ & $0.784^{+0.007}_{-0.007}$ \\
HD~169142 & $0.306^{+0.002}_{-0.002}$ & $-0.005^{+0.017}_{-0.018}$ & $0.00^{+0.00}_{-0.00}$ & $3.173^{+0.020}_{-0.021}$ & $0.567^{+0.010}_{-0.011}$ \\
HD~97048 & $0.511^{+0.012}_{-0.013}$ & $0.359^{+0.015}_{-0.015}$ & $0.81^{+0.03}_{-0.03}$ & $3.252^{+0.015}_{-0.015}$ & $0.770^{+0.008}_{-0.008}$ \\
IP~Tau & $-1.431^{+0.038}_{-0.045}$ & $-0.655^{+0.014}_{-0.015}$ & $0.57^{+0.10}_{-0.11}$ & $3.291^{+0.015}_{-0.015}$ & $0.676^{+0.027}_{-0.025}$ \\
LkCa~15 & $-0.849^{+0.044}_{-0.052}$ & $-0.574^{+0.018}_{-0.019}$ & $1.01^{+0.12}_{-0.13}$ & $3.327^{+0.019}_{-0.019}$ & $0.654^{+0.033}_{-0.032}$ \\
LkH$\alpha$~330 & $-0.354^{+0.014}_{-0.015}$ & $-0.009^{+0.011}_{-0.011}$ & $2.18^{+0.04}_{-0.04}$ & $3.204^{+0.014}_{-0.014}$ & $0.771^{+0.007}_{-0.008}$ \\
PDS~70 & $-1.189^{+0.032}_{-0.046}$ & $-0.952^{+0.045}_{-0.043}$ & $0.22^{+0.08}_{-0.10}$ & $3.325^{+0.037}_{-0.041}$ & $0.430^{+0.049}_{-0.038}$ \\
RX~J1615 & $-1.213^{+0.011}_{-0.008}$ & $-0.921^{+0.016}_{-0.018}$ & $0.02^{+0.03}_{-0.01}$ & $3.352^{+0.013}_{-0.014}$ & $0.459^{+0.012}_{-0.015}$ \\
RY~Lup & $-0.799^{+0.102}_{-0.097}$ & $-0.015^{+0.008}_{-0.009}$ & $0.46^{+0.27}_{-0.26}$ & $3.299^{+0.009}_{-0.011}$ & $0.860^{+0.025}_{-0.031}$ \\
SZ~Cha & $-0.786^{+0.069}_{-0.083}$ & $-0.362^{+0.018}_{-0.021}$ & $1.33^{+0.19}_{-0.22}$ & $3.289^{+0.023}_{-0.026}$ & $0.729^{+0.040}_{-0.039}$ \\
UX~Tau~A & $-1.086^{+0.173}_{-0.067}$ & $-0.219^{+0.008}_{-0.011}$ & $0.19^{+0.39}_{-0.14}$ & $3.397^{+0.002}_{-0.004}$ & $0.879^{+0.016}_{-0.052}$ \\
V1247~Ori & $-0.367^{+0.008}_{-0.005}$ & $-0.251^{+0.015}_{-0.015}$ & $0.02^{+0.02}_{-0.01}$ & $3.167^{+0.016}_{-0.016}$ & $0.788^{+0.006}_{-0.007}$ \\
V1366~Ori & $-0.263^{+0.012}_{-0.012}$ & $-0.321^{+0.011}_{-0.011}$ & $0.31^{+0.03}_{-0.03}$ & $3.162^{+0.012}_{-0.012}$ & $0.826^{+0.005}_{-0.006}$ \\
\hline
\end{tabular}
\end{table*}

\noindent Due to numerical reasons, we fit for the logarithms of $F_{\mathrm{s}V}$, $F_{\mathrm{d}K}$, and $T_\mathrm{dp}$ \citep{lazareff2017}.
First, we applied a shuffled complex evolution (SCE) algorithm \citep[][]{duan1993} to find the set of parameters that is minimizing the $\chi^2$ map\footnote{\url{www.github.com/stijnvanhoey/Optimization_SCE}}.
This solution was used as a starting point for a Markov chain Monte Carlo (MCMC) sampler \texttt{emcee} \citep[version 3.0.2;][]{foreman-mackey2013}, with 200 walkers and 10000 steps to estimate the variances of model parameters. We discarded the first 1000 steps of each walker as the burn-in phase of the sampler;
the convergence of the walkers was confirmed by visual inspection of the chains.
To minimize the correlation among the resulting samples, we continued using only every 40$^\mathrm{th}$ step of the final chains.
This provided 45000 modelings of posterior samples for each target.
We adopt the median and the 68\,\% confidence interval as the final estimate for the fit parameters.
An example SED fit for HD~100453 is presented in Figure~\ref{fig:sed_fit_example} and the best-fit parameters for all our targets are listed in Table~\ref{tbl:sed_fitting}. From these SED fits, we derived the flux contribution of the disk, $f_\mathrm{d}(\lambda)$ to the integrated flux at wavelength $\lambda$ as 
\begin{equation}
    f_\mathrm{d}(\lambda) = \frac{1}{1 + \frac{F_{\mathrm{S}V}}{F_{\mathrm{d}K}}
    \frac{B(\nu_K,T_\mathrm{dp})}{B(\nu_V,T_\mathrm{eff})}
    \frac{B(\lambda,T_\mathrm{eff})}{B(\lambda,T_\mathrm{dp})}}\;.
    \label{eqn:f_c}
\end{equation}
This fraction is equivalent to the contribution of the second term in Eq.~\eqref{eqn:flux_density} to the total flux density, at a given wavelength $\lambda$.
This contribution factor is required as an input parameter and initial value to our parametric models for the GRAVITY data and we evaluated Eq.~\eqref{eqn:f_c} at the approximate central wavelength of the GRAVITY instrument at $\lambda_0=2.25$\,\textmu m (Table~\ref{tbl:sed_fitting}).

\begin{table*}
\caption{
Model parameter and priors of the MCMC fit to the GRAVITY data. 
}
\label{tbl:gravity_model_priors}
\def\arraystretch{1.2}
\setlength{\tabcolsep}{10pt}
\centering
\begin{tabular}{@{}lll@{}}
\hline\hline
Parameter & Description & Uniform Prior\\
\hline
$k_\mathrm{c}$ & Spectral index of circumstellar component & $U(-6, 0)$\\
$f_\mathrm{c}$ & Fractional flux of circumstellar component & $U(0, 1)$\\
$f_\mathrm{h}$ & Fractional flux of halo component & $U(0, 1)$\\
$f_\mathrm{Lor}$ & Weighting of radial profile & $U(0, 1)$\\
$l_a$ & Logarithmic half-flux semi-major axis & $U(-2, 2)$\\
$l_\mathrm{kr}$ & Logarithmic kernel to ring ratio & $U(-3, 3)$\\
$\cos(i_\mathrm{in})$ & Cosine of disk inclination & $U(0, 1)$\\
$\mathrm{PA}_\mathrm{in}$ & Position angle of disk major axis, defined from north to east & $U(-360\degr, 360\degr)$\\
$c_j, s_j$ & Cosine and sine modulation amplitudes of order $j$ & $U(-1, 1)$\\
\hline
\end{tabular}
\end{table*}

\subsection{Parametric modeling of GRAVITY observations}
\label{subsubsec:analysis_parametric_modeling}
We modeled the complex visibilities $V(u,v,\lambda)$ using a combination of stellar (s), circumstellar (c), and halo (h) contributions.  The halo mimics the contribution from an extended, over-resolved component due to scattered light. Models are fit to both squared visibilities and closure phases simultaneously. 

We considered radial brightness distributions for the circumstellar component varying between a Gaussian profile and a Lorentzian profile. The model, of half-flux semi major axis $a$, is convolved with a kernel of semi major axis $a_\mathrm{k}$, enabling us to describe rings of different widths as well as ellipsoids. All models include modulation amplitudes to first order $m=1$. The model parameters are defined in Table~\ref{tbl:gravity_model_priors} and a detailed description of the models can be found in Appendix~\ref{subsec:appendix_inner_disk_model}. 
Following the analysis of \citet{lazareff2017}, we added an extra term of
 \begin{equation}
     \left(\frac{f_\mathrm{c}-f_\mathrm{d}}{\sigma_{f_\mathrm{d}}}\right)^2
 \end{equation}
 to the evaluation of $\chi^2$.
This assumption that $f_\mathrm{d}$ and $f_\mathrm{c}$ measure the same quantity helps to break the size-flux degeneracy of the model and to reduce the standard error of the disk half-light radii \citep[see Sect. 3.4 of][]{lazareff2017}.
$\sigma_{f_\mathrm{d}}$ refers to the uncertainty of the fractional flux contribution from the circumstellar disk at 2.25\,\textmu m (see Sect.~\ref{subsubsec:analysis_sed_modeling} and Table~\ref{tbl:sed_fitting}). 

We carry out an initial iteration of the SCE algorithm on the binned data to find the global minimum of the $\chi^2$ map. The contributions of the squared visibilities and the closure phases to the combined $\chi^2$ statistics were analyzed, and weighting factors were introduced such that (a) the $\chi^2$ contributions for both squared visibilities and closure phases are the same and (b) the reduced $\chi^2$ value is close to unity. 
After applying these weighting factors, the SCE algorithm was carried out again for both scenarios (a) and (b), and the results were used as starting point for an MCMC. We used flat priors for all parameters as indicated in Table~\ref{tbl:gravity_model_priors}. Even though $\mathrm{PA}_\mathrm{in}\in[0\degr,180\degr]$ we allowed values in the range $[-360\degr,360\degr]$ for the fitting procedure. This allows continuous posterior distributions that do not exhibit any phase jumps when the true position angle is close to 0\degr or 180\degr. The final posterior distributions for $\mathrm{PA}_\mathrm{in}$ were resampled such that the median value resided in the interval [0,180]. We used 200 walkers that are sampling 10000 steps each. We discarded the first 1000 steps of each chain as burn-in phase and continue using every 40$^\mathrm{th}$ sample from the remaining posterior distributions. This provided 45000 uncorrelated samples as our final posterior distribution. 
The MCMC was carried out for both scenarios (a) and (b) and we used the median of the posterior distribution from method~(b) as our best-fit model and derived associated uncertainties from the 68\,\% confidence intervals generated by method~(a).
This procedure is analogous to the methodology described in \citet{lazareff2017} and \citet{perraut2019}.

The physical quantities that correspond to the best-fit parameters, such as the half-flux radius, $a$, and the inner disk orientations,  are presented in Table~\ref{tbl:results_misalignments}. Appendix \ref{subsec:appendix_inner_disk_model} provides the best-fit parameters (Table~\ref{tbl:results_inner_disks_unconstrained}), best-fit model maps (Fig.~\ref{fig:gravity_fits_all}) and comparison between best-fit model and observations (Fig.~\ref{fig:gravity_observables}).


\subsection{Critical view on geometric parameters uncertainties}
\begin{figure}
\resizebox{\hsize}{!}{\includegraphics{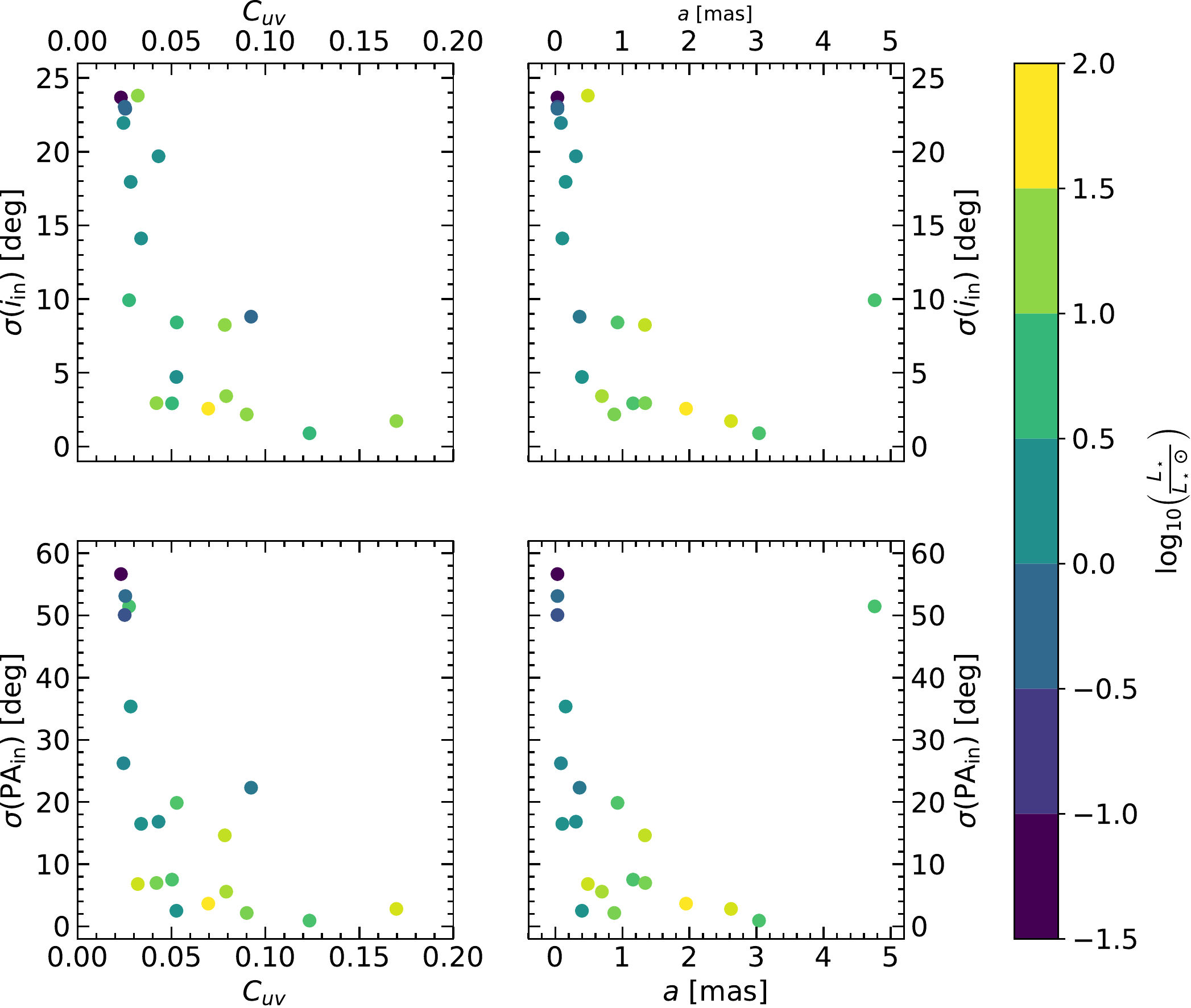}}
\caption{%
Uncertainties of the inner disk geometric parameters ($i_\mathrm{in}$, $\mathrm{PA}_\mathrm{in}$) as a function of $(u,v)$ plane coverage, $C_{uv}$, and the disk half-flux radius, $a$. The colors indicate the stellar luminosity. 
}
\label{fig:inner_disk_uncertainty_dependencies}
\end{figure}

We assessed how the uncertainties of our derived geometric parameters depend on the observational setup, in particular,  the $(u,v)$ plane coverage and the angular size of the disk. The results from this analysis are visualized in Fig.~\ref{fig:inner_disk_uncertainty_dependencies}.

The $(u,v)$ plane coverage, $C_{uv}$, was assessed geometrically. For each of the individual unbinned exposures we drew a circle for each baseline with a radius of 5\,m around each measurement in the $(u,v)$ plane. $C_{uv}$ was determined as the fraction of area that was covered by these circles to the full area of a circle with the longest available  baseline of 130.2\,m as its radius. Accordingly, $C_{uv}\in[0,1]$ and the larger its value, the better the fractional coverage of the $(u,v)$ plane. 
Both inclination and position angle are poorly constrained for the datasets with a scarcely sampled $(u,v)$ plane.
In the most extreme cases, we find uncertainties of up to 25\degr~and 60\degr~in inclination and position angle, respectively. For example, the observations of LkH$\alpha$~330 have a very scarce $(u,v)$ plane, yielding the large uncertainties of its inner disk inclination. The magnitude of these uncertainties decreases the larger $C_{uv}$, that is, the better the $(u,v)$ plane is sampled.
In the best constrained cases, we have uncertainties on the order of a few degrees.

For a few targets with $C_{uv}<0.05$ though, we obtain a similarly low uncertainty. The reason for this good performance despite the scarce coverage of the $(u,v)$ plane can be explained by the angular size of the environment that we try to resolve. As shown in the right panel of Fig.~\ref{fig:inner_disk_uncertainty_dependencies} there is a clear anticorrelation between the determined disk half-flux radius, $a$, and the uncertainties of the geometric parameters: the smaller the inner disk extent, the more difficult it is to derive its orientation. As the disk sizes depend on the stellar temperatures, the geometries for the  Herbig AeBe stars are usually better constrained than for the T Tauri stars. There is one outlier, HD 139614, for which we measure the largest half-flux radius among the sample with $4.75^{+0.55}_{-0.42}$\,mas yet the uncertainties in inclination and position angle are comparably large, probably because the disk is observed almost face-on, which makes it challenging to precisely constrain inclination and position angle, or alternatively, because it is too resolved.

\begin{table*}
\caption{
Inner and outer disk geometries of our sample. 
}
\label{tbl:results_misalignments}
\def\arraystretch{1.2}
\centering
\small
\begin{tabular}{@{}llllllllll@{}}
\hline\hline
Target & \multicolumn{2}{c}{$a$} & $i_\mathrm{in}$ & PA$_\mathrm{in}$ & $i_\mathrm{out}$ & PA$_\mathrm{out}$ & $\Delta\theta_1$ & $\Delta\theta_2$ & $D_\mathrm{KS}$\\    
& (mas) & (au) & ($\degr$) & ($\degr$) & ($\degr$) & ($\degr$) & ($\degr$) & ($\degr$) & \\
\hline
CQ~Tau & $1.16^{+0.06}_{-0.06}$ & $0.173^{+0.009}_{-0.009}$ & $29.25^{+2.80}_{-3.05}$ & $140.37^{+6.53}_{-8.50}$ & $32.25^{+1.42}_{-1.25}$ & $233.92^{+1.42}_{-1.25}$ & $44^{+4}_{-3}$ & $41^{+3}_{-4}$ & 1.00 \\
DoAr~44 & $0.36^{+0.07}_{-0.05}$ & $0.053^{+0.010}_{-0.007}$ & $25.67^{+7.91}_{-9.71}$ & $138.30^{+14.63}_{-29.97}$ & $23.20^{+1.98}_{-1.58}$ & $65.63^{+1.98}_{-1.58}$ & $27^{+9}_{-9}$ & $39^{+9}_{-9}$ & 0.68 \\
GM~Aur & $0.08^{+0.10}_{-0.05}$ & $0.013^{+0.015}_{-0.008}$ & $68.04^{+16.18}_{-27.72}$ & $36.92^{+30.65}_{-21.81}$ & $52.14^{+7.50}_{-5.18}$ & $57.18^{+7.50}_{-5.18}$ & $33^{+15}_{-14}$ & $112^{+19}_{-28}$ & 0.24 \\
HD~100453 & $3.04^{+0.05}_{-0.05}$ & $0.315^{+0.006}_{-0.006}$ & $46.05^{+0.88}_{-0.92}$ & $81.58^{+0.92}_{-0.93}$ & $33.80^{+0.77}_{-0.72}$ & $324.35^{+0.77}_{-0.72}$ & $67^{+1}_{-1}$ & $41^{+1}_{-1}$ & 1.00 \\
HD~100546 & $2.62^{+0.09}_{-0.09}$ & $0.283^{+0.010}_{-0.010}$ & $44.67^{+1.67}_{-1.78}$ & $140.65^{+2.71}_{-2.89}$ & $40.23^{+1.31}_{-1.20}$ & $324.26^{+1.31}_{-1.20}$ & $85^{+2}_{-2}$ & $5^{+2}_{-2}$ & 0.69 \\
HD~135344~B & $0.93^{+0.04}_{-0.04}$ & $0.125^{+0.006}_{-0.006}$ & $22.85^{+7.21}_{-9.62}$ & $14.45^{+18.33}_{-21.39}$ & $16.74^{+0.64}_{-0.57}$ & $241.92^{+0.64}_{-0.57}$ & $35^{+8}_{-9}$ & $17^{+8}_{-6}$ & 0.53 \\
HD~139614 & $4.76^{+0.55}_{-0.43}$ & $0.635^{+0.073}_{-0.058}$ & $22.50^{+9.70}_{-10.15}$ & $6.79^{+54.27}_{-48.64}$ & $17.94^{+0.46}_{-0.42}$ & $276.64^{+0.46}_{-0.42}$ & $27^{+13}_{-12}$ & $28^{+13}_{-12}$ & 0.50 \\
HD~142527 & $1.34^{+0.04}_{-0.04}$ & $0.213^{+0.007}_{-0.006}$ & $23.76^{+2.70}_{-3.18}$ & $15.44^{+7.44}_{-6.52}$ & $38.21^{+1.38}_{-1.25}$ & $162.72^{+1.38}_{-1.25}$ & $59^{+3}_{-3}$ & $22^{+3}_{-3}$ & 0.99 \\
HD~169142 & $1.34^{+0.47}_{-0.22}$ & $0.153^{+0.054}_{-0.025}$ & $35.20^{+6.89}_{-9.60}$ & $31.55^{+14.51}_{-14.76}$ & $12.45^{+0.58}_{-0.52}$ & $5.88^{+0.58}_{-0.52}$ & $25^{+7}_{-9}$ & $46^{+7}_{-10}$ & 0.81 \\
HD~97048 & $1.95^{+0.09}_{-0.10}$ & $0.360^{+0.017}_{-0.019}$ & $47.37^{+2.43}_{-2.70}$ & $176.04^{+3.67}_{-3.64}$ & $45.33^{+2.55}_{-2.16}$ & $2.84^{+2.55}_{-2.16}$ & $92^{+3}_{-3}$ & $6^{+3}_{-3}$ & 0.42 \\
IP~Tau & $0.03^{+0.06}_{-0.02}$ & $0.004^{+0.008}_{-0.002}$ & $61.74^{+20.07}_{-27.29}$ & $163.13^{+60.20}_{-53.10}$ & $45.00^{+5.32}_{-4.96}$ & $345.79^{+5.32}_{-4.96}$ & $93^{+23}_{-24}$ & $40^{+21}_{-19}$ & 0.24 \\
LkCa~15 & $0.31^{+0.13}_{-0.09}$ & $0.048^{+0.021}_{-0.013}$ & $61.02^{+18.59}_{-20.80}$ & $101.33^{+18.24}_{-15.41}$ & $43.95^{+2.39}_{-2.06}$ & $63.22^{+2.39}_{-2.06}$ & $37^{+16}_{-13}$ & $95^{+18}_{-20}$ & 0.53 \\
LkH$\alpha$~330 & $0.49^{+0.03}_{-0.03}$ & $0.155^{+0.009}_{-0.010}$ & $63.49^{+18.55}_{-29.06}$ & $75.66^{+8.68}_{-4.93}$ & $20.95^{+0.39}_{-0.37}$ & $234.74^{+0.39}_{-0.37}$ & $82^{+19}_{-29}$ & $45^{+18}_{-27}$ & 0.69 \\
PDS~70 & $0.03^{+0.08}_{-0.02}$ & $0.004^{+0.009}_{-0.002}$ & $65.76^{+17.64}_{-28.45}$ & $165.85^{+49.51}_{-50.64}$ & $50.19^{+0.96}_{-0.91}$ & $160.21^{+0.96}_{-0.91}$ & $37^{+22}_{-18}$ & $101^{+23}_{-25}$ & 0.21 \\
RX~J1615 & $0.03^{+0.07}_{-0.02}$ & $0.005^{+0.011}_{-0.003}$ & $65.54^{+17.92}_{-27.90}$ & $145.64^{+51.00}_{-55.24}$ & $47.12^{+7.19}_{-4.92}$ & $325.03^{+7.19}_{-4.92}$ & $99^{+23}_{-25}$ & $39^{+22}_{-19}$ & 0.25 \\
RY~Lup & $0.40^{+0.01}_{-0.01}$ & $0.061^{+0.002}_{-0.002}$ & $45.51^{+4.67}_{-4.77}$ & $71.66^{+2.71}_{-2.30}$ & $56.30^{+7.22}_{-5.10}$ & $287.47^{+7.22}_{-5.10}$ & $96^{+8}_{-7}$ & $30^{+5}_{-3}$ & 0.98 \\
SZ~Cha & $0.15^{+0.09}_{-0.04}$ & $0.029^{+0.016}_{-0.009}$ & $43.46^{+16.69}_{-19.22}$ & $173.77^{+30.32}_{-40.39}$ & $46.84^{+2.59}_{-2.22}$ & $156.80^{+2.59}_{-2.22}$ & $84^{+17}_{-20}$ & $27^{+14}_{-13}$ & 0.07 \\
UX~Tau~A & $0.10^{+0.14}_{-0.06}$ & $0.015^{+0.020}_{-0.008}$ & $73.46^{+11.76}_{-16.47}$ & $115.67^{+14.90}_{-18.07}$ & $37.96^{+0.97}_{-0.90}$ & $346.95^{+0.97}_{-0.90}$ & $96^{+13}_{-17}$ & $54^{+13}_{-15}$ & 0.86 \\
V1247~Ori & $0.69^{+0.03}_{-0.03}$ & $0.279^{+0.013}_{-0.013}$ & $35.42^{+3.17}_{-3.67}$ & $145.09^{+5.58}_{-5.58}$ & $24.96^{+1.66}_{-1.46}$ & $124.42^{+1.66}_{-1.46}$ & $15^{+3}_{-3}$ & $59^{+4}_{-4}$ & 0.93 \\
V1366~Ori & $0.88^{+0.06}_{-0.06}$ & $0.271^{+0.018}_{-0.018}$ & $63.96^{+2.13}_{-2.22}$ & $130.31^{+2.20}_{-2.11}$ & $44.93^{+7.06}_{-5.75}$ & $117.54^{+7.06}_{-5.75}$ & $22^{+6}_{-6}$ & $108^{+7}_{-6}$ & 1.00 \\
\hline
\end{tabular}
\tablefoot{The uncertainties represent the 68\,\% confidence intervals of the marginalized posterior distributions.}
\end{table*}

\section{Outer disks probed with ALMA CO line data}
\label{sec:analysis_alma}

In this section, we estimate the outer disk geometrical parameters by fitting gas velocity maps. While dust continuum observations of our sample are available, some of them show substructures with significant asymmetry that could affect our estimates of the outer disk inclination and position angle. We therefore chose to model the velocity field of the rotating outer disk, assumed to be in Keplerian motion and considering its morphology as a conical surface \citep{Teague2018}. 

\subsection{Methodology}
\label{subsec:alma_methodology}

\begin{figure*}
\resizebox{\hsize}{!}{\includegraphics{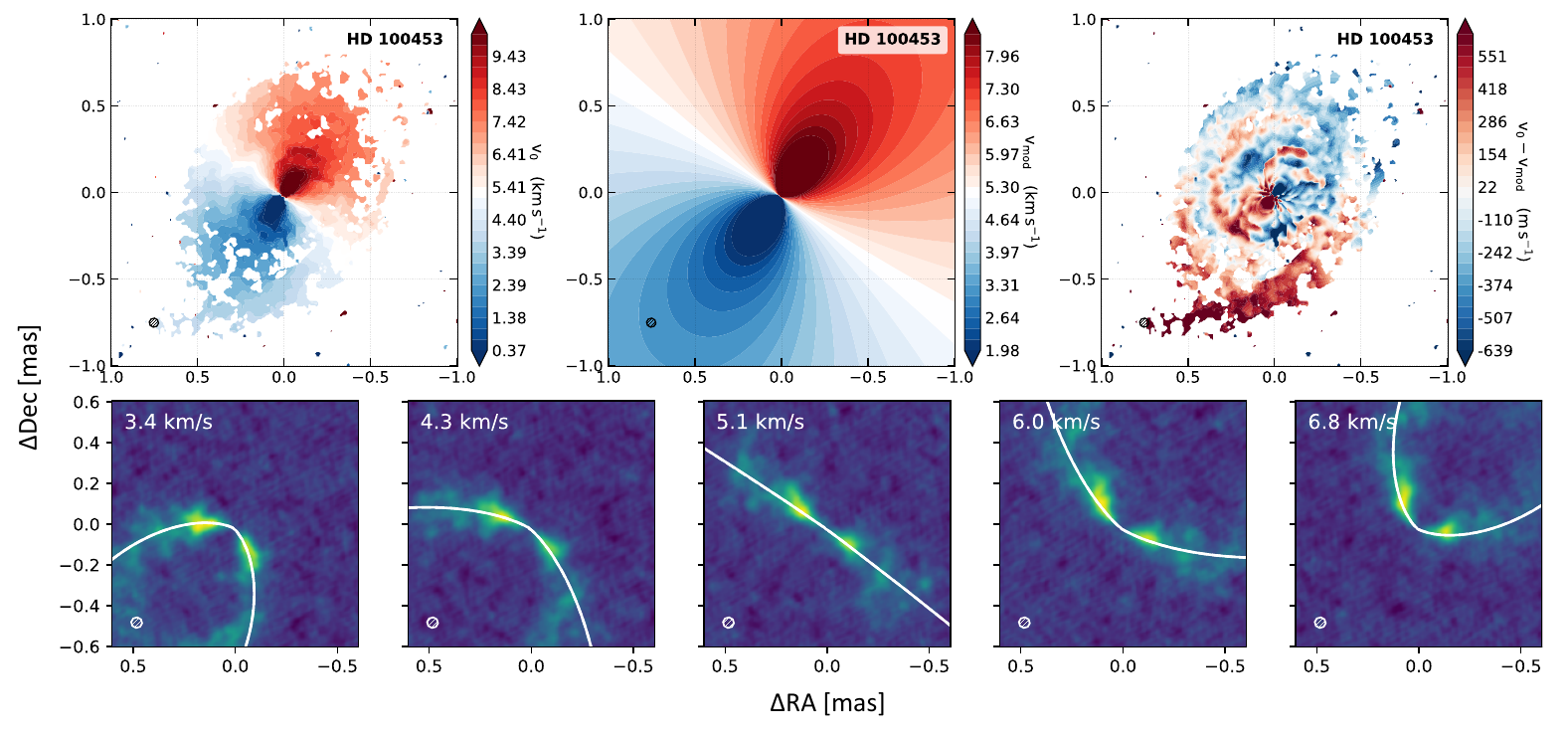}}
\caption{%
Velocity maps and model fits to the ALMA CO line data of HD~100453.
In all images north points up and east to the left.
\textit{Top~left}:
Quadratically collapsed moment map derived with \texttt{bettermoments}.
\textit{Top~center}:
Best-fit Keplerian disk model derived with the \texttt{eddy} tool.
\textit{Top~right}:
Residuals after subtraction of the model from the data.
\textit{Bottom~panels}:
Iso-velocity contours of the best-fit model plotted for the individual channel maps of the data cube.
}
\label{fig:alma_example}
\end{figure*}

We collapsed the CO line data cubes using the quadratic collapsing method as implemented in the \texttt{bettermoments} Python library \citep[][]{teague2018b}. We masked out pixels with low signal-to-noise ratios in the peak line intensity after the calculation of the line center velocity, $v_0$. The magnitude of this clipping parameter was determined by visual inspection of the $v_0$ maps. An example moment map for HD~100453 is presented in the upper right panel of Fig.~\ref{fig:alma_example}. We present the obtained velocity profiles for all targets in Fig.~\ref{fig:alma_data_all}.

For each target we fit the collapsed rotation profiles with the \texttt{eddy} Python tool \citep[][]{teague2019b}.
We utilized the thick disk model, whose projected velocity profile is parametrized by
\begin{equation}
    v_\mathrm{proj}\left(r,\phi\right)=v_\mathrm{Kep}\left(r\right)\cos\left(\phi\right)\sin\left(i_\mathrm{out}\right)+v_\mathrm{LSR}
\end{equation}
with Keplerian velocity
\begin{equation}
    v_\mathrm{Kep}\left(r\right)=\sqrt{\frac{GM_\star r^2}{\left(r^2+z\left(r\right)^2\right)^\frac{3}{2}}}\;,
\end{equation}
the radius-dependent emission surface
\begin{equation}
    z\left(r\right)=z_0\left(\frac{r}{1\arcsec}\right)^\psi\;,
\end{equation}
and the outer disk polar angle, $\phi$ in the disk-frame cylindrical coordinates. We use the local standard-of-rest (LSR) frame as a reference for radial velocities. The LSR is a point that has a velocity equal to the average velocity of stars in the solar neighborhood.

The fit parameters are: the outer disk inclination, $i_\mathrm{out}$; the position angle of the outer disk, $\mathrm{PA}_\mathrm{out}$, defined from north to the redshifted part in an easterly direction; the systemic velocity $v_\mathrm{LSR}$ (the velocity of the star along the line of sight in the LSR frame); and the emission surface parameters $z_0$ and $\psi$.
The outer position angle, $\mathrm{PA}_\mathrm{out}$, is uniquely defined within the interval $[0\degr,360\degr]$, whereas the outer inclination is bounded by $[-90\degr,90\degr]$. To facilitate a proper comparison with the inner disk geometries we projected the outer disk inclinations to $[0\degr,90\degr]$ after the fitting. 
As there is a degeneracy 
between $M_\star$ and $i_\mathrm{out}$, we fixed the stellar mass for each target to the mean value from Table~\ref{tbl:sample_selection}. The reported mass uncertainties were estimated after the fit of the outer disk by a Monte Carlo approach, exploiting the known $v_\mathrm{proj}\propto \sqrt{M_\star}\sin\left(i_\mathrm{out}\right)$ dependency.

Due to beam smearing effects, we applied an inner mask of at least one beam major axis before proceeding with the fit. 
We further performed a down-sampling of the data to only fit spatially uncorrelated pixels, which additionally accelerates the fitting procedure. The fit was implemented using \texttt{emcee}, and we used 100 walkers with 4000 steps from which the first 2000 were discarded as burn-in phase. The MCMC fitting was repeated, using the marginalized posteriors of the first iteration as a starting value and again the initial 2000 steps of each chain were discarded. This provided 200000 samples of our final posterior distribution. As the uncertainties might be underestimated, we performed a rescaling of the uncertainty map that was generated with \texttt{bettermoments}. The rescaling factor was calibrated to obtain a reduced $\chi^2$ value of 1. The full MCMC process was repeated with these rescaled uncertainties. We adopted the median of the marginalized posterior distributions from this second fit as the best-fit parameters for the outer disk and used the 68\,\% confidence intervals as corresponding uncertainties.  

We confirmed the accuracy of our best fit models by visual inspection of iso-velocity contours plotted for the individual channel maps (Fig.~\ref{fig:alma_example}, bottom), especially relevant for highly-inclined disks. 



\subsection{Outer disk geometries}
\label{subsec:results_outer_disks}

An example of the fit outer disk model and the corresponding residuals for HD~100453 is presented in Fig.~\ref{fig:alma_example}. For all targets, we report the best-fit position angles, and outer disk inclinations that in addition consider the additional source of uncertainty from the stellar masses, in Table~\ref{tbl:results_misalignments}. The full output of the \texttt{eddy} Keplerian disk models to the ALMA data is listed in Table~\ref{tbl:results_outer_disk_fits}. The uncertainties reported in this table represent the statistical errors that originate from the marginalized posterior distributions of the MCMC and do not include the mass uncertainties that need to be propagated to the derived inclinations.
For some targets significant features are present after the subtraction of the velocity profile, in particular the residuals obtained for HD~100453, UX~Tau~A and CQ~Tau, exhibit prominent spiral structures \citep[e.g.,][]{rosotti2020,menard2020,wolfer2021}. A detailed analysis of the morphology of the residuals from the fitting procedure on the full sample will be presented in a forthcoming publication (Wölfer et al. in prep.).

\section{Inner and outer disk misalignments}
\label{sec:analysis_misalignments}

\subsection{Misalignment angles}
\label{subsec:results_misalignments}
We combine the results from the GRAVITY and ALMA data to probe potential misalignments between inner and outer disk geometry.
To that end, we calculate a misalignment angle $\Delta\theta$ as a function of inner and outer inclinations, $i_\mathrm{in}$ and $i_\mathrm{out}$, and position angles, $\mathrm{PA}_\mathrm{in}$ and $\mathrm{PA}_\mathrm{out}$ \citep[see e.g.,][]{fekel1981,min2017}:
\begin{equation}
\begin{aligned}
    \Delta\theta(i_\mathrm{in},\mathrm{PA}_\mathrm{in},i_\mathrm{out},&\mathrm{PA}_\mathrm{out})=\\
    =\arccos&\left[ \sin\left(i_\mathrm{in}\right)\sin\left(i_\mathrm{out}\right)\cos\left(\mathrm{PA}_\mathrm{in} - \mathrm{PA}_\mathrm{out}\right) \right.\\
    &\left. + \cos\left(i_\mathrm{in}\right) \cos\left(i_\mathrm{out}\right)\right]\;.
\end{aligned}
\end{equation}
This misalignment angle $\Delta\theta$ corresponds to the angle between the two normal vectors defined by the planes of the inner and outer disk, respectively. As introduced in Sect.~\ref{sec:analysis_gravity}, we cannot tell which side of the inner disk is closer to the observer and which is farther away. 
Accordingly, two potential misalignment angles need to be calculated that are representing each of the two scenarios. We define these two possibilities as
\begin{equation}
    \Delta\theta_1=\Delta\theta(i_\mathrm{in},\mathrm{PA}_\mathrm{in},i_\mathrm{out},\mathrm{PA}_\mathrm{out})
\end{equation}
and
\begin{equation}
    \Delta\theta_2=\Delta\theta(i_\mathrm{in},\mathrm{PA}_\mathrm{in}+180\degr,i_\mathrm{out},\mathrm{PA}_\mathrm{out})\;.
\end{equation}
We estimated the uncertainties on the misalignment angles by a Monte Carlo approach, which randomly selected 200000 samples from the posterior distribution of each required parameter.
The median of this resulting distribution and the 68\,\% confidence interval were selected as final values and uncertainties of the misalignment angles. The results of this analysis are presented in Table~\ref{tbl:results_misalignments}. 

\begin{figure*}
\centering
\includegraphics[width=0.85\textwidth]{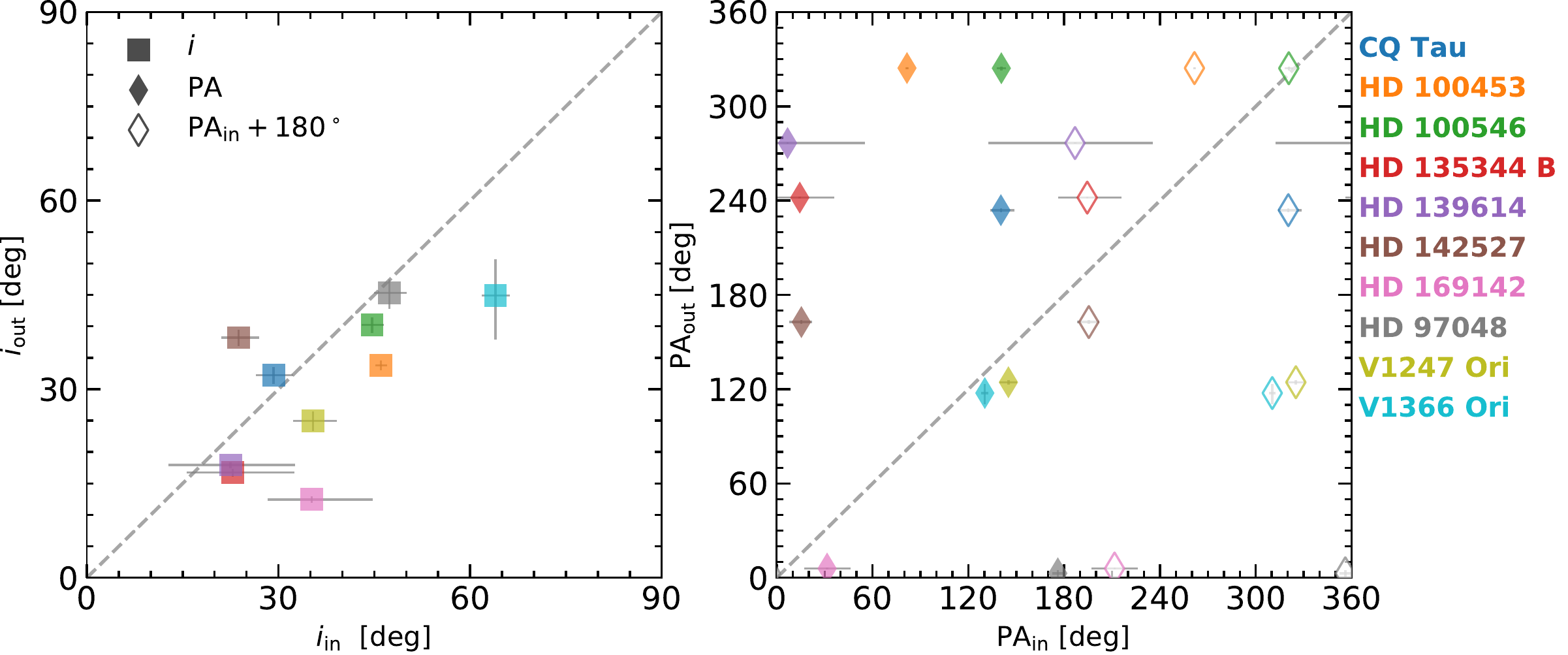}
\includegraphics[width=0.85\textwidth]{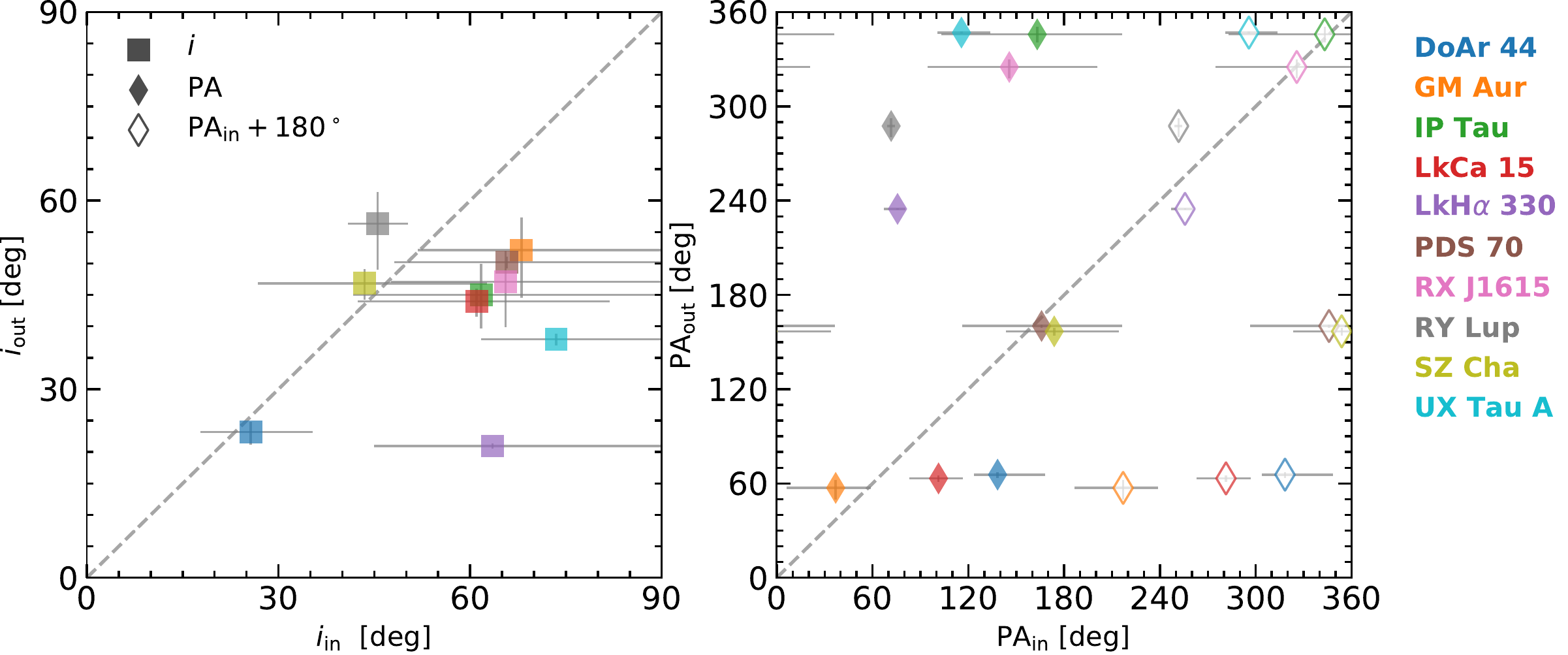}
\caption{
Misalignments between inner and outer disks in our sample of transition disks.
The upper and bottom panels show the misalignments for the most massive stars of our sample (Herbig Ae/Be and intermediate-mass T Tauri stars) and for the T Tauri stars, respectively.  \textit{Left panels}:
Comparison of outer and inner disk inclinations. The dashed line indicates perfect alignment in inclination.
\textit{Right panels}:
Same as left panel but for inner and outer disk position angles.
As we do not know the true orientation of the inner disk, we present both possibilities for the inner disk near sides.
The full-colored markers represent the position angles from the GRAVITY fits and the white markers with the colored edges have an additional offset of 180\degr. 
}
\label{fig:misalignments_2}
\end{figure*}

Figure~\ref{fig:misalignments_2} compares inner and outer inclinations (left panel) and position angles (right panel), respectively.
The dashed lines indicate the values for which the inner and outer inclination and position angle are equal. Due to the uncertainty of the inner disk position angle, we present both solutions with $\mathrm{PA}_\mathrm{in}\in[0\degr,180\degr]$ ($\Delta\theta_1$) as measured from the GRAVITY observables and $\mathrm{PA}_\mathrm{in}+180\degr$ ($\Delta\theta_2$). If both the inclination and either of the two position angle solutions overlap with the dashed lines, it is possible that inner and outer disks are well aligned: this is supported for GM~Aur, IP~Tau, PDS~70, RX~J1615, and SZ~Cha within the 68\,\% confidence intervals.
For the remaining targets, our data is indicating that inner and outer disks exhibit a possible misalignment between inner and outer disk regions, which significance will be discussed in the following section.

\subsection{Significance assessment }
As $\Delta\theta\in[0\degr,180\degr]$ and only configurations with perfectly aligned inner and outer disk geometries yield $\Delta\theta=0$, the derived misalignment angles usually deviate from zero, even within the provided uncertainties.
It is therefore difficult to conclude from these angles, whether a disk is significantly misaligned or not.
To properly test this underlying question in a statistical framework, we utilized an hypothesis testing approach.
As a null hypothesis we assumed that the inner and outer disk are perfectly aligned, that is, $i_\mathrm{in}=i_\mathrm{out}$ and $\mathrm{PA}_\mathrm{in}=\mathrm{PA}_\mathrm{out}$.
The outer disk inclination and position angle are in most cases better constrained than the inner disk geometry.
In addition to the posterior distributions of $i_\mathrm{out}$ and $\mathrm{PA}_\mathrm{out}$ we shifted the posterior distributions of $i_\mathrm{in}$ and $\mathrm{PA}_\mathrm{in}$ such that the median of the corresponding inner and outer distributions agreed.
This simulates perfectly aligned inner and outer disks while accounting for the uncertainties that arise from the data and model fitting.
From these simulated posterior distributions we calculated the misalignment angles $\Delta\theta_1^\mathrm{sim}$ and $\Delta\theta_2^\mathrm{sim}$.
These distributions describe how a perfectly aligned disk geometry would manifest in the fit results and derived parameters.

To test if our null hypothesis holds (i.e., the inner and outer parts of the analyzed disks are well aligned), we assessed how much the actual posterior distributions $\Delta\theta_1$ and $\Delta\theta_2$ and the simulations performed for the null hypothesis $\Delta\theta_1^\mathrm{sim}$ and $\Delta\theta_2^\mathrm{sim}$ were alike.
We performed a Kolmogorov-Smirnov (KS) test and used the KS distance $D_\mathrm{KS}\in[0,1]$ as a measure for the significance of the misalignment.
The details of this framework are explained in Appendix~\ref{sec:misalignment_angles}.

A value of $D_\mathrm{KS}$ close to zero indicates that inner and outer disks are almost perfectly aligned and $D_\mathrm{KS}$ close to unity indicates a significant misalignment.
This framework naturally assigns lower values of $D_\mathrm{KS}$ to targets with loosely constrained geometries (i.e., broad distributions of $\Delta\theta_1$ and $\Delta\theta_2$).
That way, it can be avoided that objects with large parameter uncertainties get misclassified as significantly misaligned with $D_\mathrm{KS}$ close to unity.
On the other hand, large uncertainties do not necessarily provide $D_\mathrm{KS}$ values close to zero as the empirical distribution functions of relatively broad distributions can vastly differ if the medians of both distributions are distinct.
We thus expect targets whose geometry is insufficiently characterized to exhibit intermediate values of $D_\mathrm{KS}\sim0.5$.
For that reason we \textit{qualitatively} identified three regimes with
\begin{enumerate}[(A)]
    \item $0.9<D_\mathrm{KS}\leq1$: targets that seem to exhibit significant misalignments between inner and outer disk structures;\label{itm:misalignment}
    \item $0.3\leq D_\mathrm{KS}\leq0.9$: ambiguous targets, whose misalignment status is difficult to evaluate based on the current data;\label{itm:ambiguous}
    \item and $0\leq D_\mathrm{KS}<0.3$: targets that show no significant signs of misalignments.\label{itm:no_misalignment}
\end{enumerate}

Our goal here is to qualitatively distribute the systems in categories of disks that are more or less likely to be misaligned than others. Whereas targets in categories~\eqref{itm:misalignment} and \eqref{itm:no_misalignment} provide strong statistical evidence to be either aligned or misaligned, we do not have conclusive evidence that significantly supports either state for the targets in the regime~\eqref{itm:ambiguous}. We note that we use the KS distance $D_\mathrm{KS}$ instead of the p-values, as the latter are all too small to be used to categorize the disks, possibly due to the size of our sample. 

\begin{figure}
\resizebox{\hsize}{!}{\includegraphics{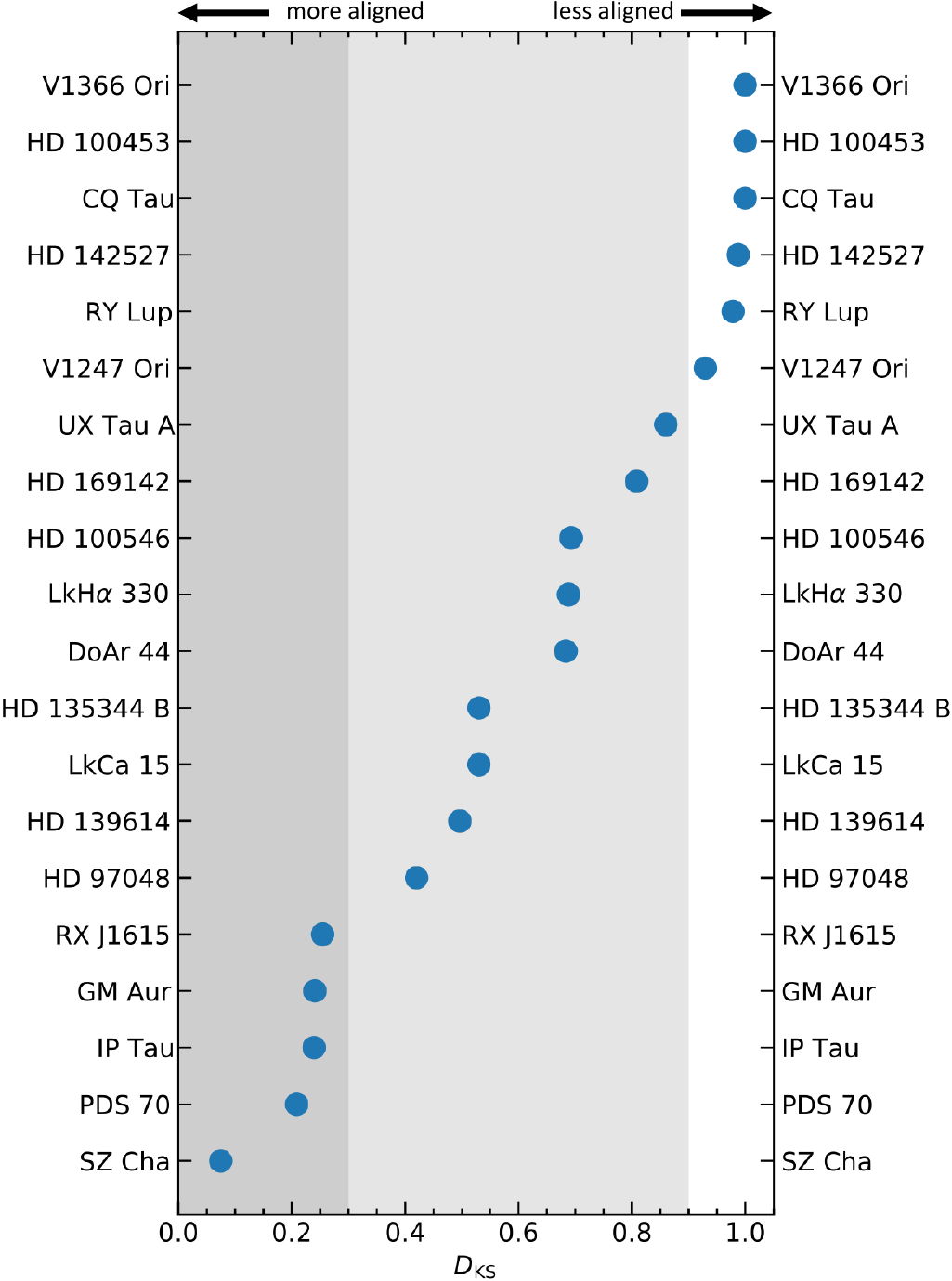}}
\caption{
KS statistics testing the null hypothesis of perfectly aligned disk geometries.
From bottom to top the targets are sorted from best ($D_\mathrm{KS}=0$) to worst agreement ($D_\mathrm{KS}=1$) with the null hypothesis.
The different shades of background color indicate targets with no significant misalignment (left), significant misalignment (right) and dubious cases in the middle.
}
\label{fig:ks_statistics}
\end{figure}

The numerical values of the Kolmogorov-Smirnov distances for all targets are listed in Table~\ref{tbl:results_misalignments} and visualized in Fig.~\ref{fig:ks_statistics}. The figure presents $D_\mathrm{KS}$ sorted from best (bottom) to least (top) agreement with the null hypothesis.  Histograms of the simulated and actual misalignment angle distributions can be found in Fig.~\ref{fig:misalignment_posterior_distributions} in Appendix~\ref{sec:misalignment_angles}.  Based on the selection criteria from Sect.~\ref{sec:analysis_misalignments} we find that SZ~Cha, PDS~70, IP~Tau, GM~Aur, and RX~J1615 show no significant signs for misalignments between inner and outer disks.
This is in good agreement with the inclinations and position angles that are presented in Fig.~\ref{fig:misalignments_2}: all of the aforementioned targets are compatible with the dashed lines that represent agreement of inner and outer disk orientations.
On the other hand, V1366~Ori, HD~100453, CQ~Tau, HD~142527, RY~Lup, and V1247~Ori exhibit values of $D_\mathrm{KS}$ close to unity, indicative of significant misalignments present in these systems.
The remaining targets fall into category~\eqref{itm:ambiguous}, for which we cannot conclusively tell whether a misalignment is present or not.

\section{Shadows in scattered light images}
\label{subsec:discussion_comparison}
Our targets were observed with VLT/SPHERE in scattered light, that is sensitive to illumination and shadowing effects. It is therefore informative to discuss these images in light of the inner disk orientations and misalignment angles derived before and check whether our predictions match the presence of shadows in scattered light images. For all targets, a gallery of this archival imagery is presented in Fig.~\ref{fig:sphere_scattered_light_all}, and subsets of images are shown in Figs.\,\ref{fig:shadow_predictions_a} and \ref{fig:shadow_predictions_b}. Each pixel in the disk images is scaled with squared radial distance to the star to account for the drop off in stellar illumination with increasing separation from the star and enhance faint features.  We note that the image scaling does not take into account the geometry of the scattering surface which would be needed to derive an accurate morphology of the observed features.  A plethora of substructures is visible for the transition disks of our sample that were discussed in the literature or are the subject of forthcoming publications (references given in Appendix~\ref{sec:scattered_light_data}). 

\subsection{Methodolody}
For targets that exhibit significant misalignments and the ambiguous cases, we further predict the locations of the shadows that depend on the morphology of the inner disk, and on the height of the scattering surface of the outer disk (Z$_{\rm{scat}}$). This analysis follows the framework proposed by \citet{min2017} and further used in \citet{benisty2018}. For given orientations of the inner and outer disk, and assuming Z$_{\rm{scat}}$, the line connecting the shadows can be defined with its position angle
\begin{equation}
\begin{aligned}
    \tan&\left(\alpha\right)=\\ &=\frac{\sin\left(i_\mathrm{in}\right)\cos\left(i_\mathrm{out}\right)\sin\left(\mathrm{PA}_\mathrm{in}\right) - \cos\left(i_\mathrm{in}\right)\sin\left(i_\mathrm{out}\right)\sin\left(\mathrm{PA}_\mathrm{out}\right)}{\sin\left(i_\mathrm{in}\right)\cos\left(i_\mathrm{out}\right)\cos\left(\mathrm{PA}_\mathrm{in}\right) - \cos\left(i_\mathrm{in}\right)\sin\left(i_\mathrm{out}\right)\cos\left(\mathrm{PA}_\mathrm{out}\right)}
\end{aligned}
\end{equation}
and offset in declination with respect to the star
\begin{equation}
    \eta = \frac{Z_{\rm{scat}} \cdot \cos\left(i_\mathrm{in}\right)}{\cos\left(i_\mathrm{out}\right)\sin\left(i_\mathrm{in}\right)\sin\left(\mathrm{PA}_\mathrm{in}\right) - \cos\left(i_\mathrm{in}\right)\sin\left(i_\mathrm{out}\right)\sin\left(\mathrm{PA}_\mathrm{out}\right)}\;.
\end{equation}

As explained previously, depending on whether the near sides of the inner and outer disk match, these equations provide two sets of solutions.  For all targets we assumed a fixed scattering height of $Z_\mathrm{scat}/R=0.1$ as done by \citet{min2017}. The radial separation to the star, $R$, at which we consider the scattering surface was determined by visual inspection of the scattered light images. We propagated the uncertainties of the inner and outer geometric disk parameters to obtain posterior distributions of $\alpha$ and $\eta$ for each target.

\subsection{Systems with significant misalignments}
\label{subsubsec:case_a}
\begin{figure}
\resizebox{\hsize}{!}{\includegraphics{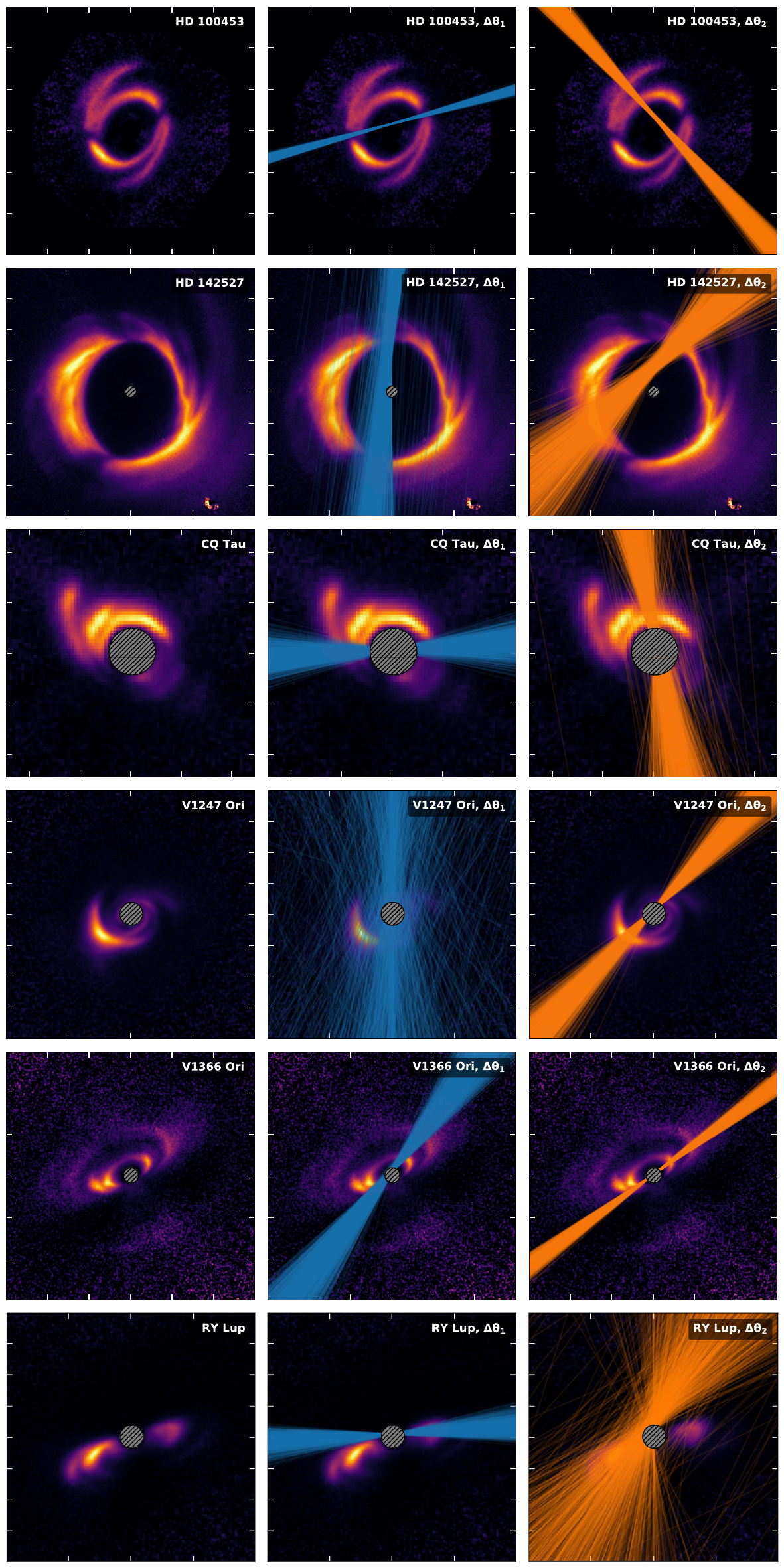}}
\caption{
Left: scattered light images. Middle, right: same, with predicted lines connecting putative shadows based on both potential misalignment configurations, $\Delta\theta_1$ and $\Delta\theta_2$, respectively. The colored lines are 1000 randomly drawn samples from our posterior distributions that describe the shadow locations. The gray circles indicate the coronagraph. 
}
\label{fig:shadow_predictions_a}
\end{figure}

Our analysis predicts misalignments for 
six targets, HD~100453, HD~142527, CQ~Tau, V1247 Ori, V1366 Ori and RY Lup, that we discuss in the following. For these, we show in Fig.~\ref{fig:shadow_predictions_a} the two families of shadows in blue and orange, with 1000 randomly drawn samples from our posterior distributions of $\alpha$ and $\eta$. 

\vskip 0.05cm
\textbf{HD~100453.} This system exhibits dark lines in the east and the west. Ellipse fitting of the scattered light ring led to an outer disk inclination of $\sim38\degr$ and position angle of $\sim142\degr$ \citep{benisty2017}. An inner disk inclination of $\sim48\degr$ and position angle of $\sim80\degr$ were considered to reproduce the observed shadows, leading to a misalignment angle of $\sim72\degr$. These values are in good agreement with our measurements: $i_\mathrm{out}\approx34\degr$;  $\mathrm{PA}_\mathrm{out}\approx324\degr$, and $i_\mathrm{in}=46\degr\pm1\degr$; $\mathrm{PA}_\mathrm{in}=82\degr\pm1\degr$, yielding misalignment angles of $\Delta\theta_1=67\degr\pm1\degr$ and $\Delta\theta_2=41\degr\pm1\degr$. 
As evident in Fig.\,\ref{fig:shadow_predictions_a}, the solutions indicated with blue  ($\Delta\theta_1$) lines perfectly fit the location of the shadows.  
For the calculation of $\alpha$ and $\eta$, we assumed $R=40$\,au, yielding $Z_\mathrm{scat}=4$\,au, as in  \citet{min2017}.

\textbf{HD~142527.}  The disk around this object also exhibits dark regions in the north and southeast of the star \citep[see e.g.,][]{fukagawa2006,avenhaus2017}. \citet{marino2015} derived that inner and outer disks must be misaligned by 70\degr, with the inner disk position angle of -8\degr, to explain the observed morphology.
This is marginally consistent with the position angle of $\mathrm{PA}_\mathrm{in}=15\degr\pm7\degr$ and the misalignment angle $\Delta\theta_1=59\degr\pm3\degr$ that we derived.
Nevertheless, our analysis clearly confirms a strong misalignment between inner and outer disks. The shadow predictions for $\Delta\theta_1$ as presented in Fig.~\ref{fig:shadow_predictions_a}, computed with $R=$175\,au, are in very good agreement with the observed shadow lanes.

\textbf{CQ~Tau.} 
The spiral structure seen in scattered light data of CQ~Tau was presented by \citet{uyama2020a}. In the SPHERE data presented in this work, two dark regions are apparent in the south and in the west (Benisty et al. in prep), at similar locations as the drop in peak intensity of the CO isotopologues seen in ALMA observations \citep{ubeira2019,wolfer2021}. We derived an inner and outer disk inclination of $23\degr\pm3\degr$ and $32\degr\pm1\degr$ with associated position angles of $140^{+7}_{-9}$\,\degr\, and $234\degr\pm1\degr$, respectively, indicating a significant misalignment. For $\Delta\theta_1=44^{+4}_{-3}$\,\degr\, the predicted shadows with $R=15$\,au partly agree with the darker regions observed in scattered light (Fig.~\ref{fig:shadow_predictions_a}).

\textbf{V1247 Ori}.
The disk shows asymmetries in scattered light \citep{ohta2016}, in particular two spiral features well seen in the SPHERE data (Kraus et al. in prep).  ALMA continuum data show an asymmetric ring with a crescent and their analysis suggests a possible misalignment between inner and outer disk casting a shadow at a position angle of approximately 25\degr \citep{kraus2017}. Our geometrical parameters suggest a significant misalignment between inner and outer disk components with $\Delta\theta_1=15^{+3}_{-4}$\,\degr\, and $\Delta\theta_2=59\degr\pm4\degr$. 
For $\Delta\theta_1$, the predicted locations of shadows (computed with $R$=100\,au) might trace some of the features observed in the scattered light image. Although the spiral structures makes it difficult to easily assess the presence of shadows, the shadows could be explaining why the spiral arms do not extend further. Neither solution of predicted shadows however agrees with a position angle of 25\degr. 

\textbf{V1366 Ori}.
The scattered light images, presented in \citet{deBoer2020}, do not show any clear signatures of shadows but it might be due to the high inclination of the system, for which shadows could too narrow to be detected. We derive misaligment angles of $\Delta\theta_1=22\degr\pm6\degr$ or $\Delta\theta_1=108\degr\pm7\degr$. 
The predicted shadow lanes, for $R=$85\,au, are presented in Fig.~\ref{fig:shadow_predictions_a}. Future observations with a higher signal-to-noise ratio and better angular resolution might reveal these predicted shadow lanes.

\textbf{RY~Lup}.
No clear shadows are visible in the scattered light images  \citep{langlois2018}, even though we derive a significant misalignment with $D_\mathrm{KS}=0.98$. Such nondetection can likely be attributed to the high inclination of the outer disk ($i_\mathrm{out}=56^{+7}_{-5}$\,\degr). Photometric and polarimetric observations of the system are indicative of a variable, highly inclined inner disk, with possible inner disk inclinations ranging from $\sim86\degr$ and $\sim55\degr$ \citep{manset2009}. The latter estimate is marginally consistent with our estimate $i_\mathrm{in}=46\degr\pm5\degr$.


\subsection{Ambiguous cases with known shadows}
\label{subsubsec:case_b}
\begin{figure}
\resizebox{\hsize}{!}{\includegraphics{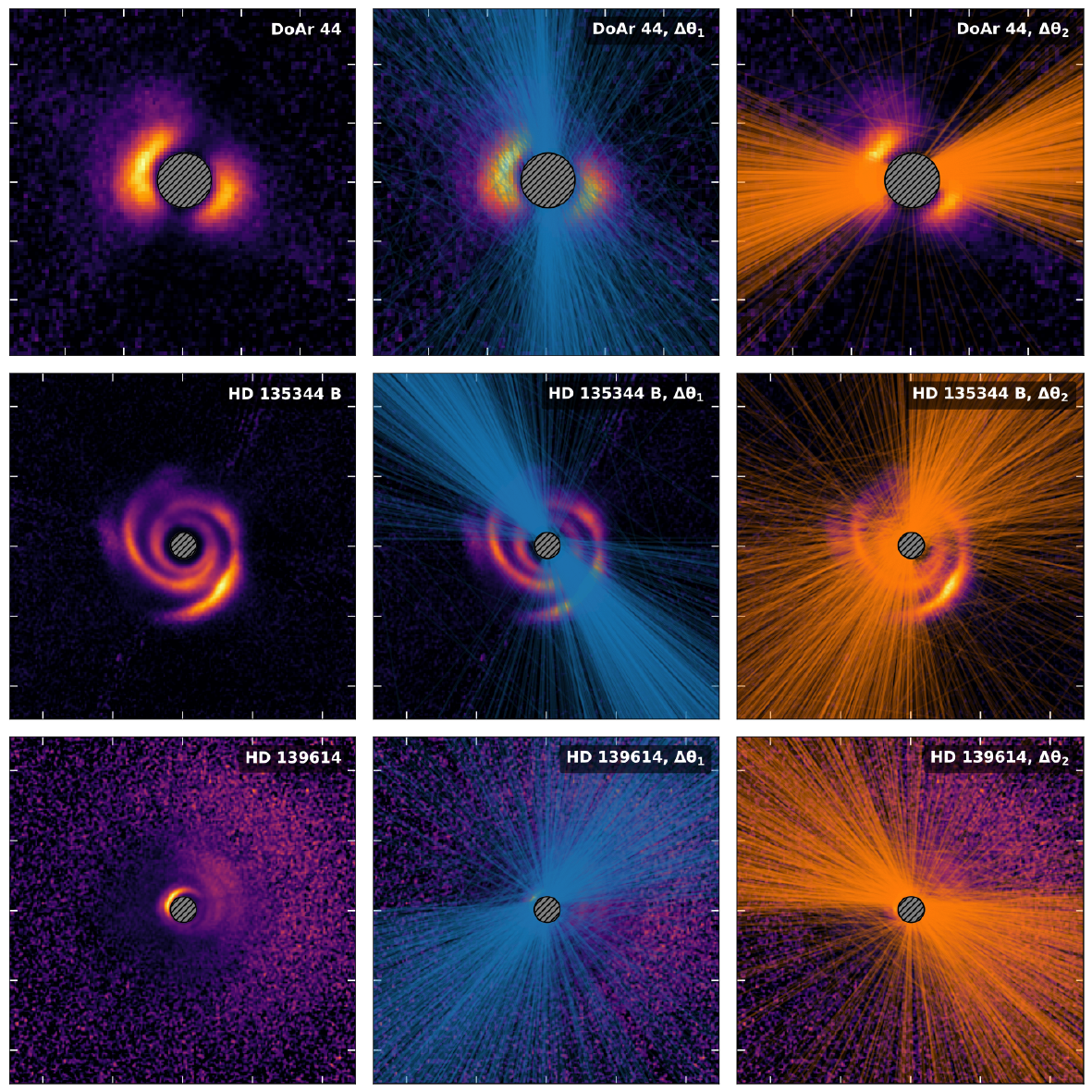}}
\caption{
Predicted shadows for ambiguous cases that exhibit shadows in scattered light.
The figure elements and image orientations are the same as for Fig.~\ref{fig:shadow_predictions_a}.
}
\label{fig:shadow_predictions_b}
\end{figure}

Three of the targets classified as ambiguous cases show shadow features in scattered light, while having intermediate values of $D_\mathrm{KS}$.
We present these cases below, and derive the predicted shadow positions in Fig.~\ref{fig:shadow_predictions_b}. 

\vskip 0.05cm
\textbf{DoAr~44}.
North-south shadows are clearly evident in the scattered light data \citep{avenhaus2018}. With an outer disk inclination and and a position angle of $20\degr$ and $60\degr$, respectively, estimated from ALMA dust continuum images, \citet{casassus2018} used  an inner disk inclination and position angle of 29.7$\degr$ and 134$\degr$, respectively, to reproduce the shadows. The geometries obtained from our analysis agree very well with these values, and we find misalignment angles of $\Delta\theta_1=27\degr\pm9\degr$ and $\Delta\theta_2=39\degr\pm9\degr$. We note that the solutions (with $R$=20\,au) for $\Delta\theta_1$ reproduce well the shadows locations.

With $D_\mathrm{KS}=0.68$, we do not rank DoAr~44 as a significantly misaligned disk, yet the strong shadowing in scattered light is clearly supporting this.
When inspecting the simulated and true posterior distributions of the misalignment angles (see Fig.~\ref{fig:misalignment_posterior_distributions} in Appendix~\ref{sec:misalignment_angles}), it is obvious that the medians of these distributions are distinct. 
The rather large uncertainties for each of the distributions, however, do not allow for a value of $D_\mathrm{KS}>0.9$.
Additional GRAVITY measurements that complete the $(u,v)$ plane coverage should allow to derive better constraints on the inner disk inclination and especially its position angle that are currently dominating the uncertainties of the misalignment angles.
These data would facilitate a confirmation of the misalignment hypothesis at higher statistical significance.

\vskip 0.05cm
\textbf{HD 135344 B}.
Even though this system shows clear signs of shadows in scattered light data, we do not find significant evidence for a misalignment with $D_\mathrm{KS}=0.53$. The large uncertainties for the inner disk parameters prohibited a higher value of $D_\mathrm{KS}$. Interestingly, for that disk, three narrow shadow lanes, instead of two, were found, and are variable on timescales of (at least) months \citep{stolker2017}. A misalignment angle of $\sim22\degr$ was inferred  \citep{stolker2016}, consistent with $\Delta\theta_2=17^{+8}_{-6}$\,\degr\, that we find. However, we find that while the shadow lanes associated with $\Delta\theta_2$ (computed with $R$=25\,au) might explain two shadows, the morphology of the inner disk is likely too complex to be captured by our rather simplistic models.


\vskip 0.05cm
\textbf{HD 139614.} More than half of the outer disk is not visible in scattered light. To generate such a broad shadow, the misalignment must be as small as 4\degr\; \citep{muro_arena2020}. The predicted inclination of the inner disk ($20.6\degr$) is in agreement with our findings $i_\mathrm{in}=22.5^{+9.7}_{-10.2}$\,$\degr$, while the predicted position angle (272\degr) differs from our estimate $\mathrm{PA}_\mathrm{in}=6.79^{+54.27}_{-48.64}$\,\degr. However, the uncertainties on the inner disk position angle are very large, due to the fact that disk is nearly face-on. While $\Delta\theta_1=27^{+13}_{-12}$\,\degr\, predicts shadow lanes (with $R$=30\,au) over the disk area that is illuminated by stellar light, $\Delta\theta_2=28^{+13}_{-12}$\,\degr\, agrees much better with the scattered light morphology.



In agreement with our derived misalignment angles and Kolmogorov-Smirnov statistics, none of the five targets in the category with aligned disks shows any significant shadow in scattered light.

\section{Discussion}
\label{sec:discussion}

In the previous sections, we found that the predictions from our joint analysis of VLTI and ALMA observations are consistent with the presence of misalignment in at least 6 out of the 20 disks of our sample while 5/20 disks appear to have inner disks well aligned with their outer disks. Unfortunately, in nearly half of our sample (9/20 disks) a possible misalignment is difficult to assess. This is partly due to the limited angular resolution of VLTI insufficient to resolve the inner disk of T Tauri stars or due to the specific geometry of some disks. While our analysis can therefore not be conclusive on the occurrence rate of misalignments in transition disks, it sets the methodology for future higher quality observations of these disks. In the following, we discuss the possible origin of disk misalignments, in the light of the nature of these inner disks in transition disks. 

\vspace{0.3cm}
\textbf{Observational evidence for misalignments.}
The presence of misaligned disk regions, or more generally of warps in disks, has been long known, in particular from the modulation present in the photometric data of young stars.  Most young stars are variable \citep{Herbst1994,Herbst2012} and a subclass of late-type objects, the so-called dippers, accounting for $\sim$30\% of the young stellar population \citep{Cody2018} show short duration extinction events, with highly variable photometric dips in both their occurrence timescales and shapes \citep[e.g.,][]{Cody2014}. These events were interpreted as being due to occulting dusty material very close to the star \citep{McGinnis2015,Bodman2017}. To hide part of the stellar surface, the dust shielding the star must be crossing our line of sight and therefore be moving on a very inclined orbit \citep{Bouvier2003,Bodman2017}. However, spatially resolved ALMA images of the outer disks of 24 dippers showed an isotropic distribution of inclinations, with some dipper disks close to face-on, indicating that disk misalignment might be quite common in young stars \citep{ansdell2018}. In addition to photometric variations and shadows in scattered light images, there are further observational indications of misalignments in disks, such as  different orientations of the dust continuum emission of transition disks \citep{francis2020} or perturbed kinematics of gas (HCO+) emission \citep{Loomis2017}. 

\vspace{0.3cm}
\textbf{Origin of the misalignments.} The  dippers with periodic light curves, can be explained by the presence of an inclined magnetic field that induces a warp or misalignment in the innermost disk regions, that regularly crosses our line of sight \citep{Bouvier1999}. In aperiodic dippers, the stochastic extinction events must result from rapid changes in the inner disk structure and orientation that can produce asymmetric stellar occultations, but it is unclear what exactly is driving these changes. An unstable accretion regime leading to a magnetically-induced warp  \citep{Kurosawa2013}, asymmetric density structures such as vortices induced by Rossby wave instabilities at the edge of a dead zone \citep{Meheut2012,Flock2017}, or dusty disk winds \citep{Bans2012} have been invoked. In variable intermediate-mass young stars (UX Ors), highly inclined dust clumps that sublimate close to the stars were suggested \citep{Grinin1996} but their origin is not clear. It is difficult to constrain such scenarios with our observations, as simultaneous photometric \textit{and} interferometric campaigns should be carefully planned to address this issue. However, we note that structural variability was already inferred in some objects, in particular in transition disks, from photometric, spectroscopic and interferometric studies \citep{CES2011,espaillat2014,Chen2019}.

Another scenario to consider is the presence of a massive companion that would induce a misalignment of specific disk radii. This other possibility appears as a natural one in transition disks as they show a dust-depleted cavity that can be carved by multiple planets \citep{Bae2019} or by a  stellar companion \citep{price2018}. When the companion angular momentum is greater than the inner disk's, the disk region within the companion's orbit will be tilted \citep{xg2013, Matsakos2017}. Using 3D simulations, \citet{Nealon2018} similarly found that, for a planet massive enough to carve a gap, the inner and outer disk will be misaligned, and \citet{Bitsch2013} found that in some cases, the disk can have a higher inclination than the planet. In the extreme case of binary systems, the inner disk can even break and precess \citep{Facchini2013}, leading to a variety of misalignment angles that could induce narrow or broad shadows in scattered light \citep{facchini2018, benisty2018}. Significant misalignments can also be induced by secular precession resonances in the case of high stellar-companion mass ratio \citep{Owen2017}. Interestingly, in the case of a triple system such as HD100453, a massive planet located within the cavity can be brought to a high inclination via the Kozai-Lidov effect and lead to an inner disk misalignment \citep{Martin2016,Gonzalez2020,Nealon2020,ballabio2021}. While our observations are not sensitive to low mass companions, stellar binary companions of high mass ratio, within the field of view of the telescopes (250\,mas for the ATs; 60\,mas for the UTs), would have been detected in our interferometric observations. The detection limits available from direct imaging data in the inner regions (within transition disk cavities) are in general limited by the complexities of the disk structures and with the uncertainties in the evolutionary models considered, the current estimates are around a few to $\sim$10 Jupiter masses \citep{AsensioTorres2021}. So far, out of our sample, only PDS70 has directly imaged planets \citep{keppler2018} while yet-unconfirmed candidate companions were claimed in the cavities of LkCa15 and HD100546 \citep{Quanz2013,Sallum2015}. 

Finally, another possibility is that the misalignment is an outcome of earlier stages of star formation \citep{Bate2018}. \citet{Bate2010} note that the inner disk could be misaligned with respect to the stellar rotation axis, due to the chaotic nature of star and disk formation. It was recently  proposed that the outer disk could be misaligned as a result of late accretion events from material accreted with a misaligned angular momentum \citep{Dullemond2019,kuffmeier2021}, a scenario possibly at play in SU\,Aur where both a large scale arm  \citep{akiyama2019} and a misaligned inner disk are observed \citep{Ginski2021}. In the sample of transition disks that we studied in this paper, two targets within the "ambiguous" subset, HD100546 and UX Tau\,A, show extended features in scattered light  \citep{Ardila2007,menard2020} but in the case of UX Tau\,A they seem well accounted for by a flyby.

\vspace{0.3cm}
\textbf{Misalignment with respect to the stellar rotation axis.} An interesting question which  could help understand the cause for the misalignment, is whether in our sample of disks, it is the inner disk or the outer disk that is misaligned with respect to the stellar rotation axis.  Within the transition disk sample analyzed in this paper, two objects are known dippers, LkCa15 \citep{Alencar2018} and DoAr44 \citep{bouvier2020b} and have been studied in great details. \citet{Alencar2018} find that LkCa15 must present an extended inner disk warp to reproduce the duration of the dips, and derive a stellar inclination $i_{\star}$ larger than 65$\degr$, a value consistent with our estimate of the inner disk inclination ($i_\mathrm{in}=61.0^{+18.6}_{-20.8}$\,\degr) although our estimate suffers large error bars. Spectro-polarimetric observations also confirm such a large inclination for the star \citep[$i_{\star}\ge70\degr$;][]{Donati2019}. It is in any case, much higher than the inclination that we derive for the outer disk ($i_\mathrm{out}=43.9^{+2.3}_{-2.1}$\,\degr) and that is obtained from continuum ALMA  observations \citep[50.16$\degr\pm$0.03$\degr$;][]{Facchini2020} and suggests that the outer disk is misaligned with respect to the star+inner disk system. Similarly, in DoAr\,44, \citet{bouvier2020b} find that the stellar inclination is $i_{\star}=30\degr\pm5\degr$ in agreement with our estimate of the inner disk inclination. Finally, in another transition disk (and dipper) not included in our sample, RXJ1604.3-2130, \citet{Davies2019} find that $i_{\star}\ge 61\degr$, while its outer disk is seen almost face-on \citep{Pinilla2018}. It is unclear whether all transition disks have an inner disk sharing the same orientation as the star, but these three examples seem to suggest that the misalignment occurred on the outer disk. In these objects, it is likely that there is, in addition to a global misalignment of the inner disk, an additional warp due to an  magnetic field inclined with respect to the stellar rotation axis \citep{Sicilia2020}. 

\vspace{0.3cm}
\textbf{Dependence on stellar properties.} The three objects discussed above are at the same time, dippers and transition disks, and are surrounding T Tauri stars, for which a strong magnetic field shaping the innermost disk region can be inferred. In contrast, it is unclear whether Herbig AeBe stars host strong magnetic fields \citep{Alecian2013} capable of warping the inner disk. However, it appears that shadows in scattered light, indirect tracers of disk misalignments, are predominant among the intermediate mass young stars, and in particular those with spiral arms and high near-infrared excess \citep{garufi2018}, possibly hinting at another mechanism than in the T Tauri case. It is interesting to note that such a high near-infrared excess must result from a larger disk surface to reprocess the stellar light. This could imply that in these disks dust grains are lifted to higher altitude, possibly through a disk wind or due to the dynamical interaction with a massive companion.  

\vspace{0.3cm}
Finally, we note that our current understanding of the nature of these inner disks in transition disk is limited by the complexity of the ongoing phenomena therein resulting in a strongly depleted cavity. In some cases, the dust in the inner disk even seems to be disappearing and rapidly replenished \citep[e.g.,][]{Sicilia2020} and variable shadows \citep{stolker2017} indicate a highly dynamical structure. Other cases still show significant accretion rates \citep{manara2014} and accretion signatures similar to continuous disks \citep{bouvier2020b}. Given the complexity and the short dynamical timescales of the inner disk regions, it is remarkable that in several cases our parametric modeling of the VLTI observations of the inner disks matches so well the shadowing exhibited by the scattered light observations. One of the next steps forward will consist in studying these inner disks and the mass flow within the cavity simultaneously through high spectral and high angular observations and if possible, follow the putative motions of shadows and/or substructures in the outer disk to constrain the origin of the misalignments. 


\section{Conclusions}
\label{sec:conclusions}
We investigated misalignments between inner and outer disk regions of transitional disk systems using VLTI/GRAVITY and ALMA observations. The analysis is conducted for a sample of 20 transitional disks around stars of various masses and luminosities, comprising T Tauri stars with masses as low as $0.6\,M_\sun$ to Herbig stars with masses of up to $3\,M_\sun$. The geometries of the inner disk regions were constrained by parametric model fits to the squared visibilities and closure phases collected with GRAVITY. We fit ALMA molecular line velocity maps using a Keplerian disk model.  From inner and outer disk orientations, we derived misalignment angles between the normal vectors of the planes of both disk components. We performed hypothesis testing in a Kolmogorov-Smirnov framework to assess whether the disks from our sample exhibited significant misalignments. For systems with significant misalignments or shadows observed in near-infrared scattered light, we simulated the positions of these shadows based on the orientations of the inner and outer disks.

\noindent Our main findings can be summarized as:
\begin{enumerate}
\item We find six systems whose outer and inner disk were significantly misaligned, 5 that appear to not have misalignments, and 9 targets for which we can not accurately evaluate this with the current data. 

\item In those for which we find that the outer and inner disk are significantly misaligned: For HD~100453 and HD~142527, the predicted shadow positions are in great agreement with scattered light observations. For CQ~Tau, the predicted shadow positions seems to probe parts of the disk that are less illuminated in scattered light. We do not find any evidence for shadows in the disks around V1247~Ori, V1366~Ori and RY~Lup.
Especially for the latter two systems this might be caused by the large inclination. Future observations at higher spatial resolution and with higher signal-to-noise ratio detections of these disks might reveal the shadow features that are predicted by our analysis.

\item Three disks around DoAr~44 and, HD~135344~B, and HD~139614 show dark regions in scattered light, even though we do not probe any significant misalignments with our analysis. This is mostly caused by the large uncertainties that we obtained for especially the inner disk geometries of these systems. In addition, the multiple shadow lanes and broad shadow regions in these disks are likely caused by complex physical processes that are not accurately modeled by our geometrical models.

\item With the current data available from VLTI, we can derive precise geometric constraints with uncertainties of a few degrees for the orientations of the inner disks of Herbig Ae/Be stars; we can measure the inner disk geometries of the T Tauri stars with marginal constraints of several tens of degrees. This correlation can be attributed to the angular size of the environment that needs to be resolved. Another important parameter for the quality of our fit results was the $(u,v)$ plane coverage. Observations that only scarcely sampled this plane provided on average worse constraints for the inner disk inclinations and position angles.
\end{enumerate}

We additionally discuss the possible origins of the misalignment in transition disks, although there is no consensus on the mechanisms responsible for it. As the nature of these inner disks in transition disk is very complex, future observing campaigns, at both high spectral and spatial resolution will be key to study the gas and dust content, structure, and dynamics therein, and help constrain the origin of the cavities and misalignments. 

\begin{acknowledgements}
We are very grateful to Jaehan Bae, Paola Pinilla and Nicolas Kurtovic for insightful discussions. We thank the referee for a very constructive report that helped improve the quality of the manuscript. 
This paper makes use of the following ALMA data: ADS/JAO.ALMA\#2011.0.00465.S, ADS/JAO.ALMA\#2012.1.00158.S, ADS/JAO.ALMA\#2012.1.00870.S, ADS/JAO.ALMA\#2013.1.00498.S,
ADS/JAO.ALMA\#2013.1.00163.S, ADS/JAO.ALMA\#2013.1.00592.S, ADS/JAO.ALMA\#2013.1.01075.S, ADS/JAO.ALMA\#2015.1.00490.S, ADS/JAO.ALMA\#2015.1.01301.S, ADS/JAO.ALMA\#2015.1.01600.S, ADS/JAO.ALMA\#2015.1.00888.S, ADS/JAO.ALMA\#2015.1.00986.S, ADS/JAO.ALMA\#2015.1.00192.S,
ADS/JAO.ALMA\#2016.A.00026.S, ADS/JAO.ALMA\#2016.1.00826.S, ADS/JAO.ALMA\#2016.1.00344.S, ADS/JAO.ALMA\#2017.A.00006.S, ADS/JAO.ALMA\#2017.1.01404.S, ADS/JAO.ALMA\#2017.1.01424.S,
ADS/JAO.ALMA\#2017.1.00449.S, ADS/JAO.ALMA\#2018.1.01055.L, ADS/JAO.ALMA\#2018.1.01255.S, ADS/JAO.ALMA\#2018.1.01302.S. 
ALMA is a partnership of ESO (representing its member states), NSF (USA), and NINS (Japan), together with NRC (Canada),  NSC and ASIAA (Taiwan), and KASI (Republic of Korea), in cooperation with the Republic of Chile. The Joint ALMA Observatory is operated by ESO, AUI/NRAO, and NAOJ. This project has received funding from the European Research Council (ERC) under the European Union’s Horizon 2020 research and innovation programme (grant agreements No. 101002188, No. 832428, No. 678194, No. 639889, No. 823823) and the
Deutsche Forschungsgemeinschaft (DFG, German Research Foundation) -
325594231, FOR 2634/2. R.G.L. acknowledges support from Science Foundation Ireland under Grant No. 18/SIRG/5597. This research has made use of the Jean-Marie Mariotti Center \texttt{Aspro} and \texttt{SearchCal}
services \footnote{Available at http://www.jmmc.fr/aspro; http://www.jmmc.fr/searchcal}. This research has used the SIMBAD database, operated at CDS, Strasbourg, France \citep{wenger2000}. This work has used data from the European Space Agency (ESA) mission {\it Gaia} (\url{https://www.cosmos.esa.int/gaia}), processed by the {\it Gaia} Data Processing and Analysis Consortium (DPAC, \url{https://www.cosmos.esa.int/web/gaia/dpac/consortium}). This publication makes use of VOSA, developed under the Spanish Virtual Observatory project supported by the Spanish MINECO through grant AyA2017-84089. To achieve the scientific results presented in this article we made use of the \emph{Python} programming language\footnote{Python Software Foundation, \url{https://www.python.org/}}, especially the \emph{SciPy} \citep{virtanen2020}, \emph{NumPy} \citep{numpy}, \emph{Matplotlib} \citep{Matplotlib}, \emph{emcee} \citep{foreman-mackey2013}, \emph{scikit-image} \citep{scikit-image}, \emph{scikit-learn} \citep{scikit-learn}, \emph{photutils} \citep{photutils}, and \emph{astropy} \citep{astropy_1,astropy_2} packages.

\end{acknowledgements}

\bibliographystyle{aa} 
\bibliography{mybib} 

\begin{appendix}

\section{Observing log}
\subsection{GRAVITY observations}
\label{sec:gravity_observations}
We present the observational setup and observing conditions for all our observations in Table~\ref{tbl:gravity_observation_setup}. All observations were carried our in single field mode mode using either the UTs or ATs. 

\begin{table*}[t]
\caption{
Setup and weather conditions of the GRAVITY observations.
}
\label{tbl:gravity_observation_setup}
\def\arraystretch{1.2}
\centering
\begin{tabular}{@{}lllllllll@{}}
\hline\hline
Target & Calibrator & Date & Configuration & $N_\mathrm{exp}$ & $R$\tablefootmark{a} & $\langle\omega\rangle$\tablefootmark{b} & $\langle\tau_0\rangle$\tablefootmark{c} & Program ID\\    
& & (yyyy-mm-dd) & & & & (\arcsec) & (ms)\\
\hline
CQ~Tau & TYC~1863-337-1 & 2018-11-18 & U1-U2-U3-U4 & 3 & MR & 0.4 & 3.7 & 0102.C-0210(A)\\
CQ~Tau & HD~40003 & 2020-01-30 & D0-G2-J3-K0 & 6 & HR & 0.9 & 2.7 & 0104.C-0567(A)\\
DoAr~44 & HD~147701 & 2019-06-22 & U1-U2-U3-U4 & 28 & HR & 0.7 & 9.7 & 0103.C-0097(A)\\
GM~Aur & BD+26~738 & 2018-10-29 & U1-U2-U3-U4 & 3 & MR & 0.5 & 8.7 & 0102.C-0210(A)\\
HD~97048 & HD~82554 & 2017-03-20 & A0-G1-J2-K0 & 6 & HR & 0.8 & 3.2 & 098.D-0488(A)\\
HD~97048 & HD~65810 & 2017-03-21 & A0-G1-J2-K0 & 6 & HR & 0.5 & 7.5 & 098.D-0488(A)\\
HD~100453 & HD~99909 & 2020-12-17 & D0-G2-J3-K0 & 3 & MR & 0.5 & 4.6 & 106.21JR.001\\
HD~100453 & HD~99909 & 2021-01-21 & D0-G2-J3-K0 & 2 & MR & 0.7 & 5.8 & 106.21JR.001\\
HD~100453 & HD~99909 & 2021-01-24 & D0-G2-J3-K0 & 2 & MR & 0.6 & 5.1 & 106.21JR.001\\
HD~100453 & HD~99909 & 2021-01-25 & D0-G2-J3-K0 & 3 & MR & 0.9 & 4.1 & 106.21JR.001\\
HD~100453 & HD~99909 & 2021-02-23 & D0-G2-J3-K0 & 2 & MR & 0.4 & 13.3 & 106.21JR.001\\
HD~100453 & HD~99909 & 2021-02-25 & D0-G2-J3-K0 & 2 & MR & 0.9 & 4.9 & 106.21JR.001\\
HD~100453 & HD~99909 & 2021-03-18 & D0-G2-J3-K0 & 2 & MR & 0.8 & 4.0 & 106.21JR.001\\
HD~100546 & HD~99264 & 2018-04-27 & U1-U2-U3-U4 & 9 & HR & 0.8 & 3.1 & 0101.C-0311(B)\\
HD~100546 & HD~101531 & 2019-01-12 & D0-G2-J3-K0 & 13 & HR & 0.9 & 3.9 & 0102.C-0408(A)\\
HD~100546 & HD~101531 & 2020-01-28 & D0-G2-J3-K0 & 6 & HR & 1.0 & 5.4 & 0104.C-0567(A)\\
HD~100546 & HD~99264 & 2020-02-04 & A0-B2-C1-D0 & 3 & HR & 1.1 & 1.7 & 0104.C-0567(C)\\
HD~135344B & HD~148703 & 2018-03-05 & A0-G1-J2-J3 & 7 & HR & 0.5 & 10.7 & 0100.C-0278(E)\\
HD~139614 & HD~148974 & 2019-03-20 & D0-G2-J3-K0 & 3 & HR & 0.6 & 4.4 & 0102.C-0408(D)\\
HD~142527 & HD~143118 & 2017-03-19 & A0-G1-J2-K0 & 7 & HR & 0.7 & 6.4 & 098.D-0488(A)\\
HD~169142 & HD~169830 & 2017-08-17 & A0-G1-J2-K0 & 4 & HR & 0.9 & 4.4 & 099.B-0162(F)\\
HD~169142 & HD~317458 & 2019-05-24 & A0-G1-J2-J3 & 8 & HR & 1.0 & 3.3 & 0103.C-0347(C)\\
IP~Tau & BD+23~734 & 2018-10-28 & U1-U2-U3-U4 & 3 & MR & 0.6 & 4.0 & 0102.C-0210(A)\\
LkCa~15 & BD+23~734 & 2018-11-19 & U1-U2-U3-U4 & 3 & MR & 0.6 & 7.0 & 0102.C-0210(A)\\
LkCa~15 & BD+23~734 & 2020-12-17 & D0-G2-J3-K0 & 3 & MR & 0.6 & 7.3 & 106.21JR.001\\
LkH$\alpha$~330 & TYC~2345-46-1 & 2018-10-28 & U1-U2-U3-U4 & 5 & MR & 0.6 & 5.4 & 0102.C-0210(A)\\
PDS~70 & HD~124058 & 2018-06-25 & U1-U2-U3-U4 & 3 & MR & 0.7 & 2.7 & 0101.C-0281(B)\\
RX~J1615 & HD~145320 & 2018-06-29 & U1-U2-U3-U4 & 3 & MR & 1.2 & 2.1 & 0101.C-0281(B)\\
RY~LUP & HD~110878 & 2017-06-11 & U1-U2-U3-U4 & 6 & HR & 0.8 & 2.7 & 099.C-0667(B)\\
SZ~Cha & TYC~9411-934-1 & 2018-06-04 & U1-U2-U3-U4 & 3 & MR & 0.4 & 4.7 & 0101.C-0281(A)\\
UX~Tau~A & HD~285803 & 2018-11-19 & U1-U2-U3-U4 & 4 & MR & 0.6 & 7.3 & 0102.C-0210(A)\\
V1247~Ori & HD~37409 & 2018-11-19 & U1-U2-U3-U4 & 3 & MR & 0.6 & 6.3 & 0102.C-0210(A)\\
V1247~Ori & HD~37409 & 2020-12-25 & A0-B2-D0-J3 & 3 & MR & 0.8 & 4.1 & 106.21JR.001\\
V1247~Ori & HD~37409 & 2021-01-18 & A0-G1-J2-J3 & 3 & MR & 0.8 & 3.5 & 106.21JR.001\\
V1366~Ori & HD~35262 & 2018-12-20 & U1-U2-U3-U4 & 4 & MR & 0.8 & 9.1 & 0102.C-0210(A)\\
V1366~Ori & HD~35262 & 2020-12-25 & A0-B2-D0-J3 & 3 & MR & 1.0 & 3.7 & 106.21JR.001\\
V1366~Ori & HD~35262 & 2021-01-17 & A0-G1-J2-J3 & 3 & MR & 0.6 & 6.5 & 106.21JR.001\\
V1366~Ori & HD~35262 & 2021-01-18 & A0-G1-J2-J3 & 3 & MR & 0.5 & 6.1 & 106.21JR.001\\
\hline
\end{tabular}
\tablefoot{
\tablefoottext{a}{Spectral resolution is either medium (MR, $R\approx500$) or high (HR, $R\approx4000$).}
\tablefoottext{b}{$\langle\omega\rangle$ denotes the average seeing conditions during the science observation.}
\tablefoottext{c}{$\tau_0$ denotes the coherence time during the science observation.}
}
\end{table*}

\subsection{ALMA observations}
\label{sec:alma_observations}

For each target we search the ALMA archive for the CO isotopologue line data with the best combination of spatial resolution, spectral resolution and sensitivity. Generally the $^{12}$CO line is the brightest and thus chosen for our analysis, but in a few cases only the $^{13}$CO line was available. The properties of the ALMA line cube observations and references is presented in Table~\ref{tbl:alma_observation_setup}. For most targets, the authors of the papers where these data were published provided the reduced fits files (PC).
For GM~Aur, HD~139614, SZ~Cha and IP~Tau, the datasets were downloaded directly from the ALMA archive (programs 2018.1.01055.L, PI Öberg; 2015.1.01600.S, PI Panic; 2013.1.00163.S, PI Simon; 2013.1.01075.S, PI Daemgen). For all these targets but GM Aur, the data were reduced using the provided calibration scripts and imaged using the \texttt{tclean} task with natural weighting. 

\begin{table*}
\caption{
Basic properties of the ALMA line observations
}
\label{tbl:alma_observation_setup}
\def\arraystretch{1.2}
\centering
\begin{tabular}{@{}llllllll@{}}
\hline\hline
Target  &  Program ID & Origin\tablefootmark{a} & Line & Beam & Vel res. & RMS & Refs \\    
&  & & & & (km s$^{-1}$)&(mJy bm$^{-1}$)&\\
\hline
CQ~Tau & 2013.1.00498.S, & PC & $^{12}$CO 2--1 & 0.12$\times$0.10" & 0.5 & 1.2 & 1 \\ 
&2016.A.00026.S, &&&&&& \\
&2017.1.01404.S &&&&&& \\
DoAr~44 & 2012.1.00158.S & PC & $^{13}$CO 3--2 & 0.25$\times$0.19" & 0.5 & 6.8 & 2 \\ 
GM~Aur & 2018.1.01055.L & ADP & $^{13}$CO 2--1 & 0.55$\times$0.36" & 0.17 & 2.7 &  - \\ 
HD~97048 & 2016.1.00826.S & PC & $^{13}$CO 3--2 & 0.11$\times$0.07 & 0.12 & 3.6 & 3\\ 
HD~100453 & 2017.1.01424.S & PC & $^{12}$CO 3--2 & 0.054$\times$0.052" & 0.42 & 0.95 & 4\\ 
HD~100546 & 2016.1.00344.S & PC & $^{12}$CO 2--1 & 0.076$\times$0.057" & 0.5 & 1.2 & 5 \\
HD~135344B & 2012.1.00158.S & PC & $^{13}$CO 3--2 & 0.26$\times$0.21" & 0.5 & 9.8 & 2 \\
HD~139614 & 2015.1.01600.S & MR & $^{13}$CO 2--1 & 0.80$\times$0.58" & 0.5 & 9.6 & - \\
HD~142527 & 2011.0.00465.S & PC & $^{12}$CO 3--2 & 0.57$\times$0.35" & 0.5 & 9.3 & 6 \\
HD~169142 & 2013.1.00592.S & PC & $^{12}$CO 2--1 & 0.18$\times$0.13" & 0.06 & 1.2 & 7 \\ 
& 2015.1.00490.S &&&&&& \\
& 2015.1.01301.S &&&&&& \\
IP~Tau & 2013.1.00163.S & MR & $^{12}$CO 2--1 & 0.25$\times$0.21" & 1.0 & 5.6 & 8 \\
LkCa~15 & 2018.1.01255.S & PC & $^{12}$CO 2--1 & 0.41$\times$0.30" & 0.04 & 5.5 & 9 \\ 
LkH$\alpha$~330 & 2018.1.01302.S & PC & $^{13}$CO 2--1 & 0.067$\times$0.046" & 1.5 & 0.32 & 10 \\ 
PDS~70 & 2017.A.00006.S & PC & $^{12}$CO 3--2 & 0.11$\times$0.098" & 0.43 & 1.1 & 11\\ 
RX~J1615.3-3255 & 2012.1.00870.S & PC & $^{12}$CO 3--2 & 0.10$\times$0.09" & 0.35 & 3.4 & 12 \\ 
RY~Lup & 2017.1.00449.S & PC & $^{12}$CO 3--2 & 0.22$\times$0.17" & 0.85 & 3.5 & 13 \\ 
SZ~Cha & 2013.1.01075.S & MR & $^{12}$CO 3--2 & 0.82$\times$0.43" & 0.5 & 26 & - \\
UX~Tau~A & 2015.1.00888.S & PC & $^{12}$CO 3--2 & 0.20$\times$0.16" & 0.21 & 3.4 & 14 \\
V1247~Ori & 2015.1.00986.S & PC & $^{12}$CO 3--2 & 0.047$\times$0.030" & 1.0 & 2.0 & 15\\
V1366~Ori & 2015.1.00192.S & PC & $^{12}$CO 2--1 & 0.70$\times$0.54" & 0.2 & 10 & 16 \\
\hline
\end{tabular}\\
\tablefoot{
\tablefoottext{a}{PC: Private communication with main author, ADP: Archival data product, MR: Manual reduction and imaging of archival data.}
}
\tablebib{
(1)~\citet{wolfer2021};
(2)~\citet{vandermarel2016};
(3)~\citet{pinte2019};
(4)~\citet{rosotti2020};
(5)~\citet{perez2020};
(6)~\citet{vandermarel2021};
(7)~\citet{Yu2021}; 
(8)~\citet{simon2017}; 
(9)~\textcolor{blue}{Facchini et al. (in prep.)};
(10)~\textcolor{blue}{Pinilla et al. (in prep.)};
(11)~\citet{keppler2019}; 
(12)~\textcolor{blue}{P\'erez et al. (in prep.)}; 
(13)~\textcolor{blue}{van der Marel et al. (in prep.)}; 
(14)~\citet{menard2020}; 
(15)~\citet{kraus2017}; 
(16)~\textcolor{blue}{van der Plas et al. (in prep.)}
}
\end{table*}

\subsection{Spectral energy distributions}
\label{sec:target_photometry}

The photometry collected from Tycho-2 (\textit{B} and \textit{V} bands), Gaia EDR3 ($G_\mathrm{BP}$, \textit{G}, and $G_\mathrm{RP}$ bands), USNO-B (\textit{R} and \textit{I} bands), 2MASS (\textit{J}, \textit{H}, and $K$ bands) catalogs is presented in Table~\ref{tbl:target_photometry}.
\begin{sidewaystable*}
\caption{
Photometric data of the targets in our sample.
}
\label{tbl:target_photometry}
\def\arraystretch{1.2}
\centering
\begin{tabular}{@{}lllllllllll@{}}
\hline\hline
Target & \textit{B} &  \textit{V} & $G_\mathrm{BP}$ (EDR3) & \textit{G} (EDR3) & $G_\mathrm{RP}$  (EDR3) & \textit{R} & \textit{I} & \textit{J} & \textit{H} & $K$\\    
& (mag) & (mag) & (mag) & (mag) & (mag) & (mag) & (mag) & (mag) & (mag) & (mag)\\
\hline
CQ TAU & $11.26\pm0.17$ & $10.59\pm0.05$ & $10.727\pm0.138$ & $10.229\pm0.038$ & $9.569\pm0.099$ & $10.14\pm0.30$ & $9.81\pm0.30$ & $7.93\pm0.02$ & $7.06\pm0.02$ & $6.17\pm0.02$ \\
DoAr 44 & $13.72\pm0.25$ & $13.06\pm0.20$ & $12.828\pm0.009$ & $11.862\pm0.004$ & $10.864\pm0.007$ & $12.35\pm0.30$ & $10.56\pm0.30$ & $9.23\pm0.02$ & $8.25\pm0.06$ & $7.61\pm0.02$ \\
GM AUR & $12.73\pm0.25$ & $12.30\pm0.20$ & $12.410\pm0.015$ & $11.727\pm0.005$ & $10.789\pm0.011$ & $11.02\pm0.30$ & $10.61\pm0.30$ & $9.34\pm0.02$ & $8.60\pm0.02$ & $8.28\pm0.02$ \\
HD 100453 & $8.07\pm0.02$ & $7.80\pm0.01$ & $7.871\pm0.003$ & $7.735\pm0.003$ & $7.458\pm0.004$ & $7.63\pm0.30$ & $7.50\pm0.30$ & $6.95\pm0.03$ & $6.39\pm0.04$ & $5.60\pm0.02$ \\
HD 100546 & $6.71\pm0.01$ & $6.70\pm0.01$ & $6.675\pm0.003$ & $6.686\pm0.003$ & $6.624\pm0.005$ & $6.70\pm0.30$ & $6.71\pm0.30$ & $6.43\pm0.02$ & $5.96\pm0.03$ & $5.42\pm0.02$ \\
HD 135344B & $9.18\pm0.03$ & $8.72\pm0.01$ & $8.786\pm0.005$ & $8.514\pm0.003$ & $8.064\pm0.005$ & $8.41\pm0.30$ & $8.18\pm0.30$ & $7.28\pm0.03$ & $6.59\pm0.03$ & $5.84\pm0.02$ \\
HD 139614 & $8.47\pm0.02$ & $8.25\pm0.01$ & $8.330\pm0.003$ & $8.219\pm0.003$ & $7.991\pm0.004$ & $8.12\pm0.30$ & $8.01\pm0.30$ & $7.67\pm0.03$ & $7.33\pm0.04$ & $6.75\pm0.03$ \\
HD 142527 & $9.01\pm0.03$ & $8.35\pm0.01$ & $8.502\pm0.004$ & $8.115\pm0.003$ & $7.498\pm0.004$ & $7.91\pm0.30$ & $7.58\pm0.30$ & $6.50\pm0.03$ & $5.72\pm0.03$ & $4.98\pm0.02$ \\
HD 169142 & $8.41\pm0.02$ & $8.16\pm0.01$ & $8.205\pm0.003$ & $8.069\pm0.003$ & $7.791\pm0.004$ & $8.00\pm0.30$ & $7.88\pm0.30$ & $7.31\pm0.02$ & $6.91\pm0.04$ & $6.41\pm0.02$ \\
HD 97048 & $8.76\pm0.02$ & $8.45\pm0.01$ & $8.556\pm0.003$ & $8.352\pm0.003$ & $7.956\pm0.004$ & $8.27\pm0.30$ & $8.12\pm0.30$ & $7.27\pm0.02$ & $6.66\pm0.05$ & $5.94\pm0.03$ \\
IP TAU & $14.28\pm0.25$ & $13.62\pm0.20$ & $13.430\pm0.019$ & $12.423\pm0.006$ & $11.407\pm0.014$ & $12.29\pm0.30$ & $10.98\pm0.30$ & $9.78\pm0.02$ & $8.89\pm0.02$ & $8.35\pm0.02$ \\
LKCA 15 & $12.99\pm0.25$ & $12.03\pm0.20$ & $12.325\pm0.028$ & $11.591\pm0.009$ & $10.686\pm0.021$ & $11.43\pm0.30$ & $10.97\pm0.30$ & $9.42\pm0.02$ & $8.60\pm0.02$ & $8.16\pm0.02$ \\
LkH$\alpha$ 330 & $13.52\pm0.25$ & $12.59\pm0.20$ & $12.294\pm0.003$ & $11.412\pm0.003$ & $10.467\pm0.004$ & $11.11\pm0.30$ & $10.29\pm0.30$ & $8.83\pm0.03$ & $7.92\pm0.03$ & $7.03\pm0.02$ \\
PDS 70 & $13.08\pm0.25$ & $12.01\pm0.20$ & $12.388\pm0.010$ & $11.606\pm0.004$ & $10.714\pm0.007$ & $11.20\pm0.30$ & $10.34\pm0.30$ & $9.55\pm0.02$ & $8.82\pm0.04$ & $8.54\pm0.02$ \\
RX J1615 3-3255 & $12.99\pm0.25$ & $12.21\pm0.20$ & $12.201\pm0.011$ & $11.494\pm0.004$ & $10.640\pm0.008$ & $11.16\pm0.30$ & $10.49\pm0.30$ & $9.44\pm0.02$ & $8.78\pm0.02$ & $8.56\pm0.02$ \\
RY LUP & $12.66\pm0.25$ & $11.17\pm0.11$ & $11.504\pm0.070$ & $10.993\pm0.020$ & $10.118\pm0.052$ & $10.29\pm0.30$ & $9.47\pm0.30$ & $8.55\pm0.02$ & $7.69\pm0.05$ & $6.98\pm0.02$ \\
SZ CHA & $12.78\pm0.25$ & $12.27\pm0.20$ & $12.473\pm0.048$ & $11.730\pm0.016$ & $10.708\pm0.036$ & $10.66\pm0.30$ & $10.24\pm0.30$ & $9.25\pm0.03$ & $8.41\pm0.04$ & $7.76\pm0.03$ \\
UX TAU A & $13.00\pm0.25$ & $11.18\pm0.11$ & $11.684\pm0.080$ & $11.245\pm0.023$ & $10.392\pm0.059$ & $10.30\pm0.30$ & $9.48\pm0.30$ & $8.62\pm0.02$ & $7.96\pm0.02$ & $7.55\pm0.02$ \\
V1247 ORI & $10.10\pm0.07$ & $9.82\pm0.02$ & $9.940\pm0.004$ & $9.770\pm0.003$ & $9.448\pm0.004$ & $9.64\pm0.30$ & $9.50\pm0.30$ & $8.88\pm0.03$ & $8.20\pm0.05$ & $7.41\pm0.03$ \\
V1366 ORI & $10.14\pm0.07$ & $9.85\pm0.02$ & $9.938\pm0.006$ & $9.860\pm0.003$ & $9.680\pm0.005$ & $9.66\pm0.30$ & $9.52\pm0.30$ & $9.26\pm0.03$ & $8.48\pm0.03$ & $7.68\pm0.02$ \\
\hline
\end{tabular}
\end{sidewaystable*}

\section{GRAVITY inner disk modeling}
\label{sec:gravity_best_fits}

In this section we present the parametric models that we use to fit the GRAVITY data. Sect.~\ref{subsec:appendix_inner_disk_model} reiterates the main components of the analytical model derived by \citet{lazareff2017} that is used to describe the observed complex visibilites.

\subsection{Model description}
\label{subsec:appendix_inner_disk_model}

\subsubsection{Complex visibilities}
\label{subsubsec:model_vis2}

We model the complex visibilities $V(u,v,\lambda)$ as described by \citet{lazareff2017} using a combination of stellar (s), circum-stellar (c), and halo (h) contributions.
For a proper derivation of the used model, the reader is referred to Sect.~3 of \citet{lazareff2017}.
The final model is provided by
\begin{equation}
    V(u,v,\lambda)=\frac{f_\mathrm{s}(\lambda_0/\lambda)^{k_\mathrm{s}}+V_\mathrm{c}(u,v,\lambda)(\lambda_0/\lambda)^{k_\mathrm{c}}}{(f_\mathrm{s}+f_\mathrm{h})(\lambda_0/\lambda)^{k_\mathrm{s}}+f_\mathrm{c}(\lambda_0/\lambda)^{k_\mathrm{c}}}\;,
    \label{eqn:complex_visibilities}
\end{equation}
where $\lambda_0=2.25$\,\textmu m is defined as the GRAVITY reference wavelength as before; $f_\mathrm{s}$, $f_\mathrm{c}$, and $f_\mathrm{h}$ denote the fraction of the total flux at wavelength $\lambda_0$ within the VLTI field of view that originate from the star, the circumstellar material, and the halo component, respectively; and, $k_\mathrm{s}$ and $k_\mathrm{c}$ represent the spectral indices of stellar and circumstellar components, respectively.

Eq.~\ref{eqn:complex_visibilities} assumes that the star is an unresolved point source for all baselines. $V_\mathrm{c}(u,v,\lambda)$ are the visibilities that are associated with the circumstellar dust. All three quantities $f_\mathrm{s}$, $f_\mathrm{c}$, and $f_\mathrm{h}$ are positive semi-definite and related by
\begin{equation}
    f_\mathrm{s}+f_\mathrm{c}+f_\mathrm{h}=1\;.
\end{equation}
The fractional flux contribution $f_\mathrm{h}$ originates from a halo that is fully resolved at all the baseline configurations.
This extended component mimics scattered light \citep[e.g.,][]{pinte2008}. 

The spectral dependence of the star and the circumstellar component is assumed to follow a power law whose spectral index at frequency $\nu$ is defined as
\begin{equation}
    k=\frac{\mathrm{d}\log_{10}(F_\nu)}{\mathrm{d}\log_{10}(\nu)}\;.
    \label{eqn:spectral_index}
\end{equation}
As the halo component is assumed to originate from scattered starlight, this component inhibits the same wavelength dependence -- and therefore the identical spectral index -- as the star.
We derived the stellar spectral indices, $k_\mathrm{s}$, at GRAVITY reference wavelength $\lambda_0=2.25$\,\textmu m, assuming blackbody emission with the host star temperatures reported in Table~\ref{tbl:sample_selection}.

The closure phases that are corresponding to the disk model of  Eq.~\eqref{eqn:complex_visibilities} can be derived as
\begin{equation}
    \Phi(u_1, v_1, u_2, v_2, \lambda) = \arg\left(B_{123}\right)= \arctan\left(\frac{\Re\left(B_{123}\right)}{\Im\left(B_{123}\right)}\right)
\end{equation}
with the triple product
\begin{equation}
    B_{123}(u_1, v_1, u_2, v_2, \lambda)=V(u_1, v_1,\lambda) V(u_2, v_2, \lambda) V^*(u_1+u_2, v_1+v_2, \lambda)\;.
\end{equation}
and with $\Re$ and $\Im$ refer to a complex number's real and imaginary part, respectively, and $^*$ denotes the complex conjugate.

\subsubsection{Disk model}
\label{subsubsec:model_inner_disk}
A detailed analytical derivation of the complex visibilities associated with the models is presented in Sect.~3.6 of \citet{lazareff2017} and we recall the main parameters in Table~\ref{tbl:gravity_model_priors}. With this prescription, we describe asymmetric ellipsoid and ring-like morphologies, and discuss in the following key parameters. 

The parameter $f_\mathrm{Lor}\in[0,1]$ describes the underlying radial profile of the emission: $f_\mathrm{Lor}=0$ corresponds to a Gaussian intensity profile that decays exponentially as $\exp(-r^2)$, $f_\mathrm{Lor}=1$ describes a pseudo-Lorentzian profile that is proportional to $r^{-3}$. Intermediate values of $f_\mathrm{Lor}$ refer to a combination of Gaussian and Lorentzian brightness distribution.

The parameter $l_a$ is a measure of the half-flux extent of the disk. The parameter is connected via
\begin{equation}
    l_a=\log_{10}\left(\frac{a}{1\,\mathrm{mas}}\right)
\end{equation}
to the half-flux radius
\begin{equation}
    a=\left(a_r^2+a_k^2\right)\;,
\end{equation}
which is derived from the geometrical parameters $a_\mathrm{r}$ and $a_\mathrm{k}$.
$a_\mathrm{r}$ denotes the angular radius of the ring and $a_\mathrm{k}$ describes the angular radius of the kernel that is convolved with the disk geometry.

The parameter 
\begin{equation}
   l_\mathrm{kr}=\log_{10}\left(\frac{a_\mathrm{k}}{a_\mathrm{r}}\right)
\end{equation}
captures the logarithmic ratio between these two variables.
Accordingly, a value of $l_\mathrm{kr}<0$ corresponds to ring-like geometries with $a_\mathrm{k}\ll a_\mathrm{r}$ and $l_\mathrm{kr}>0$ creates ellipsoids with $a_\mathrm{k}\gg a_\mathrm{r}$.

The pair of parameters $c_j$ and $s_j$ describe azimuthal modulation amplitudes that create azimuthal brightness asymmetries in the disk models.
Those terms are required to describe well-resolved disks that are viewed at nonzero inclination. In cases such as these, the far side of the dust sublimation rim appears more luminous than the front side and azimuthal modulations are required for a proper description of the observed geometry.
For each order $m$ of modulation amplitudes that are included in the parametrization, one additional cosine $c_j$ and a sine $s_j$ term need to be considered. 

The parameters that are most important within the scope of this work are $\cos\left(i_\mathrm{in}\right)$ and $\mathrm{PA}_\mathrm{in}$, which describe the inclination and position angle of the inner disk.
From the GRAVITY observations, we cannot in general tell which side of the disk is closer to the observer (i.e., the front side) and which is farther away (i.e., the back side). Therefore, the inner disk inclination is defined as a positive angle in the range $[0\degr,90\degr]$ with $i_\mathrm{in}=0\degr$ and $i_\mathrm{in}=90\degr$ describing face-on and edge-on disk geometries, respectively.
The position angle is defined on the interval $\mathrm{PA}_\mathrm{in}\in[0\degr,180\degr]$ and describes the positive rotation of the disk's major axis with respect to north. 
One needs to keep in mind that a tuple of $i_\mathrm{in}$ and $\mathrm{PA}_\mathrm{in}$ always refers to two potential physical configurations, which are mathematically represented by ($i_\mathrm{in}$, $\mathrm{PA}_\mathrm{in}$) and ($-i_\mathrm{in}$, $\mathrm{PA}_\mathrm{in}$).
Without further information on the inner disk orientation it is not possible to break this degeneracy.
This is considered when calculating misalignment angles as detailed in Sect.~\ref{sec:analysis_misalignments}.

\begin{figure*}
\resizebox{\hsize}{!}{\includegraphics{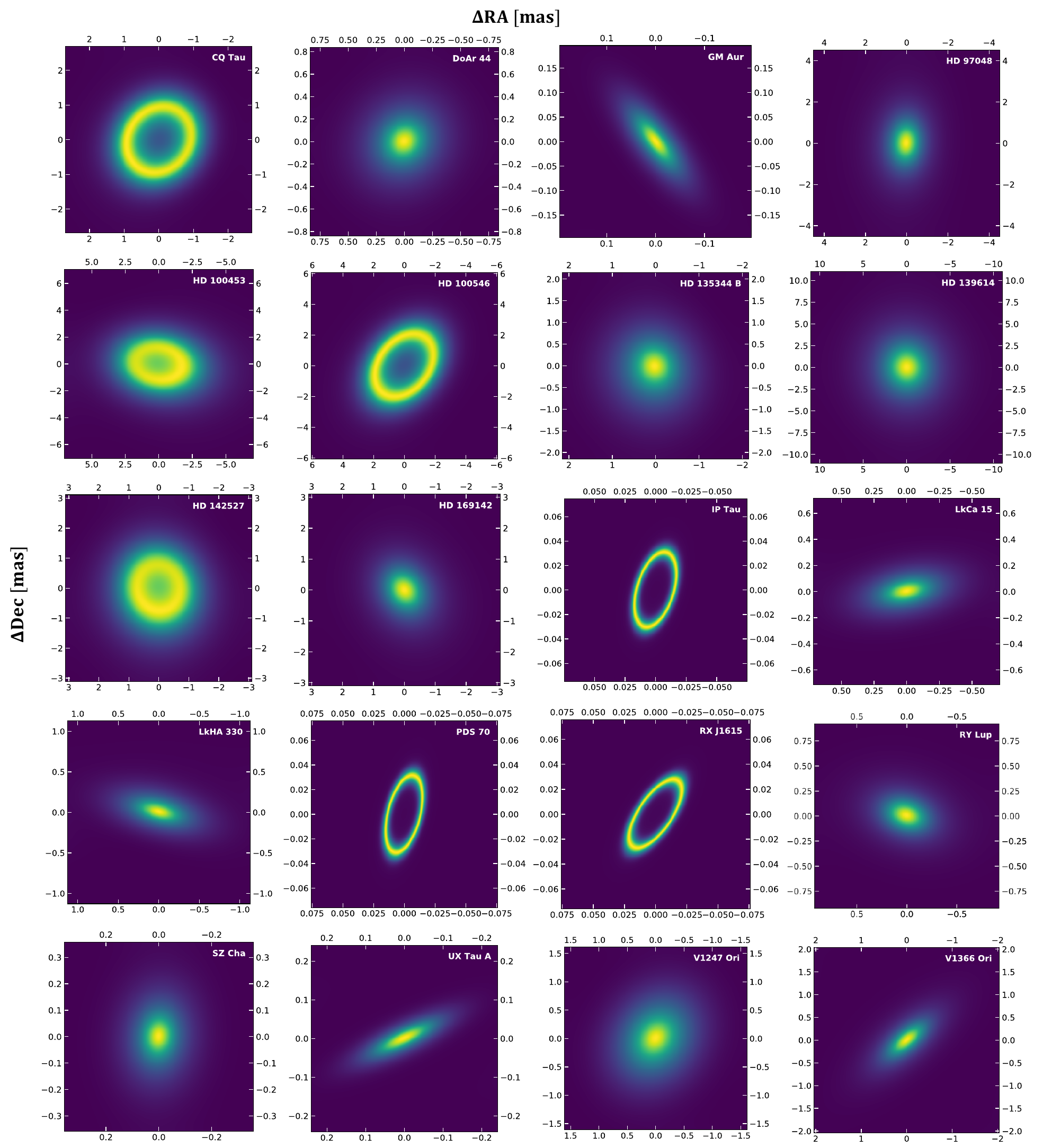}}
\caption{
Best-fit inner disk models from the GRAVITY observables.
In all images north points up and east toward the left.
}
\label{fig:gravity_fits_all}
\end{figure*}

The model fitting results are given in Table~~\ref{tbl:results_inner_disks_unconstrained}, with the best models shown in Fig.~\ref{fig:gravity_fits_all}. Fig.~\ref{fig:gravity_observables} shows the observations and best-fit models.

\begin{sidewaystable*}
\caption{
Results of the model fits to the GRAVITY data.
}
\label{tbl:results_inner_disks_unconstrained}
\def\arraystretch{1.2}
\centering
\begin{tabular}{@{}llllllllllll@{}}
\hline\hline
Target & $k_\mathrm{c}$ & $f_\mathrm{c}$ & $f_\mathrm{h}$ & $f_\mathrm{Lor}$ & $l_\mathrm{a}$ & $l_\mathrm{kr}$ & $\cos\left(i_\mathrm{in}\right)$ & $\mathrm{PA}_\mathrm{in}$ & $c_1$ & $s_1$ & $\chi^2_\mathrm{red}$\tablefootmark{a} \\    
& & & & & & & & (\degr) & & & \\
\hline
CQ~Tau & $-4.39^{+1.00}_{-1.16}$ & $0.80^{+0.03}_{-0.03}$ & $0.03^{+0.01}_{-0.01}$ & $0.73^{+0.25}_{-0.18}$ & $0.06^{+0.02}_{-0.02}$ & $-0.28^{+0.13}_{-0.23}$ & $0.87^{+0.02}_{-0.02}$ & $140.37^{+8.50}_{-6.53}$ & $0.09^{+0.03}_{-0.03}$ & $0.05^{+0.02}_{-0.02}$ & 3.58 \\
DoAr~44 & $-2.54^{+1.96}_{-1.65}$ & $0.70^{+0.01}_{-0.01}$ & $0.04^{+0.01}_{-0.01}$ & $0.74^{+0.18}_{-0.17}$ & $-0.44^{+0.06}_{-0.07}$ & $1.51^{+1.02}_{-1.00}$ & $0.90^{+0.07}_{-0.06}$ & $138.30^{+29.97}_{-14.63}$ & $-0.00^{+0.01}_{-0.01}$ & $-0.01^{+0.01}_{-0.02}$ & 2.27 \\
GM~Aur & $-1.61^{+1.10}_{-0.93}$ & $0.77^{+0.01}_{-0.01}$ & $0.05^{+0.01}_{-0.01}$ & $0.55^{+0.34}_{-0.31}$ & $-1.07^{+0.36}_{-0.33}$ & $0.70^{+2.15}_{-1.58}$ & $0.37^{+0.27}_{-0.39}$ & $36.92^{+21.81}_{-30.65}$ & $0.00^{+0.02}_{-0.02}$ & $-0.00^{+0.03}_{-0.03}$ & 10.25 \\
HD~100453 & $-5.30^{+0.41}_{-0.37}$ & $0.75^{+0.01}_{-0.01}$ & $0.01^{+0.01}_{-0.01}$ & $0.81^{+0.07}_{-0.07}$ & $0.48^{+0.01}_{-0.01}$ & $0.05^{+0.03}_{-0.04}$ & $0.69^{+0.01}_{-0.01}$ & $81.58^{+0.93}_{-0.92}$ & $-0.14^{+0.01}_{-0.01}$ & $-0.04^{+0.01}_{-0.01}$ & 2.06 \\
HD~100546 & $-1.79^{+0.99}_{-0.94}$ & $0.64^{+0.01}_{-0.01}$ & $0.13^{+0.01}_{-0.01}$ & $0.79^{+0.22}_{-0.15}$ & $0.42^{+0.02}_{-0.02}$ & $-0.29^{+0.08}_{-0.09}$ & $0.71^{+0.02}_{-0.02}$ & $140.65^{+2.89}_{-2.71}$ & $0.05^{+0.03}_{-0.03}$ & $0.03^{+0.01}_{-0.01}$ & 1.66 \\
HD~135344~B & $-4.59^{+0.96}_{-1.28}$ & $0.78^{+0.01}_{-0.01}$ & $0.02^{+0.01}_{-0.01}$ & $0.62^{+0.11}_{-0.13}$ & $-0.03^{+0.02}_{-0.02}$ & $1.46^{+1.09}_{-1.05}$ & $0.92^{+0.06}_{-0.05}$ & $14.45^{+21.39}_{-18.33}$ & $-0.10^{+0.06}_{-0.07}$ & $-0.17^{+0.04}_{-0.06}$ & 4.48 \\
HD~139614 & $-3.75^{+0.68}_{-0.58}$ & $0.55^{+0.01}_{-0.01}$ & $0.07^{+0.02}_{-0.02}$ & $0.85^{+0.18}_{-0.11}$ & $0.68^{+0.04}_{-0.05}$ & $1.75^{+0.90}_{-0.84}$ & $0.92^{+0.08}_{-0.05}$ & $6.79^{+48.64}_{-54.27}$ & $0.00^{+0.26}_{-0.26}$ & $0.01^{+0.24}_{-0.24}$ & 6.66 \\
HD~142527 & $-4.42^{+0.57}_{-0.52}$ & $0.78^{+0.01}_{-0.01}$ & $0.01^{+0.00}_{-0.01}$ & $0.83^{+0.21}_{-0.12}$ & $0.13^{+0.01}_{-0.01}$ & $0.05^{+0.09}_{-0.36}$ & $0.92^{+0.02}_{-0.02}$ & $15.44^{+6.52}_{-7.44}$ & $-0.12^{+0.01}_{-0.01}$ & $-0.05^{+0.02}_{-0.02}$ & 1.27 \\
HD~169142 & $-5.30^{+0.55}_{-1.81}$ & $0.55^{+0.01}_{-0.01}$ & $0.07^{+0.05}_{-0.03}$ & $0.99^{+0.02}_{-0.01}$ & $0.13^{+0.08}_{-0.13}$ & $2.00^{+0.74}_{-0.68}$ & $0.82^{+0.08}_{-0.08}$ & $31.55^{+14.76}_{-14.51}$ & $-0.09^{+0.04}_{-0.05}$ & $0.03^{+0.06}_{-0.05}$ & 11.00 \\
HD~97048 & $-5.25^{+0.51}_{-0.69}$ & $0.77^{+0.01}_{-0.01}$ & $0.01^{+0.01}_{-0.01}$ & $0.97^{+0.04}_{-0.02}$ & $0.29^{+0.02}_{-0.02}$ & $1.80^{+0.83}_{-0.81}$ & $0.68^{+0.03}_{-0.03}$ & $176.04^{+3.64}_{-3.67}$ & $0.02^{+0.03}_{-0.03}$ & $0.09^{+0.03}_{-0.03}$ & 2.98 \\
IP~Tau & $-0.71^{+0.85}_{-0.50}$ & $0.67^{+0.03}_{-0.03}$ & $0.08^{+0.00}_{-0.00}$ & $0.44^{+0.31}_{-0.37}$ & $-1.49^{+0.35}_{-0.45}$ & $-0.77^{+1.54}_{-2.45}$ & $0.47^{+0.33}_{-0.35}$ & $163.13^{+53.10}_{-60.20}$ & $0.00^{+0.03}_{-0.03}$ & $0.00^{+0.03}_{-0.03}$ & 2.28 \\
LkCa~15 & $-2.71^{+2.10}_{-1.86}$ & $0.65^{+0.03}_{-0.03}$ & $0.04^{+0.01}_{-0.01}$ & $0.64^{+0.33}_{-0.26}$ & $-0.51^{+0.14}_{-0.16}$ & $1.09^{+1.29}_{-1.29}$ & $0.48^{+0.30}_{-0.28}$ & $101.33^{+15.41}_{-18.24}$ & $-0.01^{+0.02}_{-0.02}$ & $0.02^{+0.03}_{-0.02}$ & 2.49 \\
LkH$\alpha$~330 & $-4.47^{+0.96}_{-1.20}$ & $0.77^{+0.01}_{-0.01}$ & $0.01^{+0.00}_{-0.01}$ & $0.62^{+0.10}_{-0.14}$ & $-0.31^{+0.03}_{-0.03}$ & $1.35^{+1.10}_{-1.15}$ & $0.45^{+0.31}_{-0.38}$ & $75.66^{+4.93}_{-8.68}$ & $-0.03^{+0.02}_{-0.01}$ & $0.09^{+0.04}_{-0.04}$ & 1.69 \\
PDS~70 & $-4.87^{+0.82}_{-1.25}$ & $0.43^{+0.04}_{-0.04}$ & $0.05^{+0.00}_{-0.00}$ & $0.43^{+0.31}_{-0.36}$ & $-1.49^{+0.36}_{-0.54}$ & $-0.76^{+1.52}_{-2.37}$ & $0.41^{+0.30}_{-0.38}$ & $165.85^{+50.64}_{-49.51}$ & $0.00^{+0.08}_{-0.08}$ & $0.00^{+0.06}_{-0.07}$ & 4.85 \\
RX~J1615 & $-1.79^{+1.35}_{-1.12}$ & $0.46^{+0.01}_{-0.01}$ & $0.11^{+0.01}_{-0.00}$ & $0.43^{+0.31}_{-0.37}$ & $-1.49^{+0.36}_{-0.50}$ & $-0.72^{+1.57}_{-2.41}$ & $0.41^{+0.30}_{-0.38}$ & $145.64^{+55.24}_{-51.00}$ & $0.00^{+0.05}_{-0.05}$ & $0.00^{+0.05}_{-0.05}$ & 15.37 \\
RY~Lup & $-5.61^{+0.28}_{-0.54}$ & $0.68^{+0.02}_{-0.02}$ & $0.04^{+0.00}_{-0.00}$ & $0.98^{+0.03}_{-0.01}$ & $-0.40^{+0.01}_{-0.01}$ & $1.91^{+0.76}_{-0.74}$ & $0.70^{+0.06}_{-0.06}$ & $71.66^{+2.30}_{-2.71}$ & $0.01^{+0.01}_{-0.01}$ & $0.05^{+0.01}_{-0.01}$ & 0.56 \\
SZ~Cha & $-3.33^{+1.85}_{-2.11}$ & $0.72^{+0.04}_{-0.04}$ & $0.01^{+0.00}_{-0.01}$ & $0.68^{+0.28}_{-0.22}$ & $-0.81^{+0.15}_{-0.19}$ & $1.32^{+1.11}_{-1.13}$ & $0.73^{+0.23}_{-0.19}$ & $173.77^{+40.39}_{-30.32}$ & $0.00^{+0.04}_{-0.04}$ & $-0.00^{+0.03}_{-0.03}$ & 1.91 \\
UX~Tau~A & $-2.24^{+1.88}_{-1.44}$ & $0.88^{+0.03}_{-0.03}$ & $0.02^{+0.00}_{-0.00}$ & $0.56^{+0.34}_{-0.31}$ & $-0.98^{+0.34}_{-0.37}$ & $0.54^{+2.01}_{-1.65}$ & $0.28^{+0.20}_{-0.26}$ & $115.67^{+18.07}_{-14.90}$ & $0.00^{+0.02}_{-0.02}$ & $-0.00^{+0.02}_{-0.01}$ & 1.65 \\
V1247~Ori & $-5.12^{+0.62}_{-1.02}$ & $0.79^{+0.01}_{-0.01}$ & $0.01^{+0.01}_{-0.01}$ & $0.42^{+0.10}_{-0.16}$ & $-0.16^{+0.02}_{-0.02}$ & $1.15^{+1.19}_{-1.25}$ & $0.81^{+0.03}_{-0.04}$ & $145.09^{+5.58}_{-5.58}$ & $-0.02^{+0.01}_{-0.01}$ & $-0.02^{+0.01}_{-0.01}$ & 7.00 \\
V1366~Ori & $-0.11^{+0.18}_{-0.08}$ & $0.82^{+0.01}_{-0.01}$ & $0.15^{+0.01}_{-0.01}$ & $0.87^{+0.14}_{-0.09}$ & $-0.06^{+0.03}_{-0.03}$ & $1.53^{+1.06}_{-1.02}$ & $0.44^{+0.03}_{-0.03}$ & $130.31^{+2.11}_{-2.20}$ & $0.00^{+0.01}_{-0.01}$ & $0.03^{+0.00}_{-0.00}$ & 8.61 \\
\hline
\end{tabular}
\end{sidewaystable*}

\begin{figure*}
\centering
\resizebox{0.95\hsize}{!}{\includegraphics{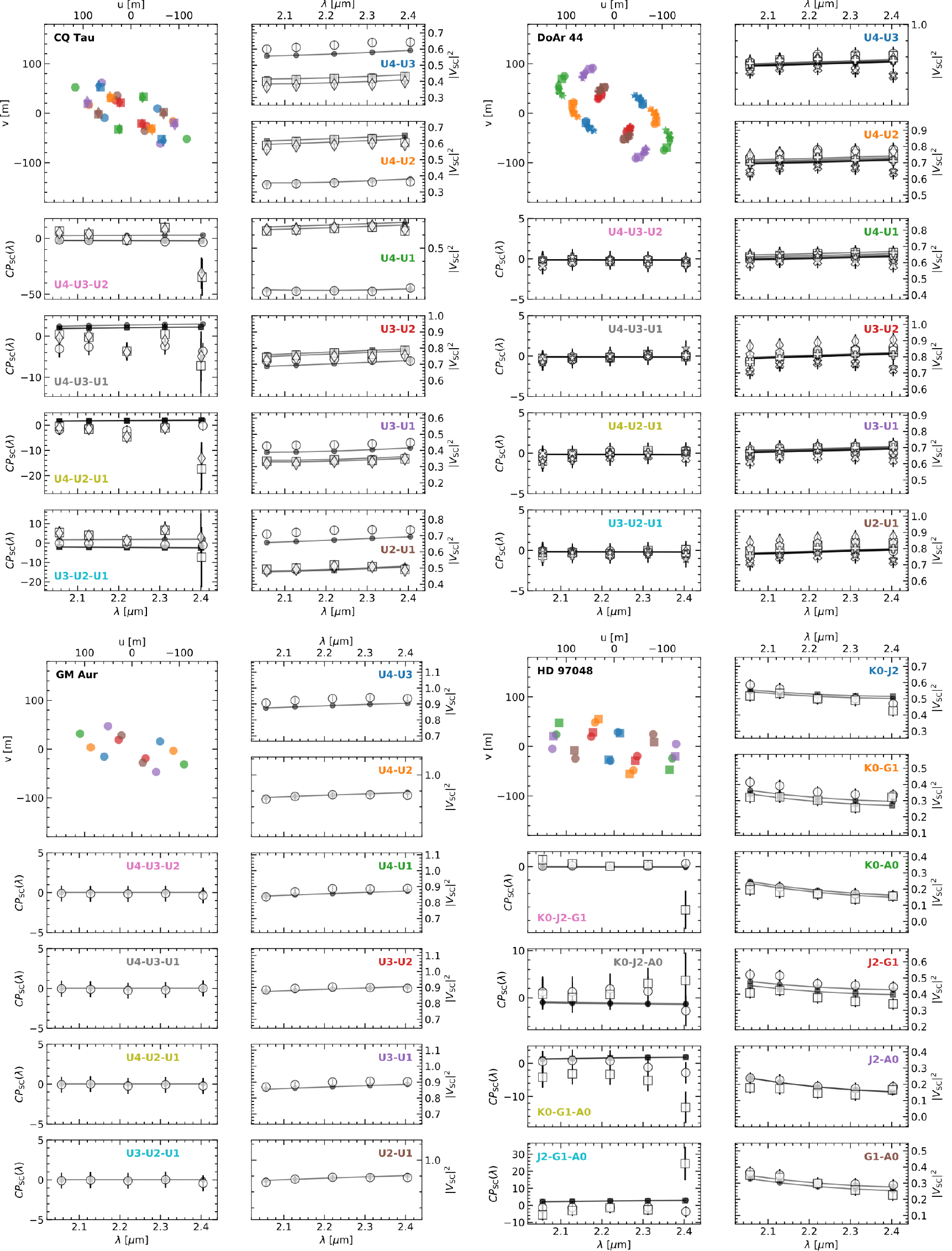}}
\caption{
GRAVITY observables and model fits.
In the upper left of each panel, the $(u,v)$ plane coverage is shown.
Each symbol corresponds to an individual exposures and the colors refer to the interferometric baselines.
The other panels of the plot show the squared visibilities (right) and closure phases (left) for each baseline or triplet as a function of wavelength.
The white markers with black contours represent the observational data and the black lines correspond to our best-fit model.
The black markers on top of the lines indicate which curve corresponds to which exposure.
}
\label{fig:gravity_observables}
\end{figure*}
\setcounter{figure}{\the\numexpr\value{figure}-1\relax}
\begin{figure*}
\captionsetup*{labelsep=space, list=no}
\resizebox{\hsize}{!}{\includegraphics{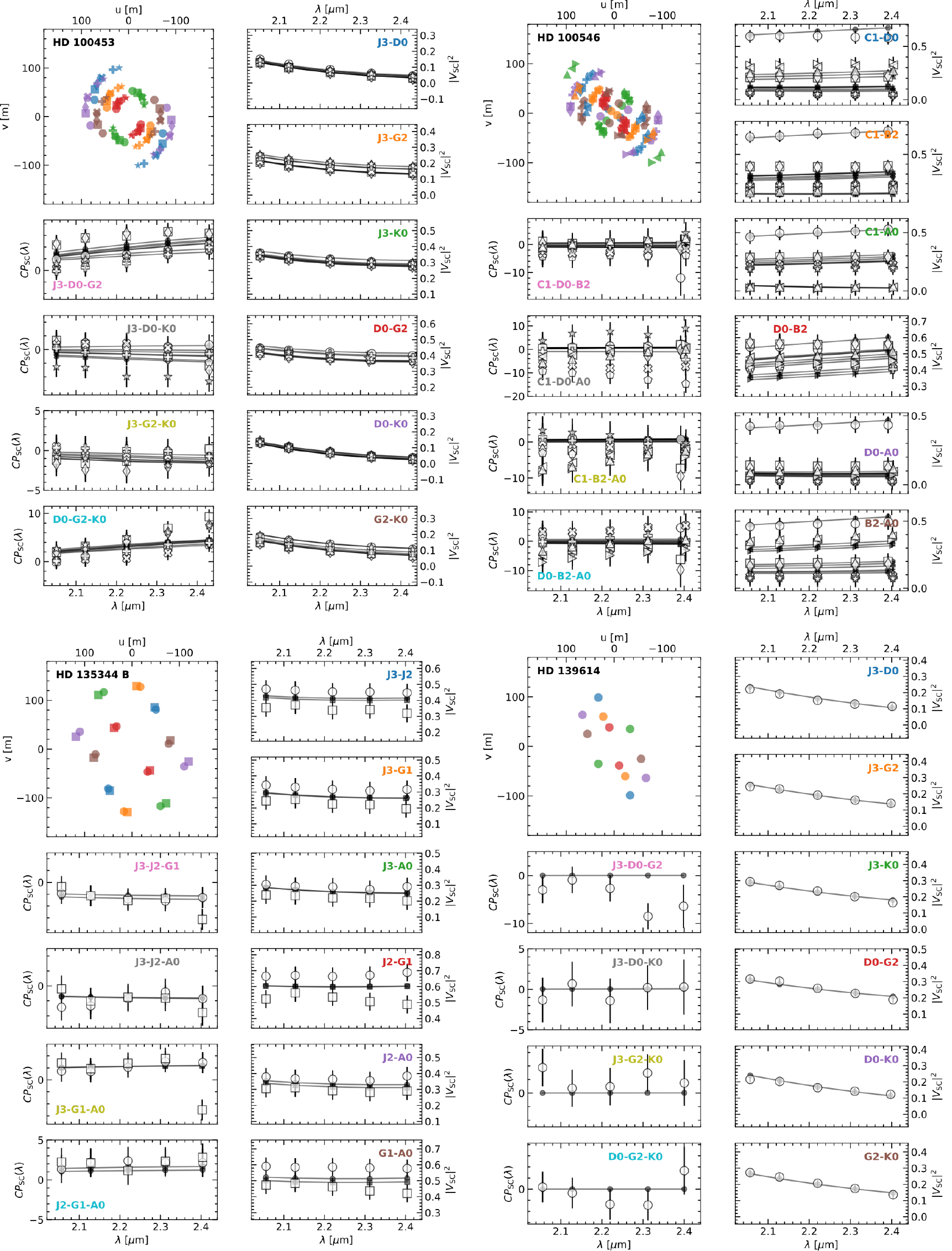}}
\caption
{
(continued).
}
\end{figure*}
\setcounter{figure}{\the\numexpr\value{figure}-1\relax}
\begin{figure*}
\captionsetup*{labelsep=space, list=no}
\resizebox{\hsize}{!}{\includegraphics{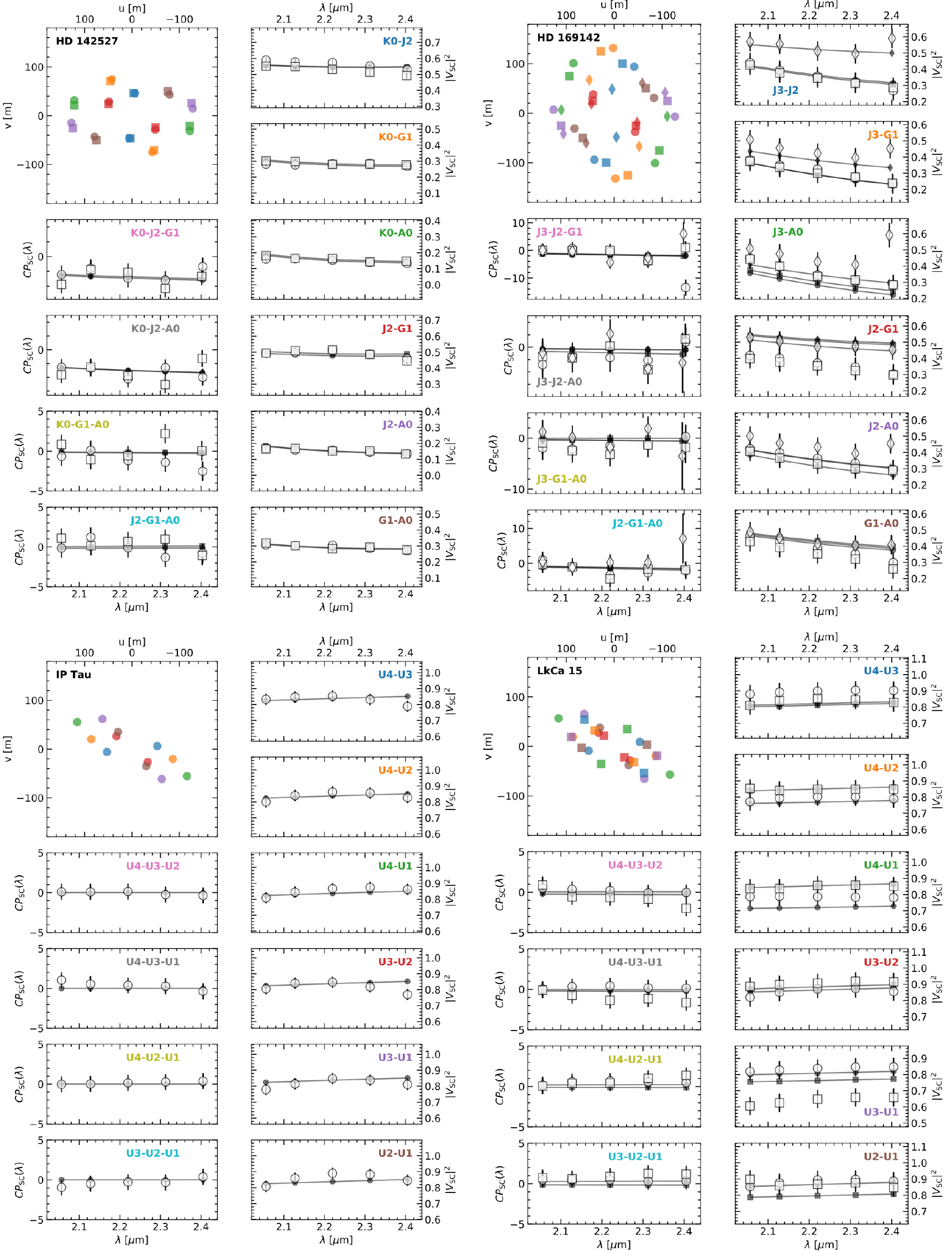}}
\caption
{
(continued).
}
\end{figure*}
\setcounter{figure}{\the\numexpr\value{figure}-1\relax}
\begin{figure*}
\captionsetup*{labelsep=space, list=no}
\resizebox{\hsize}{!}{\includegraphics{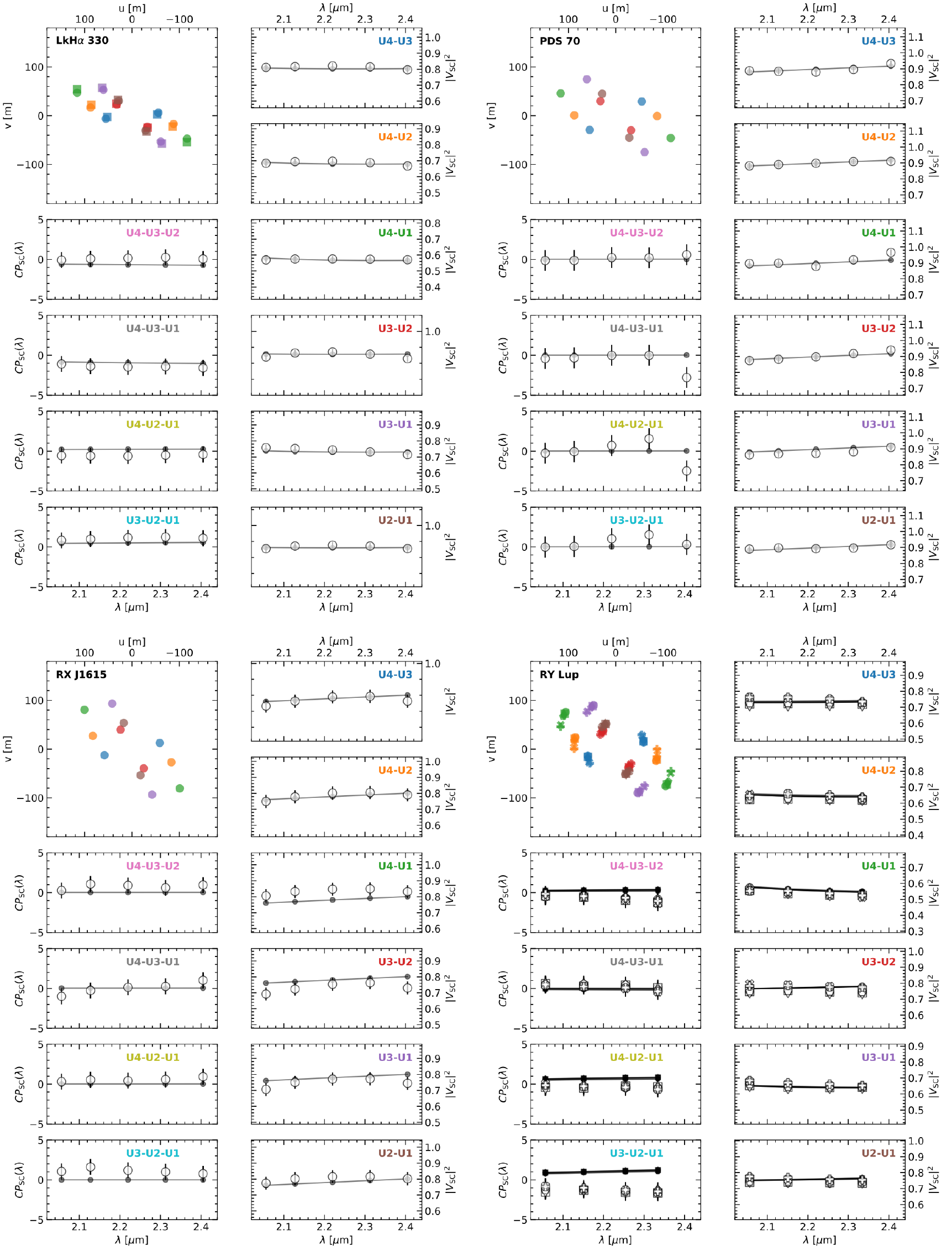}}
\caption
{
(continued).
}
\end{figure*}
\setcounter{figure}{\the\numexpr\value{figure}-1\relax}
\begin{figure*}
\captionsetup*{labelsep=space, list=no}
\resizebox{\hsize}{!}{\includegraphics{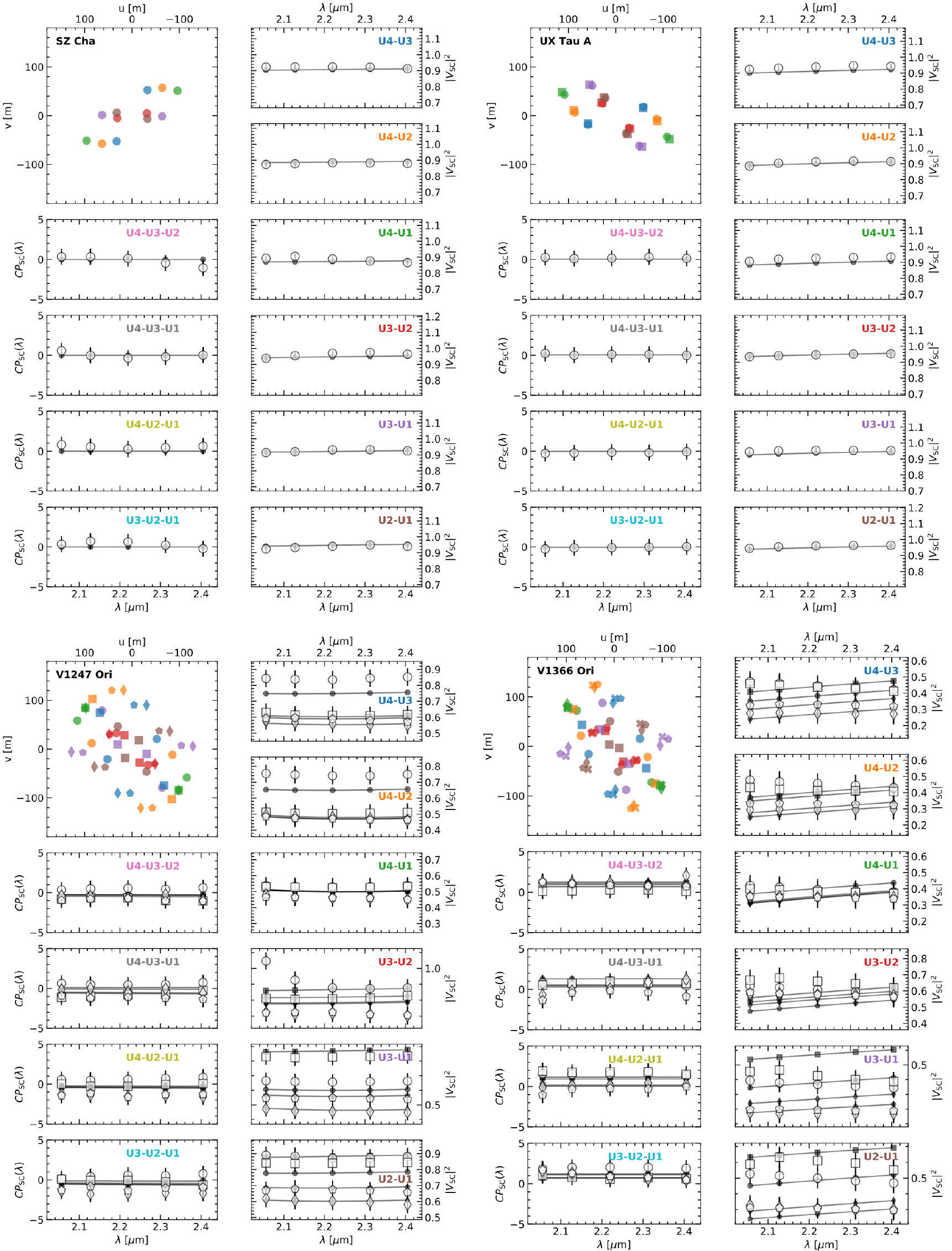}}
\caption
{
(continued).
}
\end{figure*}

\section{ALMA data and best fit parameters}
\label{sec:alma_data_and_residuals}

\begin{figure*}
\resizebox{\hsize}{!}{\includegraphics{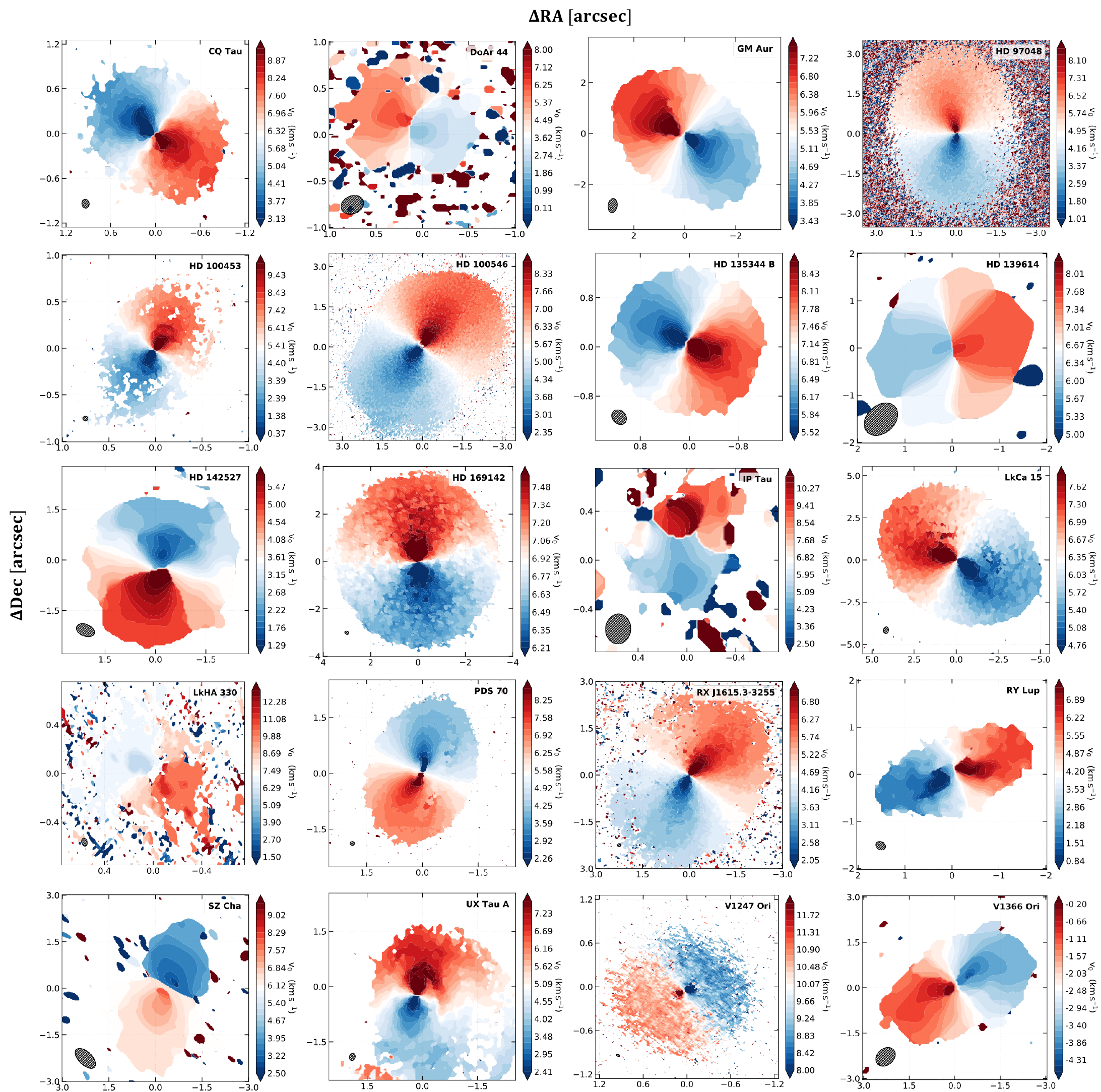}}
\caption{
ALMA line data of our targets.
We present the quadratically collapsed moment maps that were created with \texttt{bettermoments} by quadratic collapsing of the frequency channels.
The ellipse in the lower left of each panel indicates the beam size of the observation.
For all images north points up and east to the left.
}
\label{fig:alma_data_all}
\end{figure*}

We present the velocity maps that were generated with the \texttt{bettermoments} tool in Fig.~\ref{fig:alma_data_all}. The best-fit parameters of our Keplerian disk models as introduced in Sect.~\ref{sec:analysis_alma} are presented in Table~\ref{tbl:results_outer_disk_fits}. The corresponding residuals of the data and the best-fit \texttt{eddy} models will be discussed in a forthcoming publication (Wölfer et al. in prep.).

\begin{table*}
\caption{
Results from the outer disk model fits of the ALMA line data cubes.
}
\label{tbl:results_outer_disk_fits}
\def\arraystretch{1.2}
\centering
\begin{tabular}{@{}llllll@{}}
\hline\hline
Target & $i_\mathrm{out}$ & $\mathrm{PA}_\mathrm{out}$ & $v_\mathrm{LSR}$ & $z_0$ & $\psi$ \\    
& (\degr) & (\degr) & m\,s$^{-1}$ & &\\
\hline
CQ~Tau & $32.25^{+0.12}_{-0.12}$ & $233.92^{+0.19}_{-0.19}$ & $6169.02^{+5.08}_{-5.12}$ & $0.17^{+0.06}_{-0.07}$ & $3.32^{+0.95}_{-0.95}$ \\
DoAr~44 & $23.19^{+0.29}_{-0.31}$ & $65.63^{+0.22}_{-0.21}$ & $4286.37^{+3.68}_{-3.79}$ & $0.13^{+0.01}_{-0.01}$ & $0.03^{+0.03}_{-0.06}$ \\
GM~Aur & $52.25^{+0.60}_{-0.63}$ & $57.18^{+0.58}_{-0.58}$ & $5613.20^{+8.65}_{-8.82}$ & $0.00^{+0.01}_{-0.03}$ & $0.39^{+0.52}_{-1.12}$ \\
HD~100453 & $33.81^{+0.20}_{-0.20}$ & $324.35^{+0.28}_{-0.27}$ & $5191.11^{+11.22}_{-11.54}$ & $0.00^{+0.01}_{-0.03}$ & $3.92^{+2.01}_{-0.83}$ \\
HD~100546 & $40.23^{+0.11}_{-0.11}$ & $324.26^{+0.11}_{-0.11}$ & $5639.70^{+2.82}_{-2.72}$ & $0.01^{+0.00}_{-0.02}$ & $0.01^{+0.12}_{-1.17}$ \\
HD~135344~B & $16.74^{+0.04}_{-0.04}$ & $241.92^{+0.12}_{-0.12}$ & $7080.68^{+1.68}_{-1.60}$ & $0.12^{+0.03}_{-0.02}$ & $4.93^{+0.33}_{-0.16}$ \\
HD~139614 & $17.93^{+0.24}_{-0.27}$ & $276.64^{+0.19}_{-0.19}$ & $6759.12^{+1.69}_{-1.70}$ & $0.54^{+0.02}_{-0.02}$ & $0.01^{+0.02}_{-0.05}$ \\
HD~142527 & $38.21^{+0.34}_{-0.38}$ & $162.72^{+0.40}_{-0.40}$ & $3710.47^{+8.39}_{-8.41}$ & $0.12^{+0.02}_{-0.05}$ & $0.00^{+1.09}_{-1.50}$ \\
HD~169142 & $12.45^{+0.35}_{-0.38}$ & $5.88^{+0.23}_{-0.23}$ & $6883.12^{+1.57}_{-1.65}$ & $0.31^{+0.08}_{-0.07}$ & $1.10^{+0.08}_{-0.14}$ \\
HD~97048 & $45.34^{+0.24}_{-0.24}$ & $2.84^{+0.16}_{-0.15}$ & $4750.24^{+4.26}_{-4.26}$ & $0.17^{+0.01}_{-0.01}$ & $1.32^{+0.06}_{-0.07}$ \\
IP~Tau & $44.99^{+4.89}_{-5.21}$ & $345.79^{+6.80}_{-6.39}$ & $6825.76^{+208.01}_{-215.39}$ & $4.92^{+1.34}_{-0.78}$ & $4.03^{+0.59}_{-0.58}$ \\
LkCa~15 & $43.95^{+0.44}_{-0.47}$ & $63.22^{+0.31}_{-0.30}$ & $6280.38^{+4.70}_{-4.84}$ & $0.08^{+0.05}_{-0.06}$ & $1.08^{+0.31}_{-0.49}$ \\
LkH$\alpha$~330 & $20.95^{+0.28}_{-0.29}$ & $234.74^{+0.85}_{-0.84}$ & $8317.99^{+21.14}_{-20.77}$ & $0.27^{+0.93}_{-1.99}$ & $4.41^{+0.97}_{-0.48}$ \\
PDS~70 & $50.19^{+0.24}_{-0.25}$ & $160.21^{+0.20}_{-0.20}$ & $5475.55^{+4.95}_{-5.01}$ & $0.00^{+0.00}_{-0.00}$ & $4.91^{+1.75}_{-2.33}$ \\
RX~J1615 & $47.15^{+0.26}_{-0.26}$ & $325.03^{+0.15}_{-0.16}$ & $4740.76^{+2.95}_{-2.99}$ & $0.16^{+0.01}_{-0.01}$ & $1.35^{+0.07}_{-0.07}$ \\
RY~Lup & $56.44^{+0.51}_{-0.53}$ & $287.47^{+0.30}_{-0.28}$ & $3911.05^{+8.42}_{-8.99}$ & $0.09^{+0.02}_{-0.07}$ & $0.00^{+0.10}_{-1.46}$ \\
SZ~Cha & $46.84^{+0.63}_{-0.62}$ & $156.80^{+0.89}_{-0.88}$ & $5167.57^{+15.99}_{-16.25}$ & $0.00^{+0.01}_{-0.02}$ & $0.64^{+0.97}_{-1.75}$ \\
UX~Tau~A & $37.95^{+0.44}_{-0.49}$ & $346.95^{+0.42}_{-0.42}$ & $5447.99^{+8.94}_{-8.92}$ & $0.18^{+0.03}_{-0.03}$ & $1.77^{+0.27}_{-0.34}$ \\
V1247~Ori & $24.94^{+1.13}_{-1.25}$ & $124.42^{+1.00}_{-1.00}$ & $9766.64^{+14.25}_{-14.56}$ & $0.26^{+0.05}_{-0.05}$ & $0.26^{+0.08}_{-0.07}$ \\
V1366~Ori & $44.93^{+5.69}_{-7.00}$ & $117.54^{+3.50}_{-3.47}$ & $-2323.36^{+70.01}_{-70.91}$ & $0.31^{+0.16}_{-0.13}$ & $0.23^{+0.24}_{-0.47}$ \\
\hline
\end{tabular}
\tablefoot{
The uncertainties reported in this table represent the statistical errors from the marginalized posterior distributions of the MCMC fit.
These do not include the propagated mass uncertainties that have to be considered when calculating the disk inclinations, $i_\mathrm{out}$.
}
\end{table*}

\section{Misalignment angles}
\label{sec:misalignment_angles}
To quantify the alikeness of the true and simulated  misalignment distributions, we used Kolmogorov–Smirnov statistics.  The Kolmogorov–Smirnov distance
\begin{equation}
    D(F,G)=\underset{x}{\max}\left[\lvert F(x)-G(x)\rvert\right]
\end{equation}
measures the parameter $x$ that maximizes the difference between two empirical distribution functions $F$ and $G$, representing the samples $X=\{X_1,\dots X_n\}\sim F$ and $Y=\{Y_1,\dots Y_m\}\sim G$.
Accordingly, $D(F,G)\in[0,1]$ with $D(F,G)=0$ indicates that both samples $X$ and $Y$ originate from the same underlying distribution.
The stronger $D(F,G)$ deviates from zero, the higher the likeliness that $X$ and $Y$ are not drawn from the same distribution.

This statistical framework can be applied to our problem as follows.
Let $\Delta\theta_1^\mathrm{sim}\sim F_1$, $\Delta\theta_2^\mathrm{sim}\sim F_2$, $\Delta\theta_1\sim G_1$, and $\Delta\theta_2\sim G_2$ the (simulated) posterior distributions and corresponding empirical distribution functions. We define the final Kolmogorov–Smirnov distance between simulated and measured parameters as
\begin{equation}
\begin{aligned}
D_\mathrm{KS}:=\min\{&\max\left[D(F_1,G_1),D(F_2,G_2)\right],\\ &\max\left[D(F_1,G_2),D(F_2,G_1)\right]\}\in[0,1]\;.
\end{aligned}
\end{equation}

It is insufficient if only one of the misalignment distributions $\Delta\theta_1$ or $\Delta\theta_2$ agrees with its simulated counterpart, that is, for perfect alignment it is not sufficient if for instance  $D(F_1,G_1)=0$ and $D(F_2,G_2)=1$.
For that reason, we calculate the maximum of both corresponding Kolmogorov–Smirnov distances.

\begin{figure*}
\resizebox{\hsize}{!}{\includegraphics{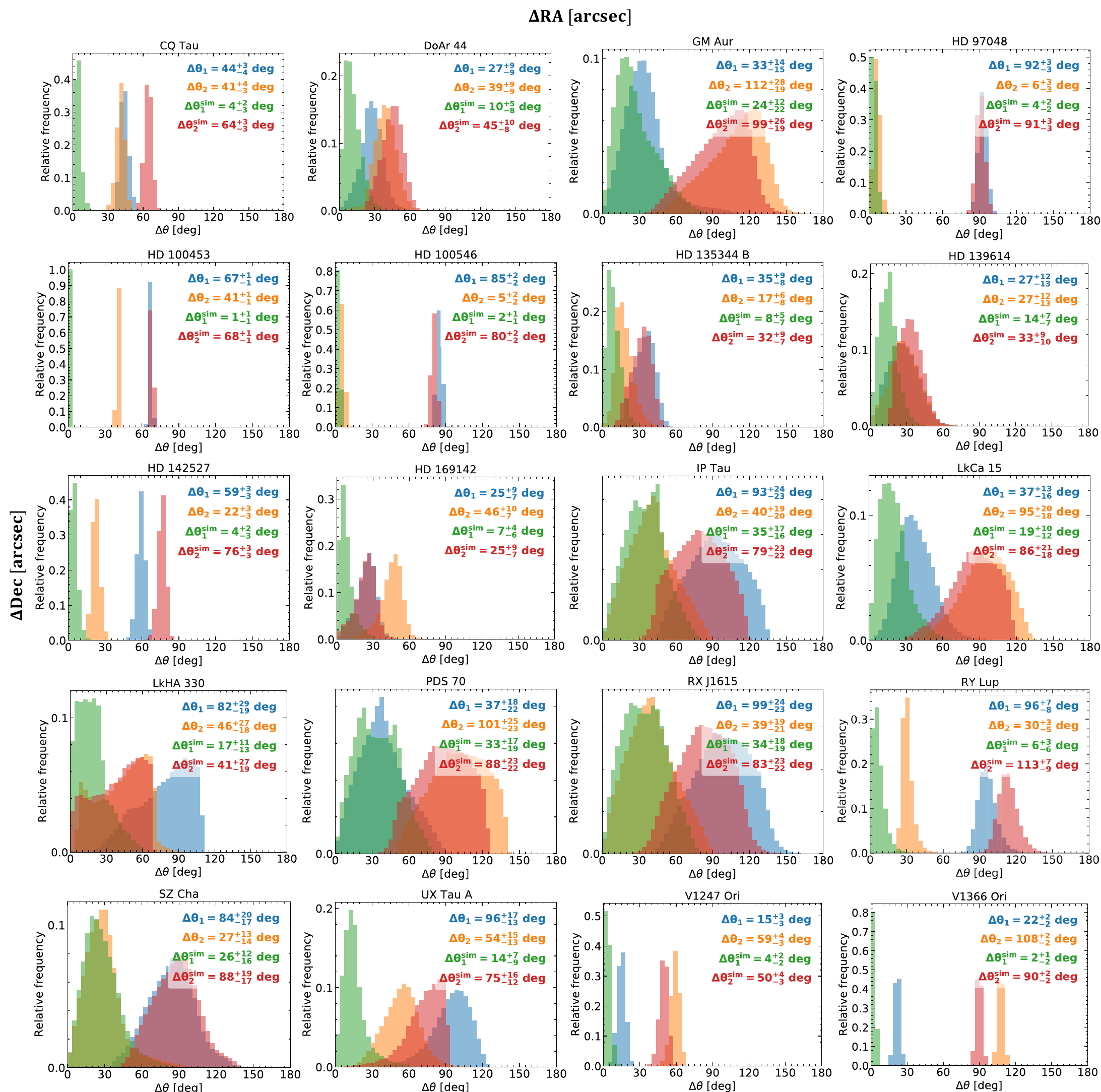}}
\caption{
Posterior distributions of the derived misalignment angles. We also present the simulated distributions for a perfectly aligned disk geometry. Good agreement of the true and simulated distributions indicates that inner and outer disk are well aligned. Deviations from this agreement are signs of potential misalignment.
}
\label{fig:misalignment_posterior_distributions}
\end{figure*}

Since the true orientation of the disk is unknown, we have to test both potential combinations, that is, $\Delta\theta_1^\mathrm{sim}$ might either correspond to $\Delta\theta_1$ or $\Delta\theta_2$ and $\Delta\theta_2^\mathrm{sim}$ to the respective remaining parameter. The minimum of these potential permutations allows to find the orientation that prefers our null hypothesis most. $D_\mathrm{KS}$ behaves as $D(F,G)$ and a value close to zero can be interpreted as a confirmation of our null hypothesis, whereas a deviation from zero indicates a misalignment between inner and outer disk components. The magnitude of the deviation from zero further indicates the significance at which the null hypothesis must be rejected. The posterior distributions of our derived misalignment angles are presented in Fig.~\ref{fig:misalignment_posterior_distributions}. The figure also shows the misalignment distributions that were simulated for a perfect alignment between inner and outer disk. The more both distributions deviate the higher the likelihood that our null hypothesis (i.e., the inner and outer disks are well aligned) needs to be rejected. The figure can be used to understand the associated values of $D_\mathrm{KS}$ that are presented in Table~\ref{tbl:results_misalignments}.

\section{Scattered-light images}
\label{sec:scattered_light_data}

\begin{figure*}
\resizebox{\hsize}{!}{\includegraphics{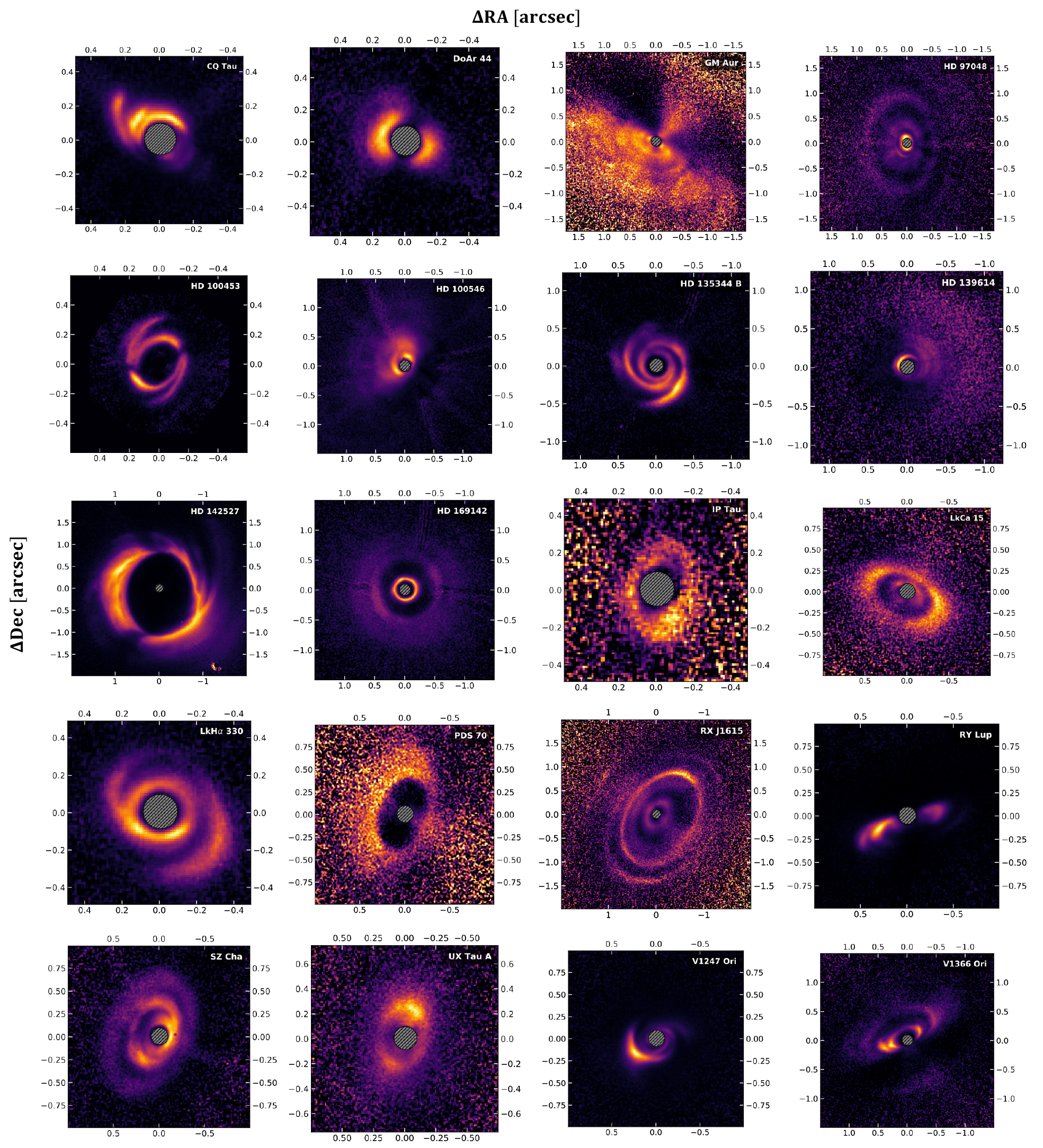}}
\caption{
SPHERE scattered light images.
We present the $Q_\phi$ images that are scaled with $r^2$.
The gray hatched circles indicate the size of the applied coronagraph.
In all frames north points up and east to the left.
}
\label{fig:sphere_scattered_light_all}
\end{figure*}

We present the corresponding SPHERE scattered light imagery in Fig.~\ref{fig:sphere_scattered_light_all}. The figure shows the polarized $Q_\phi$ images that were obtained with the polarimetric modes of SPHERE \citep[][]{deBoer2020,schmid2018}. The data reduction was performed with IRDAP \citep[][]{vanHolstein2020}. The resulting images were scaled with $r^2$ to account for the radial decrease in scattered light intensity as a function of physical separation from the star. These images were previously discussed in \cite{thalmann2016,ginski2016,stolker2016,avenhaus2017, pohl2017, benisty2017,keppler2018, avenhaus2018, casassus2018,langlois2018,muro_arena2020, menard2020, deBoer2020} and some will be presented in forthcoming papers (Benisty et al. in prep,  Ginski et al. in prep., Kraus et al. in prep, Pinilla et al. in prep.).


\end{appendix}

\end{document}